\documentclass[review, 3p, sort&compress]{elsarticle}

\usepackage{amsfonts}
\usepackage{graphicx}
\usepackage{epstopdf}
\usepackage{mhchem}
\usepackage{fullpage}
\usepackage{color}
\usepackage{subfigure}
\usepackage{threeparttable}
\usepackage{multirow}
\usepackage{url}

\usepackage[normalem]{ulem}

\graphicspath{{ink/},{figs/}}

\makeatletter
\providecommand*{\diff}%
    {\@ifnextchar^{\DIfF}{\DIfF^{}}}
\def\DIfF^#1{%
    \mathop{\mathrm{\mathstrut d}}%
        \nolimits^{#1}\gobblespace}
\def\gobblespace{%
    \futurelet\diffarg\opspace}
\def\opspace{%
    \let\DiffSpace\!%
    \ifx\diffarg(%
        \let\DiffSpace\relax
    \else
        \ifx\diffarg[%
            \let\DiffSpace\relax
    \else
        \ifx\diffarg\{%
            \let\DiffSpace\relax
        \fi\fi\fi\DiffSpace}

\providecommand*{\deriv}[3][]{%
    \frac{\diff^{#1}#2}{\diff #3^{#1}}}

\providecommand*{\pderiv}[3][]{%
\frac{\partial^{#1}#2}%
{\partial #3^{#1}}}

\newcommand{\tten}[1]{\times 10^{#1}}
\newcommand{\Ord}[1]{\mathcal{O}\left(#1\right)}

\newcommand{\e}{\mathrm{e}}

\newcommand{\etal}{\emph{et al.~}}
\newcommand{\unit}[1]{$^{#1}$}

\newenvironment{subeq}[1]{
\begin{subequations}
#1
\end{subequations}
}

\newcommand{\mn}[3]{#1_{#2,\text{{#3}}}}
\newcommand{\mnn}[2]{#1_{\text{{#2}}}}

\newcommand{\clp}{\mn{c}{L}{p}}
\newcommand{\cln}{\mn{c}{L}{n}}
\newcommand{\cls}{\mn{c}{L}{s}}
\newcommand{\cli}{\mn{c}{L}{i}}

\newcommand{\clihat}{\mn{\hat{c}}{L}{i}}

\newcommand{\clibar}{\mn{\bar{c}}{L}{i}}

\newcommand{\csp}{\mn{c}{a}{p}}
\newcommand{\csn}{\mn{c}{a}{n}}
\newcommand{\csi}{\mn{c}{a}{i}}

\newcommand{\csphat}{\mn{\hat{c}}{a}{p}}
\newcommand{\csnhat}{\mn{\hat{c}}{a}{n}}
\newcommand{\csihat}{\mn{\hat{c}}{a}{i}}

\newcommand{\csimax}{\mn{c}{a}{i}^\text{max}}
\newcommand{\csnmax}{\mn{c}{a}{n}^\text{max}}
\newcommand{\cspmax}{\mn{c}{a}{p}^\text{max}}

\newcommand{\yi}{\mnn{y}{i}}
\newcommand{\yp}{\mnn{y}{p}}
\newcommand{\yn}{\mnn{y}{n}}

\newcommand{\phiep}{\mn{\phi}{e}{p}}
\newcommand{\phien}{\mn{\phi}{e}{n}}
\newcommand{\phies}{\mn{\phi}{e}{s}}
\newcommand{\phiei}{\mn{\phi}{e}{i}}

\newcommand{\phisp}{\mn{\phi}{a}{p}}
\newcommand{\phisn}{\mn{\phi}{a}{n}}
\newcommand{\phisi}{\mn{\phi}{a}{i}}

\newcommand{\phie}{\mnn{\phi}{e}}

\newcommand{\Phiep}{\mn{\Phi}{e}{p}}
\newcommand{\Phien}{\mn{\Phi}{e}{n}}
\newcommand{\Phies}{\mn{\Phi}{e}{s}}
\newcommand{\Phiei}{\mn{\Phi}{e}{i}}

\newcommand{\Phiephat}{\mn{\hat\Phi}{e}{p}}
\newcommand{\Phienhat}{\mn{\hat\Phi}{e}{n}}

\newcommand{\Phieihat}{\mn{\hat\Phi}{e}{i}}

\newcommand{\Phiepbar}{\mn{\bar\Phi}{e}{p}}
\newcommand{\Phienbar}{\mn{\bar\Phi}{e}{n}}

\newcommand{\Phieibar}{\mn{\bar\Phi}{e}{i}}

\newcommand{\Phisp}{\mn{\Phi}{a}{p}}
\newcommand{\Phisn}{\mn{\Phi}{a}{n}}
\newcommand{\Phisi}{\mn{\Phi}{a}{i}}
\newcommand{\Phisphat}{\mn{\hat\Phi}{a}{p}}
\newcommand{\Phisnhat}{\mn{\hat\Phi}{a}{n}}
\newcommand{\Phisihat}{\mn{\hat\Phi}{a}{i}}

\newcommand{\etai}{\mnn{\eta}{i}}

\newcommand{\etaihat}{\mnn{\hat{\eta}}{i}}

\newcommand{\Up}{\mnn{\mathcal{U}}{p}}
\newcommand{\Un}{\mnn{\mathcal{U}}{n}}
\newcommand{\Ui}{\mnn{\mathcal{U}}{i}}

\newcommand{\Vi}{\mnn{\mathcal{V}}{i}}

\newcommand{\gnbar}{\mnn{g}{n}}%{\mnn{\bar{g}}{n}}
\newcommand{\gpbar}{\mnn{g}{p}}%{\mnn{\bar{g}}{p}}
\newcommand{\gibar}{\mnn{g}{i}}%{\mnn{\bar{g}}{i}}

\newcommand{\iep}{\mn{i}{e}{p}}
\newcommand{\ien}{\mn{i}{e}{n}}
\newcommand{\ies}{\mn{i}{e}{s}}
\newcommand{\iei}{\mn{i}{e}{i}}

\newcommand{\isp}{\mn{i}{a}{p}}

\newcommand{\isi}{\mn{i}{a}{i}}

\newcommand{\isphat}{\mn{\hat{\imath}}{a}{p}}
\newcommand{\isnhat}{\mn{\hat{\imath}}{a}{n}}
\newcommand{\isihat}{\mn{\hat{\imath}}{a}{i}}

\newcommand{\iephat}{\mn{\hat{\imath}}{e}{p}}
\newcommand{\ienhat}{\mn{\hat{\imath}}{e}{n}}
\newcommand{\ieshat}{\mn{\hat{\imath}}{e}{s}}
\newcommand{\ieihat}{\mn{\hat{\imath}}{e}{i}}

\newcommand{\I}{\mathcal{I}}
\newcommand{\Ihat}{\hat{\mathcal{I}}}

\newcommand{\Tp}{\mnn{T}{p}}
\newcommand{\Tn}{\mnn{T}{n}}
\newcommand{\Ts}{\mnn{T}{s}}
\newcommand{\Ti}{\mnn{T}{i}}

\newcommand{\Rcpp}{\mnn{\rho}{p}\mnn{\mathsf{c}}{p}}

\newcommand{\Rcps}{\mnn{\rho}{s}\mnn{\mathsf{c}}{s}}
\newcommand{\Rcpi}{\mnn{\rho}{i}\mnn{\mathsf{c}}{i}}

\newcommand{\Rcpe}{\mnn{\rho}{e}\mnn{\mathsf{c}}{e}}
\newcommand{\Rcpsoli}{\mn{\rho}{\text{solid}}{i}\mn{\mathsf{c}}{\text{solid}}{i}}

\newcommand{\Ke}{k_e}

\newcommand{\Ksoli}{\mn{k}{\text{solid}}{i}}

\newcommand{\kp}{\mnn{k}{p}}
\newcommand{\kn}{\mnn{k}{n}}
\newcommand{\ks}{\mnn{k}{s}}
\newcommand{\ki}{\mnn{k}{i}}

\newcommand{\qpa}{\mn{q}{a}{p}}
\newcommand{\qna}{\mn{q}{a}{n}}
\newcommand{\qia}{\mn{q}{a}{i}}
\newcommand{\qpe}{\mn{q}{e}{p}}
\newcommand{\qne}{\mn{q}{e}{n}}
\newcommand{\qie}{\mn{q}{e}{i}}
\newcommand{\qpr}{\mn{q}{r}{p}}
\newcommand{\qnr}{\mn{q}{r}{n}}
\newcommand{\qir}{\mn{q}{r}{i}}
\newcommand{\qse}{\mn{q}{e}{s}}

 \newcommand{\hp}{\mnn{h}{p}}
 \newcommand{\hn}{\mnn{h}{n}}
 
 \newcommand{\hi}{\mnn{h}{i}}

\newcommand{\vrcn}{\mnn{\varrho}{n}}
\newcommand{\vrcs}{\mnn{\varrho}{s}}
\newcommand{\vrci}{\mnn{\varrho}{i}}

\newcommand{\vKn}{\mnn{\mathcal{K}}{n}}
\newcommand{\vKs}{\mnn{\mathcal{K}}{s}}
\newcommand{\vKi}{\mnn{\mathcal{K}}{i}}

\newcommand{\vHn}{\mnn{\mathcal{H}}{n}}

\newcommand{\Qsp}{\mn{\mathcal{Q}}{a}{p}}
\newcommand{\Qsn}{\mn{\mathcal{Q}}{a}{n}}
\newcommand{\Qsi}{\mn{\mathcal{Q}}{a}{i}}
\newcommand{\Qe}{\mnn{\mathcal{Q}}{e}}

\newcommand{\alphan}{\mnn{\alpha}{n}}

\newcommand{\Cp}{\mnn{\mathcal{C}}{p}}
\newcommand{\Cn}{\mnn{\mathcal{C}}{n}}
\newcommand{\Ci}{\mnn{\mathcal{C}}{i}}

\newcommand{\Gp}{\mnn{\mathcal{G}}{p}}
\newcommand{\Gn}{\mnn{\mathcal{G}}{n}}
\newcommand{\Gi}{\mnn{\mathcal{G}}{i}}

\newcommand{\DL}{\mnn{\mathcal{D}}{L}}
\newcommand{\Da}{\mnn{\mathcal{D}}{A}}
\newcommand{\Di}{\mnn{\mathcal{D}}{i}}

\newcommand{\nue}{\mnn{\nu}{e}}
\newcommand{\nusn}{\mn{\nu}{a}{n}}
\newcommand{\nusp}{\mn{\nu}{a}{p}}
\newcommand{\nusi}{\mn{\nu}{a}{i}}

\newcommand{\sigmasi}{\mn{\sigma}{s}{i}}
\newcommand{\sigmasp}{\mn{\sigma}{a}{p}}
\newcommand{\sigmasn}{\mn{\sigma}{a}{n}}

\newcommand{\betai}{\mnn{\beta}{i}}

\newcommand{\ji}{\mnn{j}{i}}

\newcommand{\deltap}{\mnn{\delta}{p}}
\newcommand{\deltan}{\mnn{\delta}{n}}
\newcommand{\deltai}{\mnn{\delta}{i}}

\newcommand{\xip}{\mnn{\xi}{p}}
\newcommand{\xin}{\mnn{\xi}{n}}
\newcommand{\xii}{\mnn{\xi}{i}}

\newcommand{\gammac}{\gamma_c}
\newcommand{\gammat}{\gamma_T}

\newcommand{\DelGi}{\Delta\mnn{\Gamma}{i}}
\newcommand{\DelGp}{\Delta\mnn{\Gamma}{p}}
\newcommand{\DelGn}{\Delta\mnn{\Gamma}{n}}

\newcommand{\lambdap}{\mnn{\lambda}{p}^{\text{cc}}}
\newcommand{\lambdan}{\mnn{\lambda}{n}^{\text{cc}}}

\newcommand{\rhocci}{\mnn{\varrho}{i}^{\text{cc}}}

\newcommand{\Kcci}{\mnn{\mathcal{K}}{i}^{\text{cc}}}

\newcommand{\oLe}{\overline{\Le}}
\newcommand{\oBi}{\overline{\Bi}}

\newcommand{\oK}{\overline{\mathcal{K}}}
\newcommand{\ovrho}{\overline{\varrho}}
\newcommand{\oQ}{\overline{Q}}

\newcommand{\Keffi}{\mn{K}{\text{eff}}{i}}
\newcommand{\Eeffi}{\mn{E}{\text{eff}}{i}}

\newcommand{\Gammaeffi}{\mn{\Gamma}{\text{eff}}{i}}
\newcommand{\Gammaeffp}{\mn{\Gamma}{\text{eff}}{p}}
\newcommand{\Gammaeffn}{\mn{\Gamma}{\text{eff}}{n}}

\newcommand{\cellav}[1]{\langle#1\rangle_\text{cell}}
\newcommand{\posav}[1]{\langle#1\rangle}

\newcommand{\negav}[1]{\langle#1\rangle}

\newcommand{\Le}{\mathrm{Le}}
\newcommand{\Bi}{\mathrm{Bi}}
\renewcommand{\d}{\mathrm{d}}

\begin{document}

\begin{frontmatter}

  \title{Asymptotic reduction, solution, and homogenisation of a thermo-electrochemical model for a lithium-ion battery}

  \author[OX]{Matthew~G.~Hennessy\corref{cor1}}
  \ead{hennessy@maths.ox.ac.uk}
  
  \author[Y]{Iain~R.~Moyles}
  \ead{imoyles@yorku.ca}
  
  \cortext[cor1]{Corresponding author}
  
  \address[OX]{Mathematical Institute, University of Oxford, Andrew Wiles Building, Woodstock Road, Oxford OX2 6GG, United~Kingdom}
  
  \address[Y]{Department of Mathematics and Statistics, York University, Toronto, Canada}
  
  \begin{abstract}
  We study two thermo-electrochemical models for lithium-ion batteries. The first is based on volume averaging the electrode microstructure whereas the second is based on the pseudo-two-dimensional (P2D) approach which treats the electrode as a collection of spherical particles. A scaling analysis is used to reduce the volume-averaged model and show that the electrochemical reactions are the dominant source of heat. Matched asymptotic expansions are used to compute solutions of the volume-averaged model for the cases of constant applied current, oscillating applied current, and constant cell potential. The asymptotic and numerical solutions of the volume-averaged model are in remarkable agreement with numerical solutions of the thermal P2D model for (dis)charge rates up to 2C, and reasonable agreement is found at 4C. Homogenisation is then used to derive a thermal model for a battery consisting of several connected lithium-ion cells. Despite accounting for the Arrhenius dependence of the reaction coefficients, we show that thermal runaway does not occur in the model. Instead, the cell potential is simply pushed closer to the open-circuit potential. We also show that in many cases, the homogenised battery model can be solved analytically, making it ideal for use in on-board thermal management systems.
\end{abstract}


\end{frontmatter}

\section{Introduction}

Lithium-ion batteries (LIBs) are ubiquitous in modern society and are the primary energy source for portable electronic devices and electric cars.  Research and development of LIBs has been driven by their high energy density, low self-discharge rate, and lack of memory effects \cite{Tarascon2001}.  Despite the attractive features of LIBs, there are major safety concerns due to repeated incidents involving overheating, combusting, or exploding LIB-powered devices \cite{Abada2016}.  Aside from safety concerns, heat generation also plays an important role in battery ageing \cite{waldmann2014} and degradation \cite{ramadass2002a, ramadass2002b}, both of which are accelerated at high temperatures. Ensuring safe, reliable, and optimal battery usage requires detailed understanding of the link between the electrochemical processes that drive battery operation and the thermal processes that result in heat generation, distribution, and dissipation \cite{bandhauer2011}.

One of the primary causes of rapid heat generation in an LIB is from internal short circuiting, which can result from mechanical abuse, poor battery design, or the formation of lithium dendrites \cite{wen2012}.
%. There are a number of reasons why a short circuit may form, with the most common being mechanical abuse, poor battery design, or the formation of lithium dendrites \cite{wen2012}. 
Large temperature rises can also occur at high discharge rates if a battery is insufficiently cooled, triggering the onset of thermal runaway due to the combination of exothermic chemical reactions and temperature-dependent material parameters.
%If a battery is insufficiently cooled, then high discharge rates can also lead to large temperature rises, which can then trigger the onset of thermal runaway due to the combination of exothermic chemical reactions and temperature-dependent material parameters. 
The decomposition of the electrolyte and active material at high temperatures can not only produce heat, but also highly flammable gases \cite{lisbona2011}. 
%Heat generation also plays an important role in battery ageing \cite{waldmann2014} and degradation \cite{ramadass2002a, ramadass2002b}, both of which are accelerated at high temperatures. Thus, ensuring safe, reliable, and optimal battery usage requires detailed understanding of the link between the electrochemical processes that drive battery operation and the thermal processes that result in heat generation, distribution, and dissipation \cite{bandhauer2011}.

In order to combat the negative effects of heat generation, many battery packs are equipped with an on-board thermal management system (TMS) \cite{Zhao2015}. The TMS provides real-time monitoring of the temperature and makes adjustments to the battery performance and the cooling system to maintain safe and optimal operation.  The TMS is guided by numerical simulations of thermal-electrochemical LIB models, which must be physically accurate yet simple enough to be simulated in real time.  The need for rapid simulations is especially important for electric vehicles, as their battery packs consist of thousands of LIBs \cite{tesla1,tesla2} that need to be monitored individually and as a unit. Simultaneously achieving physical accuracy and low computational complexity remains a major challenge in the design of management systems \cite{Seaman2014}. 
%\im{An additional need for low computational complexity comes from large-scale applications such as electric vehicles, for example, a Tesla can have over 7000 LiBs in its battery pack \cite{tesla1,tesla2}. Each of these needs to be modelled and simulated.}

Mathematical modelling plays an important role in LIB research because it provides cost-effective insights that can be difficult to obtain through experimental measurement.  Modelling has led to improved understanding of many aspects of batteries, including intercalation kinetics \cite{Smith2017b}, active-material utilisation \cite{Dargaville2010} and distribution \cite{hosseinzadeh2018}, mechanics \cite{Chakraborty2015, Foster2017}, phase separation \cite{Smith2017, Ferguson2012, Ferguson2014, Orvananos2014, Li2014, Lim2016}, electrode fabrication \cite{Font2018}, and model parameter estimation \cite{Sethurajan2015, Krachkovskiy2018}. The development of porous electrode theory by Newman \cite{Newman1962,Newman1975} laid the foundation on which nearly every battery model is now built upon. This theory rests on the assumption that the complex microscale geometry of porous electrodes can be simplified through a volume averaging procedure. Doyle \etal\cite{Doyle1993} later argued that diffusion of intercalated lithium across the microscale can be important during battery operation and developed the pseudo-two-dimensional (P2D) model to couple microscale diffusion to macroscale electochemistry. The P2D framework, which treats the electrode as a regular array of spherical particles, is now one of the most common modelling approaches \cite{Jokar2016, santhanagopalan2006}. 

A wide range of thermo-electrochemical models have been developed \cite{pals1995, gu2000, kumaresan2008} in order to investigate short circuiting \cite{Zhao2016, lee2015, zavalis2012}, thermal runaway \cite{feng2016, Melcher2016, Wang2012}, thermal abuse \cite{Kim2007}, dominant heat generation mechanisms \cite{Li2014}, optimal battery design \cite{lee2013}, and heat distribution in battery packs \cite{An2018, smyshlyaev2011}. Many studies are highly computational in nature and make use of large-scale simulations using commercial finite-element software \cite{Cai2011}, which can be time consuming and offer limited insights. The computational burden of solving a comprehensive thermal-electrochemical model based on the P2D framework can be reduced through cutting-edge numerical implementations or model reduction. As summarised by Jokar \etal\cite{Jokar2016}, the electrochemical problem can be simplied by neglecting spatial gradients \cite{Di2008, Speltino2009} or the use of polynomial approximations \cite{Wang1998, Subramanian2005, Subramanian2007}, the Galerkin projection method \cite{Fan2018, Dao2012, Subramanian2009}, Kalman filtering \cite{Smith2010, bizeray2015}, and Pad\'e approximations \cite{Marcicki2013, Tran2018}. The thermal problem is often simplified by treating the temperature as spatially uniform and integrating the source terms in the heat equation, resulting in a so-called lumped thermal model \cite{bizeray2015, Forgez2010, wu2013, al1999}. Many of these reduction techniques are immediately applied without justification, leading to a lack of robustness for their applicability and range of use. A more systematic approach based on exploitation of the vastly different time scales of various physical processes can lead to a more organic simplification.
%giving consideration as to whether the underlying thermo-electrochemical model can be systematically simplified by exploiting, for example, the fact that some physical processes occur over vastly different time scales. Furthermore, the range of applicability of these reduction techniques is not always clear. 

Asymptotic techniques provide a systematic means of carrying out model reduction and solution construction with clear ranges of validity, but have received relatively little attention from the battery community. Richardson \etal\cite{Richardson2012} uses homogenisation to upscale a P2D model. Sulzer \etal\cite{sulzer2019} apply asymptotic methods to simplify a model for a lead-acid battery while Marquis \etal\cite{marquis2019} use similar techniques to obtain simplifications of the P2D model known as the single-particle model. Moyles \etal\cite{Moyles2019} use matched asymptotic expansions to obtain an equivalent circuit model from a volume-averaged model and show that it can reproduce experimental discharge curves for a variety of discharge rates. All of these works are based on isothermal situations, and therefore cannot account for the heat generation mechanisms and battery failures previously discussed. A systematic asymptotic reduction of a thermo-electrochemical model is therefore both novel and timely.
%which leads to the natural aim of applying asymptotic methods to a thermo-electrochemical model as a means of simplification and gaining new insights into the link between electrochemistry and heat generation.

In this paper, we use asymptotic methods to reduce a thermo-electrochemical model and obtain solutions for common modes of battery operation (e.g., fixed current and fixed potential). This reduction is based on a volume-averaged model, similar in nature to the original model proposed by Newman, rather than the P2D model. We first consider the thermo-electrochemical problem within a single cell of an LIB. We account for Arrhenius reaction kinetics in order to explore whether thermal runaway can occur. We then validate our volume-averaged model and the asymptotic solutions against numerical simulations of a thermal P2D model. Due to rapid thermal diffusion, we find that a single-cell model does not predict the large temperature increases that are seen experimentally. Therefore, we apply homogenisation to obtain a model for an LIB that consists of many cells. The increase in thermal diffusion time leads to greater temperature increases, but we find that thermal runaway does not occur, even when Arrhenius reaction kinetics are accounted for. We also find that for realistic parameter values, the homogenised battery model can be solved analytically, making it ideal for use in a TMS. 

The outline of the paper is as follows. In Sec.~\ref{sec:model} we introduce a model for a single cell of an LIB and non-dimensionalise it. This model is reduced in Sec.~\ref{sec:reduction} by neglecting non-singular terms. Asymptotic solutions are then constructed in Sec.~\ref{sec:asymptotics} and compared against the P2D model. The cell model is upscaled to a battery model in Sec.~\ref{sec:homogenisation}. Discussions and conclusions follow in Sec.~\ref{sec:conclusions}.

% \begin{itemize}
% \item{Thermal model}
% \item{Parameter selection: showing $\Delta E$ important, selecting $\beta_i$ to satisfy conditions on $E_\text{eff}$.}
% \item{Non-dim and reduction}
% \item{Showing no $\exp(T)$ terms in heat equation}
% \item{Steady-state analysis, no thermal runaway}
% \item{Li-ion batteries safe unless material failure occurs, more research into material changes during battery use needed}
% \end{itemize}

\section{Cell-scale modelling}
\label{sec:model}
\begin{figure}
  \centering
  \includegraphics[width=0.4\textwidth]{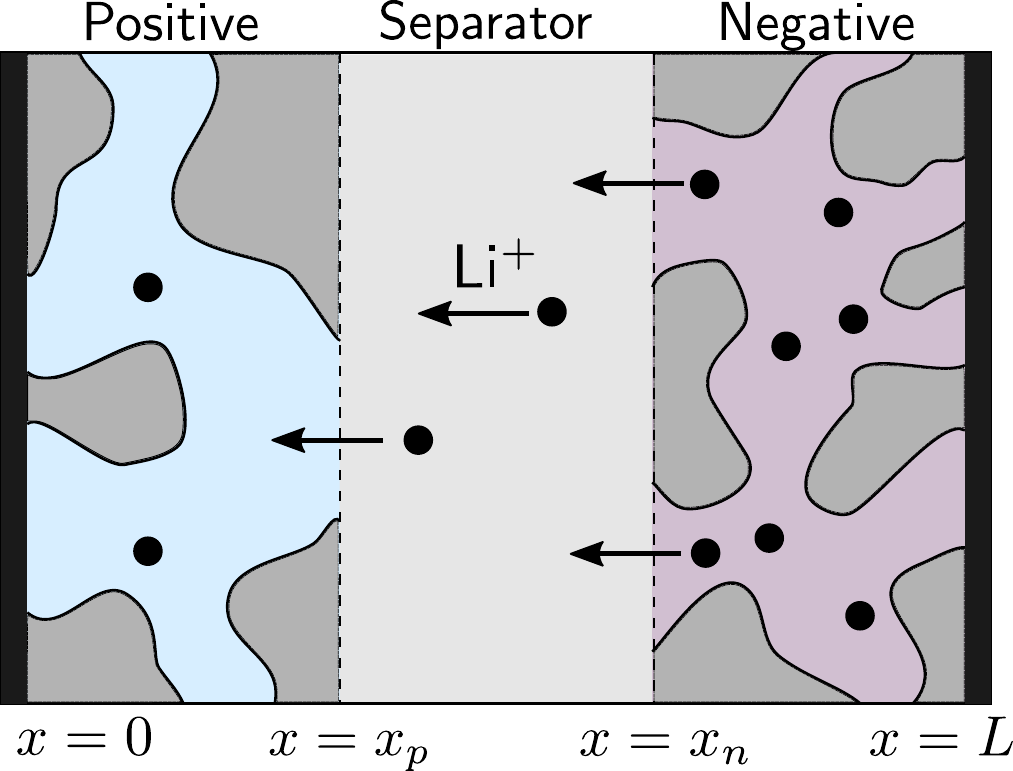}
  \caption{Schematic of a lithium-ion cell consisting of positive and negative electrodes and a separator. During discharge (depicted here), lithium ions move from the negative electrode to the positive electrode.}
  \label{fig:cell_model}
\end{figure}

We consider the thermal-electrochemical processes that occur within the cell of an LIB composed of a positive ($P$) electrode, a separator ($S$), and a negative ($N$) electrode; see Fig.~\ref{fig:cell_model}. The domain is partitioned so that the positive electrode exists on $0\leq x\leq x_p$, the separator on $x_p\leq x\leq x_n$, and the negative electrode on $x_n\leq x\leq L$.  %The electrochemical model is based on porous electrode theory \cite{Newman1962,Newman1975}, which rests on the
The electrode volume can be decomposed into three subdomains corresponding to void space occupied by an electrolyte and solid space occupied by active and inactive materials. The separator is also decomposed into a void and solid fraction but the solid only serves as an electrical insulator that permits ion flow.  In each of the phases, conservation of mass, charge, and thermal energy is imposed. A one-dimensional model is employed because of the large aspect ratio between the height and length of a cell, consistent with a number of previous studies \cite{Wang2012,Li2014,Amiribavandpour2015,Moyles2019}. In the equations that follow, the roman subscript $\text{i} = \text{n}$, $\text{p}$, $\text{s}$ is used to denote the negative electrode, positive electrode, and separator. Thus, the quantity $\psi_{j}$ in component i of the LIB cell is written as $\mn{\psi}{j}{i}$. If there is no roman subscript on a variable, then it is assumed that this quantity is uniform across the LIB cell.

\subsection{Electrochemical model}

The bulk equations of the electrochemical model have been derived from a volume-averaging procedure by Moyles \etal\cite{Moyles2019}, where the underlying assumptions and limitations of the model are discussed in detail, and a validation against experimental data is made. In summary, the model is based on the following assumptions:
\begin{itemize}
    %\item The material properties are independent of composition
    \item Dilute solution theory is used to describe the electrolyte
    \item Phase separation in the electrodes is not explicitly considered
    \item The electrostatic double layer that forms at the matrix-pore interface follows the Helmholtz model.
\end{itemize}
%\mh{A further assumption we had for the previous model was that the temperature remains constant which is no longer 
We now extend this model to account for temperature-dependent material properties and discuss how the microstucture of the electrode can be incorporated into the volume-averaged macroscale equations using the P2D framework proposed by Doyle \etal\cite{Doyle1993}.
%\im{We justify the inclusion of temperature in the parameters while excluding composition dependence with two reasons. Firstly, as shown by Moyles \etal\cite{Moyles2019} and will be summarised here, the electrolyte does not substantially deviate from its initial value and therefore the compositional change and parameter effect is minimal. 
Temperature induces an autocatalytic effect in the system whereby the reactions generate heat which speed up the reaction rates. Temperature-dependent parameters allow us to explicitly model this effect.

\subsubsection{Charge conservation in the electrode solid phase}
Ohm's law relates the current density to the gradient in the electrical potential in the solid phase of the electrode. Following Li \etal\cite{Li2014}, we account for the capacitance of electronic double layers that form at the surface of the solid matrix of the electrodes. Thus, the conservation of electric charge in the solid phase of the electrodes (i = p, n) is governed by
\begin{subequations}\label{sys:solid}
\begin{align}
% \pderiv{}{t}(\phisi \csi)&=\pderiv{}{x}\left(\phisi \mn{D}{a}{i}(\Ti)\pderiv{\csi}{x}\right)+\frac{1}{F}\pderiv{}{x}\left(\phisi \isi\right),\\
  \isi&=-\sigmasi(\Ti)\pderiv{\Phisi}{x}, \\
\pderiv{}{x}\left(\phisi\isi\right)&=-\mnn{a}{i}\left(\gibar+\mn{C}{\Gamma}{i}\pderiv{}{t}(\Phisi-\Phiei)\right),\label{eqn:divisavg}
\end{align}
\end{subequations}
where $t$ is time, $\Ti$ is temperature, $\isi$ is the current density in the active solid phase, and $\Phisi$ and $\Phiei$ are the electric potential in the active solid and electrolyte, respectively. Furthermore, $\sigmasi$ is the electrical conductivity, $\mn{C}{\Gamma}{i}$ is the capacitance per unit area of the active solid, $\phisi$ is the volume fraction of active solid material, and $\mnn{a}{i}$ is the specific area of active electrode material per unit volume, $\mnn{a}{i}$
%=\mn{A}{ae}{i} / V$, where $A_{ae}$ 
is the surface area of the interface formed between active solid material and the electrolyte. Surface-averaged electrochemical currents produced at the interface between electrolyte and active material are denoted $\gibar$ and will be specified in Sec.~\ref{sec:electrochem_kinetics} using Butler--Volmer kinetics.

\subsubsection{Charge conservation in the electrode liquid phase}
Charge transport in the liquid phase of the electrode (\emph{i.e.,} the electrolyte) occurs via molecular diffusion of lithium ions and anions as well as drift due to the presence of an electric field. Thus, conservation of electronic charge in the liquid phase of the electrodes is described by
\begin{subequations}\label{sys:fluid}
\begin{align}
  \iei&= -F [D_L(\Ti) - D_A(\Ti)] \pderiv{\cli}{x} - \sigma_{e}(\cli, \Ti)\pderiv{\Phiei}{x}, \\
\pderiv{}{x}\left(\phiei\iei\right)&=\mnn{a}{i}\left(\gibar+\mn{C}{\Gamma}{i}\pderiv{}{t}(\Phisi-\Phiei)\right),\label{eqn:divieavg}
  % \mn{N}{j}{i}&=-D_j(\Ti)\pderiv{\mn{c}{j}{i}}{x}-z_j\mu_j(\Ti)F\mn{c}{j}{i}\pderiv{\Phiei}{x}, \label{eqn:vecN}
\end{align}
\end{subequations}
where $F$ is Faraday's constant; $\iei$ is the current density; and $\cli$ and $D_L$ are the concentration and diffusivity of lithium ions, respectively; and $D_A$ is the diffusivity of the anion. The quantity $\phiei$ represents the porosity of the electrode, which is also the volume fraction of electrolyte, and is defined as the ratio of the electrolyte volume to the total volume. The ionic conductivity $\sigma_e$ can be written as
\begin{align}
  \sigma_e = F^2 [\mu_L(\Ti) + \mu_A(\Ti)] \cli,
  \label{eqn:sigma_e}
\end{align}
where $\mu_L$ and $\mu_A$ are the mobility of the lithium ions and anions. Assuming the Nernst--Einstein equation applies, the mobility and the diffusivity are linked through 
\begin{align}
  \mu_L(\Ti) = \frac{D_L(\Ti)}{R \Ti}, \quad  \mu_A(\Ti) = \frac{D_A(\Ti)}{R \Ti},
  \label{eqn:NE}
\end{align}
where $R$ is the ideal gas constant.

Within the electrode there must be global conservation of charge, which arises from adding \eqref{eqn:divisavg} and \eqref{eqn:divieavg} to find
\begin{align}
\pderiv{}{x}\left(\phisi \isi + \phiei\iei\right)=0.\label{eqn:phaseneutral}
\end{align}
From this point forward, \eqref{eqn:divieavg} will be replaced with \eqref{eqn:phaseneutral}.

%The charge of the ions is denoted $z_j$ with $z_L = +1$ for lithium and $z_A = -1$ for anions. The molar flux, $\mn{N}{j}{i}$, includes contributions from Fickian diffusion as well as drift along a potential gradient and the subscripts $j = L$ and $j=A$ denote lithium and anions respectively.  Finally, $\phiei$ is the porosity of the electrode, which is also the volume fraction of electrolyte, and is defined as the ratio of the electrolyte volume to the total volume.  

\subsubsection{Lithium conservation in the electrode solid phase}

We consider two approaches for describing lithium conservation in the solid phase of the electrodes. The first is based on volume averaging \cite{Moyles2019}. In this case, the (volume-averaged) concentration of intercalated lithium $\csi$ changes due to electrochemical reactions according to
\begin{align}
  \pderiv{}{t}\left(\phisi \csi\right) = \frac{1}{F}\pderiv{}{x}\left(\phisi \isi\right),
  \label{eqn:csi}
\end{align}
where diffusion of intercalated lithium across the macroscale has been neglected, which is a valid approximation \cite{Moyles2019}.

The second approach is based on the P2D framework, which treats the electrode as a regular array of spherical particles of radius $\mn{R}{p}{i}$.  During lithiation, lithium enters the surface of each particle and is then transported into the bulk via diffusion. Similarly, during delithiation, the lithium near the surface of the particles is depleted and then replenished by diffusion. Thus, accounting for conservation of lithium in the P2D framework involves solving the radial diffusion equation at each point $x$ in the spatial domain of the electrodes:
\subeq{
  \label{eqn:p2d}
\begin{align}
  \pderiv{\csi^*}{t} = \frac{1}{r^2}\pderiv{}{r}\left(r^2 \mn{D}{a}{i}\pderiv{\csi^*}{r}\right),
  \label{eqn:p2d_cs}
\end{align}
where $r$ is the radial pseudo-dimensional, $\mn{D}{a}{i}$ is the diffusivity of lithium in active solid material, and $\csi^* = \csi^*(r,t;x)$ is the (non-volume-averaged) concentration of intercalated lithium. The ``microscale'' equation \eqref{eqn:p2d_cs} is coupled to the macroscale through the flux boundary condition
\begin{align}
  -\mn{D}{a}{i}\pderiv{\csi^*}{r} = \frac{1}{F}\left(\gibar + \mn{C}{\Gamma}{i}\,\pderiv{}{t}\left(\Phisi - \Phiei\right)\right), \quad r = \mn{R}{p}{i},
  \label{eqn:p2d_bc}
\end{align}
A no-flux condition is imposed at the origin:
\begin{align}
  \pderiv{\csi^*}{r} = 0, \quad r = 0.
  \label{eqn:p2d_noflux}
\end{align}
}

The volume-averaged model \eqref{eqn:csi} and the P2D model \eqref{eqn:p2d} are related through the limit of fast microscale diffusion. 
% \im{We define a volume averaged quantity over particle volume $V_s$,
% \begin{align}
% \csi=\frac{1}{V_s}\int\csi^*\diff V=\frac{3}{R_p^3}\int_0^{R_p}\csi^*r^2\diff r,
% \end{align}
% which is simplified due to the radial symmetry of the particles. Taking a volume average of \eqref{eqn:p2d_cs} over a representative volume, $V$, including particle and electrolyte (so that $V_s/V=\phisi$) yields,
% \begin{align}
% \pderiv{}{t}(\phisi\csi)=-\frac{4\pi R_p^2}{V}\left(\gibar + \mn{C}{\Gamma}{i}\,\pderiv{}{t}\left(\Phisi - \Phiei\right)\right)=-\frac{3\phisi}{R_pF}\left(\gibar + \mn{C}{\Gamma}{i}\,\pderiv{}{t}\left(\Phisi - \Phiei\right)\right),\label{eqn:volavg}
% \end{align}
% which is precisely the volume averaged equation \eqref{eqn:csi} taking into account the conservation law \eqref{eqn:divisavg} and $\mnn{a}{i}=3\phisi/R_p$.}
Integrating \eqref{eqn:p2d_cs} across the particle domain, using the flux conditions \eqref{eqn:p2d_bc} and \eqref{eqn:p2d_noflux}, and assuming that $\csi^*$ is spatially uniform leads to the simplified equation
\begin{align}
  \pderiv{\csi^*}{t} = -\frac{3}{\mn{R}{p}{i} F}\left(\gibar + \mn{C}{\Gamma}{i}\,\pderiv{}{t}\left(\Phisi - \Phiei\right)\right). \label{eqn:p2d_simple}
\end{align}
Volume averaging \eqref{eqn:p2d_simple} and then using the relation $\mnn{a}{i} = 3 \phisi / \mn{R}{p}{i}$ as well as the conservation law \eqref{eqn:divisavg} yields \eqref{eqn:csi}.

Asymptotic solutions will be derived from the volume-averaged form of the mass conservation given by \eqref{eqn:csi}. We then compare the asymptotic solutions to those obtained via numerical simulations of the volume-averaged (VA) and a P2D model based on \eqref{eqn:p2d}. 

\subsubsection{Lithium conservation in the electrode liquid phase}

Lithium conservation in the liquid phase of the electrode leads to a reaction-diffusion equation:
\begin{align}
\pderiv{}{t}(\phiei \cli)&=\pderiv{}{x}\left[\phiei D_{e}(\Ti)\pderiv{\cli}{x}\right] + \frac{1}{F}\pderiv{}{x}\left\{\phiei\left[1 - \theta(\Ti)\right]\iei\right\}.\label{eqn:avgci}  
  % \pderiv{}{t}(\phiei \cli)&=\pderiv{}{x}\left(\phiei D_L(\Ti)\pderiv{\cli}{x} +\phiei\mu_L(\Ti)F\cli\pderiv{\Phiei}{x}\right)+\frac{1}{F}\pderiv{}{x}\left(\phiei\iei\right),\label{eqn:avgci}  
\end{align}
In this equation, the effective ionic diffusivity $D_e$ in the electrolyte is given by 
\begin{align}
  D_e(\Ti) = D_L(\Ti) - \theta(\Ti) \left[D_L(\Ti) - D_A(\Ti)\right].
  \label{eqn:De}
\end{align}
The parameter $\theta$ plays the role of a transference number and is given by
\begin{align}
  \theta(\Ti) = \frac{\mu_L(\Ti)}{\mu_L(\Ti) + \mu_A(\Ti)} = \frac{D_L(\Ti)}{D_L(\Ti) + D_A(\Ti)},
  \label{eqn:theta}
\end{align}
where the second equality arises from using the Nernst--Einstein equations \eqref{eqn:NE}.

\subsubsection{Charge transport and mass conservation in the separator}
Conservation of charge and lithium in the separator is given by
\begin{subequations}\label{sys:separator}
  \begin{align}
    \pderiv{}{t}(\phies \cls)&=\pderiv{}{x}\left[\phies D_{e}(\Ts)\pderiv{\cls}{x}\right]-\frac{1}{F}\pderiv{}{x}\{\phies\theta(\Ts)\ies\}, \\
    \ies&= -F [D_L(\Ts) - D_A(\Ts)] \pderiv{\cls}{x} - \sigma_{e}(\cls, \Ts)\pderiv{\Phies}{x}, \\
    \pderiv{}{x}\left(\phies\ies\right)&= 0.
  \end{align} 
\end{subequations}

\subsubsection{Electrochemical kinetics}
\label{sec:electrochem_kinetics}

The electrochemical reactions that are responsible for lithiation and delithiation are described using the Bulter--Volmer kinetics \cite{Newman2004} in the Helmholtz limit of thin electric double layers.  The electrochemical current produced at the electrode-electrolyte interface is given by
\subeq{
  \label{eqn:bv}
\begin{align}
\gibar=\ji\left(\exp\left[\frac{(1-\betai)F}{R\Ti}\etai\right]-\exp\left[\frac{-\betai F}{R\Ti}\etai\right]\right),\label{eqn:fullBV}
\end{align}
where $\mnn{\eta}{i} = \Phisi-\Phiei-\mnn{U}{i}$ is the surface overpotential, 
$\mnn{U}{i}$ is the open-circuit potential, $\ji$ is the exchange current density
\begin{align}
  \ji&=F\Keffi(\Ti)\csi^{\betai}\left(\frac{\csimax-\csi}{\csimax}\right)^{1-\betai}\cli^{1-\betai},\label{eqn:j0}
\end{align}
$\betai<1$ represents the symmetry between anodic and cathodic reaction currents,  and $\csimax$ is the maximal density of intercalated lithium in each electrode. The effective reaction constant $\Keffi$ can be written in terms of the heterogeneous reaction constants for the cathodic and anodic reactions as
\begin{align}
  \Keffi(\Ti) = [\mn{K}{a}{i}(\Ti)]^{\betai} [\mn{K}{L}{i}(\Ti)]^{1-\betai}.
  \label{eqn:Keffi}
\end{align}
}
The reaction constants are assumed to have an Arrhenius form
\begin{align}\label{eqn:KT}
  \mn{K}{y}{i}(\Ti) = \mn{K}{y}{i}^0\exp\left[\frac{\mn{E}{y}{i}}{R}\left(\frac{1}{T_a} - \frac{1}{\Ti}\right)\right],
\end{align}
where $\mn{E}{y}{i}$ is the activation energy of the reaction. By inserting \eqref{eqn:KT} into \eqref{eqn:Keffi}, the effective reaction constant can also be written in Arrhenius form,
\begin{align}
  \Keffi = \Keffi^0 \exp\left[\frac{\Eeffi}{R}\left(\frac{1}{T_a} - \frac{1}{\Ti}\right)\right],
\end{align}
where $\Keffi^0 = (\mn{K}{y}{i}^0)^{\betai}(\mn{K}{L}{i}^0)^{1-\betai}$ and $\Eeffi = \betai \mn{E}{a}{i} + (1 - \betai) \mn{E}{L}{i}$. Aside from the terms involving $\csimax$, the electrochemical kinetics described by \eqref{eqn:fullBV}--\eqref{eqn:KT} are consistent with those derived by Ferguson and Bazant \cite{Ferguson2012} using non-equilibrium thermodynamics.
% The open-circuit potential $\mnn{U}{i}$ is usually given by an empirical expression that has been fitted to experimental data. To retain generality and carry out the model reduction, we use an analytical expression for $\mnn{U}{i}$ that is consistent with the Bulter--Volumer kinetics \eqref{eqn:bc}:
The theoretical open-circuit potential, which is consistent with the Butler--Volmer kinetics \eqref{eqn:bv}, is given by
\begin{align}
  % \mnn{U}{i}&=\frac{R \Ti}{F}\log\left[\frac{\mn{K}{L}{i}(\Ti)\cli(\csimax-\csi)}{\mn{K}{a}{i}(\Ti)\csi\csimax}\right].
\mnn{U}{i}&=\frac{R \Ti}{F}\log\left[\frac{\mn{K}{L}{i}^0\cli(\csimax-\csi)}{\mn{K}{a}{i}^0\csi\csimax}\right] + \left(\frac{\Ti}{T_a} - 1\right)\left(\frac{\mn{E}{L}{i}-\mn{E}{a}{i}}{F}\right).
\label{eqn:Ui}
\end{align}
% However, when comparing the reduced model to experimental data, \eqref{eqn:Ui} will be replaced with an empirial expression.
Extended forms of the open-circuit potential can be used to account for additional physics such as phase separation, relevant for LFP electrodes, and multiple lithiation stages, relevant for graphite electrodes \cite{Ferguson2012, Ferguson2014, Orvananos2014, Thomas2017}.

%\im{We do not need to consider conservation of mass and charge in the solid inactive material because electrochemical reactions do not occur there.}

\subsection{Thermal model}
% \im{The model presented thus far, aside from the thermal dependence on the parameters, is the same as that derived by Moyles \etal\cite{Moyles2019}. To include temperature we volume average the conservation of energy equation leading to}
%Electrochemical reactions do not occur on the inactive solid material and consequently, conservation of mass and charge do not need to be imposed there.  However, the inactive material can still transport heat produced by the LIB and therefore conservation of energy needs to consider all three phases (active and inactive solid and electrolyte).  Through similar volume averaging as the electrochemical equations, 
Conservation of energy in each electrode takes the form
\begin{align}
\Rcpi\pderiv{\Ti}{t}=\pderiv{}{x}\left(\ki\pderiv{\Ti}{x}\right)+\phiei\qie+\phisi\qia+\qir,\label{eqn:heat}
\end{align}
where 
\begin{align}
  \Rcpi=\phiei\Rcpe+(1-\phiei)\Rcpsoli,\qquad \ki=\phiei\Ke+(1-\phiei)\Ksoli,
  \label{eqn:phase_av}
\end{align}
are the phase-averaged volumetric heat capacity and thermal conductivity, respectively. We note that \eqref{eqn:heat} includes the inactive material which can still transport heat. Furthermore, the equation is derived by assuming that (i) there is local thermal equilibrium in the three phases so that a common temperature is well defined and (ii) the inactive and active solid material have the same thermal properties.  The source terms in \eqref{eqn:heat} are given by
\begin{subequations}\label{sys:heatterms}
\begin{align}
  \qia&= -\isi \pderiv{\Phisi}{x} = \frac{\isi^2}{\sigmasi(\Ti)},\label{eqn:qia}\\
  \qie&= -\iei \pderiv{\Phiei}{x} = \frac{\iei^2}{\sigma_e(\cli,\Ti)} + \frac{F[D_L(\Ti)-D_A(\Ti)]}{\sigma_e(\cli, \Ti)}\,\iei\pderiv{\cli}{x},\label{eqn:qie}\\
\qir&=\mnn{a}{i}\gibar\left(\etai + \Ti\,\pderiv{\mnn{U}{i}}{\Ti}\right),\label{eqn:qir}
\end{align}
\end{subequations}
representing Ohmic heating in the active solid material, ionic Ohmic heating in the electrolyte, and heat generation due to surface electrochemical reactions, respectively.  Full derivations of \eqref{eqn:heat} can be found in \cite{Kaviany2012,Wang1998} while discussion of the electrochemical surface reaction heat can be found in \cite{Li2014,Wang2012, Newman2004}. Often, the electrochemical heat generation \eqref{eqn:qir} is decomposed into irreversible and reversible contributions given by
\begin{align}
  \qir^{\text{irr}} = \mnn{a}{i}\gibar \etai, \quad \qir^\text{rev} = \mnn{a}{i}\gibar \Ti\pderiv{\mnn{U}{i}}{\Ti},
\end{align}
respectively. The reversible contribution $\qir^{\text{rev}}$ can cool the battery
%\im{, because $\pderiv{U}{T}$ can be negative \cite{Li2014}}
whereas the irreversible contribution $\qir^{\text{irr}}$ always heats the battery. By inserting the expression for the open-circuit potential \eqref{eqn:Ui} into the equation for the electrochemical heat generation \eqref{eqn:qir}, we find that
\begin{align}
  %\qir &= \gibar \left\{\etai + \frac{R \Ti}{F}\left[\log\left(\frac{\mn{K}{L0}{i}\cli(\csimax - \csi)}{\mn{K}{a0}{i}\csi\csimax}\right) + \frac{\mn{E}{L}{i}-\mn{E}{a}{i}}{R T_a}\right]\right\} \\
  \qir = \mnn{a}{i}\gibar \left(\Phisi - \Phiei + \frac{\mn{E}{L}{i} - \mn{E}{a}{i}}{F}\right).
\end{align}

The electrochemically inert separator can also transport heat where conservation of energy is given by
%in the separator is given by
\begin{align}
\Rcps\pderiv{\Ts}{t}=\pderiv{}{x}\left(\ks\pderiv{\Ts}{x}\right)+\phies\qse,\label{eqn:heats}
\end{align}
where the phase-averaged volumetric heat capacity and thermal conductivity are given by \eqref{eqn:phase_av} and $\qse$ is given by \eqref{eqn:qie}.

\subsection{Boundary and initial conditions}
The electrolyte is free to flow between the voids of the electrodes and separator. Therefore, we impose continuity of concentration, molar flux of lithium ions, electrolyte current density, electrolyte potential at the separator-electrode interfaces $x = x_p$ and $x=x_n$. Furthermore, we also require the temperature and heat flux to be continuous. These stipulations result in the following boundary conditions
 % Therefore, we require the concentration and molar flux of lithium ions and the current density in the electrolyte, as well as the electrolyte potential, to be continuous. This yields the conditions
\begin{subequations}
\label{bc:se_fluid}
\begin{alignat}{2}
  \cli - \cls &=  0, &\quad  x&=x_p,\,x_n; \label{bc:cl_cont}\\
  \phiei \pderiv{\cli}{x} - \phies \pderiv{\cls}{x} &=  0, &\quad  x&=x_p,\,x_n; \label{bc:cx_cont} \\
  \phiei\iei-\phies\ies&=0, &\quad  x&=x_p,\,x_n; \label{bc:ie_cont} \\
  \Phiei - \Phies &=  0, &\quad  x&=x_p,\,x_n; \label{bc:phi_cont}\\
  \phiei \pderiv{\Phiei}{x} - \phies \pderiv{\Phies}{x} &=  0, &\quad  x&=x_p,\,x_n;\label{bc:phix_cont}\\
  \Ti-\Ts&=0,&\quad x&=x_p,\,x_n;\label{bc:T_cont}\\
  \ki\pderiv{\Ti}{x}-\ks\pderiv{\Ts}{x}&=0,& \quad x&=x_p,\,x_n.\label{bc:Tx_cont}
\end{alignat}
\end{subequations}
The process of volume averaging microscopic boundary conditions for flux-like quantities introduces volume fractions in \eqref{bc:cx_cont}, \eqref{bc:ie_cont}, and \eqref{bc:phix_cont} accounting for porosity difference in the materials.  The volume fractions are absent in \eqref{bc:Tx_cont} because the temperature has been phase averaged under the conditions of local thermal equilibrium.

No current can pass through the solid material of the separator because it does not conduct electricity.  Therefore, we impose
\begin{align}
\isi=0, \quad  x = x_p,\,x_n.
\end{align}
The electrode surfaces at $x=0$ and $x=L$ are in contact with current collectors which allow for the flow of electrons into and out of the battery. Ionic flow through the current collectors is not permitted and therefore at the electrode-collector interface, the ionic molar fluxes and electrolyte current must vanish, leading to
\begin{subequations}
\label{bc:ec}
\begin{alignat}{2}
  \iep&=0, &\quad &x=0,\\
  \ien&=0, &\quad &x=L,\\
  \pderiv{\cli}{x}&=0, &\quad &x=0, L.
  \end{alignat}
\end{subequations}
We require a grounding condition and without loss of generality, set the electrolyte potential at the electrode-collector interface of the negative electrode to zero, yielding
\begin{align}
  \Phien=0, \quad x=L.
\end{align}

There are two common modes of operation: constant current and constant voltage. The constant current operation
%A common mode of operation
involves drawing a prescribed current from the cell and measuring the change in electric potential across the cell as a function on time. To model this scenario, we impose a draw current at the positive electrode given by
\begin{align}
  \phisp\isp=-\mathcal{I}(t)i_0, \quad x = 0, \label{bc:iapp}
\end{align}
where $\mathcal{I}$ is the C-rate of the battery and $i_0$ the draw current density at a discharge rate of 1C (corresponding to $\mathcal{I} = 1$).  The C-rate is a commonly used measure of how quickly a battery is being charged or discharged.  The negative on the right-hand side of \eqref{bc:iapp} means that positive C-rates correspond to a discharge process. Alternatively, in the constant voltage operation, the cell voltage $\Delta V$ defined by
\begin{align}
\Delta V=\Phisp(0,t)-\Phisn(L,t).\label{eqn:cellvolt}
\end{align}
is prescribed instead. When \eqref{bc:iapp} is prescribed then the cell voltage becomes an output of the system. When \eqref{eqn:cellvolt} is used, then \eqref{bc:iapp} is instead used to determine the C-rate $\mathcal{I}(t)$.
% The draw current density $i_0$ is related to the draw current $I_{\rm app}$ via
% \begin{align}
% i_0=\frac{I_{\rm app}}{A_{\rm cell}},\label{eqn:i0def}
% \end{align}
% where $A_{\rm cell}$ is the electrode area.
%The boundary conditions given by \eqref{bc:se_fluid}--\eqref{bc:iapp} are enough to close the electrochemical problem when the C-rate $\mathcal{I}$ is prescribed. Having solved this problem, the potential difference across the cell, $\Delta V$, is then determined by
%\begin{align}
%\Delta V=\Phisp(0,t)-\Phisn(L,t).\label{eqn:cellvolt}
%\end{align}
%An alternative mode of operation involves prescribing the cell potential $\Delta V$ and then measuring the current that leaves or enters the cell.  This is a slightly more difficult situation to study because now the C-rate $\mathcal{I}(t)$ that appears in \eqref{bc:iapp} is unknown and must be solved as part of the problem. 

We assume that the current collectors are in contact with the exterior environment and admit heat transfer with an ambient temperature, $T_a$.  We model this using Newton cooling conditions
\begin{alignat}{2}
\kp\pderiv{\Tp}{x}&=-\hp(T_a-\Tp), &\quad &x=0,\\
\kn\pderiv{\Tn}{x}&=\hn(T_a-\Tn ), &\quad &x=L,
\end{alignat}
where denote heat transfer coefficients.

The initial concentrations are assumed to be spatially uniform throughout each of the battery components and are given by $\csi(x,0)=\csi^0$ and $\cli(x,0)=c_{L}^0$. The electric potentials satisfy $\Phisi(x,0)-\Phiei(x,0)=\mnn{U}{i}$, that is, the overpotential is initially zero. The initial temperature is $\Ti=T_a$.

\subsection{Parameter determination}

The parameter values are adapted from Refs.~\cite{Amiribavandpour2015, Li2014}, in which thermal P2D models are developed for ANR26650m1-a lithium-ion cells \cite{nanobattery2}. Tables of parameters values along with descriptions of how they were obtained are given in \ref{app:parameters}.

\subsection{Numerical implementation}

The VA and P2D models are solved numerically using a semi-implicit finite-difference method. At each time step, a fixed-point method is used to solve the full nonlinear system, which is decomposed into simpler subproblems that are solved sequentially.  Each fixed-point iteration begins by solving for the solid-phase potentials and current densities using Newton's method. The liquid-phase variables are then obtained by first calculating the current density, followed by the lithium concentration, and then the electrical potential. The last step of the fixed-point iteration is to update the concentration of intercalated lithium using either the VA or P2D approach. The fixed-point iterations repeat until a stopping condition is reached. For a prescribed applied current, the stopping condition is met when the change in cell potential is less than 1~mV. In the case of a prescribed cell potential, the stopping condition is met when then change in the C-rate is less than $10^{-4}$. Upon satisfying the stopping condition, the electrical problem is assumed to be solved, the temperature is updated, and then the process repeats. 

\subsection{Non-dimensionalisation}\label{sec:scaling}
We introduce characteristic scales for each of the variables to write the model in dimensionless terms. The motivation behind the choice of scales for the electrochemical model is discussed in detail in Moyles \etal\cite{Moyles2019}. The electric potentials are written as $\Phisi = (R T_a / F) [\log \Ui + \Phisi']$, $\Phiei = (R T_a / F) \Phiei'$, and $\mnn{U}{i} = (R T_a/F)[\log \Ui + \mnn{U}{i}']$, where $\log \Ui$ is the initial value of the open-circuit potential normalised by the thermal voltage. Similarly, the temperature is written as $\Ti = T_a + (\Delta T) \Ti'$.  The temperature scale $\Delta T$ will be discussed below. Each temperature-dependent parameter $\psi(\Ti)$ is rescaled according to $\psi(\Ti) = \psi^0 \psi'(\Ti')$ where $\psi^0 = \psi(T_a)$. The only exception to this is the transference number $\theta$, which is written as $\theta(\Ti) = \theta'(\Ti')$.  Spatial variables are scaled according to $x=Lx'$, $x_p=Lx_p'$, and $x_n = L x_n'$. Time is written as $t = (L^2 / D_e^0)t'$. The current densities are scaled as $\isi = i_0 \isi'$ and $\iei = i_0 \iei'$. The solid-phase and liquid-phase lithium concentrations are written as $\csi = \csi^0 + (\Delta c) \csi'$ and $\cli = \cli^0 + (\Delta c) \cli'$, where $\Delta c = (i_0 L) / (F D_e^0)$. The overpotential is written in terms of the thermal voltage, $\etai = (R T_a / F)\etai'$. The electrochemical surface current is scaled as $\gibar = \gibar^0 \gibar'$ where
% \begin{align}
% \begin{aligned}
%  \qquad t=(L^2/D_{L0})t',& \qquad \mnn{i}{j}=i_0\mnn{i}{j}',\\
% \csi-\mn{c}{a0}{i}=(i_0L)/(FD_{L0})\csi', \qquad \cli-\mnn{c}{L0}=(i_0L)/(FD_{L0})\cli',\\
% \mnn{U}{i} = (RT_a/ F)\mnn{U}{i}', \qquad \etai=(RT_a/F)\etai', \qquad g_i=\mn{g}{0}{i}g_i',
% \end{aligned}\label{eqn:escales}
% \end{align}
% where from non-dimensionalising the Butler-Volmer kinetics \eqref{eqn:j0},
\begin{align}
  \mnn{g}{i}^0&=F\Keffi^0 (\csi^0)^{\betai}\left(1 - \csi^0/\csimax\right)^{1-\betai}(\cli^0)^{1-\betai}.\label{eqn:g0def}
% \mn{g}{0}{i}&=F\mn{K}{a0}{i}^{\betai}\mn{K}{L0}{i}^{1-\betai}\mn{c}{a0}{i}^{\betai}\left(\frac{\csimax-\mn{c}{a0}{i}}{\csimax}\right)^{1-\betai}\mn{c}{L0}{i}^{1-\betai}.\label{eqn:g0def}
\end{align}
We select a common, constant porosity $\phien=\phiep=\phies=\phie$ for each of the cell components. We also assume that the volume fractions of active solid are constant in time and uniform in space.

\subsubsection{Electrochemical model}
Upon dropping the primes, the following bulk equations for the electrode active material are obtained:
\begin{subequations} \label{sys:electrode_solid}
\begin{align}
  \pderiv{\csi}{t}&=%\Di\pderiv{}{x}\left(\mn{D}{a}{i}(\Ti)\pderiv{\csn}{x}\right) +
                    \pderiv{\isi}{x}, \label{eqn:nd_csi}\\
\mn{\nu}{a}{i} \isi&=-\sigmasi(\Ti) \pderiv{\Phisi}{x},\label{eqn:ohma}\\
\phisi\pderiv{\isi}{x} &=-\Gi \left(\gibar +\Ci \pderiv{}{t}(\Phisi-\Phiei)\right), \label{eqn:div_isi}
\end{align}
\end{subequations}
The governing equations for the electrolyte in the electrode are given by
\begin{subequations}
\label{sys:electrode_liquid}
\begin{align}
%  \pderiv{\cli}{t}&=\pderiv{}{x}\left(\mathcal{R}(\Ti)\pderiv{\cli}{x}\right) + \pderiv{}{x}\big(\left[1 - \Theta(\Ti)\right]\iei\big), \label{eqn:nd_cli2}\\
  \pderiv{\cli}{t}&=\pderiv{}{x}\left(D_e(\Ti)\pderiv{\cli}{x}\right) + \pderiv{}{x}\left\{[1 - \theta(\Ti)]\iei\right\}, \label{eqn:nd_cli2}\\
\iei&=-[\DL D_L(\Ti)-\Da D_A(\Ti)]\pderiv{\cli}{x}-\nue^{-1}\sigma_e(\cli, \Ti) \pderiv{\Phiei}{x},\label{eqn:ohme}\\
\pderiv{}{x}&\left(\phisi \isi + \phiei \iei\right) = 0,
\end{align}
\end{subequations}
where the non-dimensional effective diffusivity and ionic conductivity are given by
\subeq{
  \begin{align}
    D_e(\Ti) &= \DL D_L(\Ti) - \theta(\Ti)\left[\DL D_L(\Ti) - \Da D_A(\Ti)\right], \\
    \sigma_e(\cli, \Ti) &= \left[(1 - \theta^0) \mu_A(\Ti) + \theta^0 \mu_L(\Ti)\right](1 + \gamma_c \cli). \label{nd:sigma_e}
\end{align}
}
The bulk equations in the separator are
\begin{subequations}\label{sys:separator2}
  \begin{align}
    % \pderiv{\cls}{t}&= \pderiv{}{x}\left(\R(\Ti)\pderiv{\cls}{x}\right), \label{eqn:nd_cls2}\\
% \ies&=-\Deff(\Ti)\pderiv{\cls}{x}-\nue^{-1}\Meff(\Ti)(1+\gammac\cls)\pderiv{\Phies}{x},\label{eqn:ohmes}\\
% \pderiv{\ies}{x}&=0.\label{eqn:ohmas}
 \pderiv{\cls}{t}&=\pderiv{}{x}\left(D_e(\Ts)\pderiv{\cls}{x}\right)-\pderiv{}{x}\left(\theta\ies\right),\label{eqn:nd_cls2}\\
 \ies&=-[\DL D_L(\Ts)-\Da D_A(\Ts)]\pderiv{\cls}{x}-\nue^{-1}\sigma_e(\cls, \Ts) \pderiv{\Phies}{x}\label{eqn:ohmes}\\
\pderiv{\ies}{x}&=0.\label{eqn:ohmas}
\end{align}
\end{subequations}
The non-dimensional Butler--Volmer kinetics can be written as
\begin{subequations}\label{sys:nondimbutler}
\begin{align}
\gibar&=\ji\left(\exp\left[\frac{(1-\betai)\etai}{1 + \gamma_T \Ti}\right]-\exp\left[-\frac{\betai\etai}{1 + \gamma_T \Ti}\right]\right),\\
  \ji&=\Keffi(\Ti)\left(1+\deltai\gammac\csi\right)^{\betai}\left(1-\deltai\xii(1-\xii)^{-1}\gammac\csi\right)^{1-\betai}(1+\gammac\cli)^{1-\betai}, \label{eqn:nd_j0i} \\
  \Keffi &= \exp\left\{(\Gammaeffi/\gammat)\left[1 - (1 + \gamma_T \Ti)^{-1}\right]\right\}, \label{eqn:nd_Keff}
\end{align}
\end{subequations}
with $\Gammaeffi = \gammat\Eeffi / (R T_a)$.  The definitions of $\Gammaeffi$ and $\Keffi$ lead to a simplified form in the limit $\gammat \Ti \ll 1$ as shown in \eqref{red:Keffi}.  The non-dimensional overpotential can be written as
\begin{align}
  % \etai=\Phisi-\Phiei - (1 + \gammat T)\log(\Vi \Wi) - \gammat \Ti \log(\Ui), \label{eqn:butleretadef}
  \etai=\Phisi-\Phiei - (1 + \gammat \Ti)\log\Vi - \gammat \Ti [\log\Ui + \DelGi], \label{eqn:butleretadef}
\end{align}
where %$\DelGi = (\mn{E}{L}{i} - \mn{E}{a}{i}) / (R T_a)$ is the difference in non-dimensional activation energy and 
the constant ($\Ui$) and composition-dependent ($\Vi$) %, and temperature-dependent ($\Wi$)
contributions to the 
open-circuit potential are defined according to
\begin{align}
\Ui=\frac{\deltai \mn{K}{L}{i}^0(1-\xii)}{\mn{K}{a}{i}^0}, \quad
  \Vi=\frac{(1+\gammac\cli)[1-\deltai\xii(1-\xii)^{-1}\gammac\csi]}{(1+\deltai\gammac\csi)},
  % \quad
%\Wi = \frac{\mn{K}{L}{i}(\Ti)}{\mn{K}{a}{i}(\Ti)}.
\label{eqn:UVdef}
\end{align}
The non-dimensional numbers are described in \cite{Moyles2019} and given by
\begin{align}
\begin{aligned}
  \Di = \frac{\mnn{D}{i}^0}{D_{e}^0 }, \quad
  \mn{\nu}{a}{i} = \frac{i_0 L F}{R T_a \sigmasi^0}, \quad
  \nue = \frac{i_0 L F}{R T_a \sigma_e^0}, \quad 
  \Gi = \frac{\mnn{a}{i}\mnn{g}{i}^0L}{i_0}, \quad
  \Ci = \frac{\mn{C}{\Gamma}{i} R T_a D_{e}^0}{\mnn{g}{i}^0 F L^2}, \quad \\
  \theta^0 = \frac{\mu_{L}^0}{\mu_{L}^0 + \mu_{A}^0}, \quad
  \deltai = \frac{c_{L}^0}{\csi^0}, \quad
  \xii = \frac{\csi^0}{\csimax}, \quad
  \gammac = \frac{\Delta c}{c_{L}^0}, \quad
  \gammat = \frac{\Delta T}{T_a},\quad \DelGi = \frac{\mn{E}{L}{i} - \mn{E}{a}{i}}{R T_a},
  \end{aligned}
  \label{eqn:nondim}
\end{align}
where $\sigma_e^0 = F^2 (\mu_{L}^0 + \mu_{A}^0)c_{L}^{0}$ and $D_e^0 = D_L^0 - \theta^0 (D_L^0 - D_A^0)$.

\subsubsection{Thermal model}
We choose a temperature scale based on the heat generated from the electrochemical reactions in the positive electrode,
\begin{align}
  \Delta T = \left(\frac{\mnn{a}{p} \mnn{g}{p}^0 L^2}{\Rcpp D_{e}^0}\right)\left(\frac{R T_a}{F}\right),
  %\frac{\mnn{a}{p}\mn{g}{0}{p}RT_aL^2  \mh{\log \Up}}{D_{L0}\rcpp F}.
  \label{eqn:Del_T_scale}
\end{align}
anticipating this to be the dominant heat generation mechanism. Indeed, using parameters from Tables \ref{tab:params2} and \ref{tab:thermal}, we estimate the heat generation from Ohmic heating to be $\qpa \sim 86$~W~m\unit{-3}, $\qna \sim 1.8$~W~m\unit{-3}, and $\qie \simeq 73$~W~m\unit{-3} while for the electrochemical heat, we find that $\qpr \sim 1.4\tten{4}$~W~m\unit{-3} and $\qnr \sim 1.3\tten{4}$~W~m\unit{-3}. Evaluating \eqref{eqn:Del_T_scale} gives a temperature scale of $\Delta T \sim 0.65$~K.
The resulting dimensionless heat equations become (upon dropping primes):
\begin{subequations}\label{sys:heat}
\begin{align}
\pderiv{\Tp}{t}&=\Le\pderiv[2]{\Tp}{x}+\phie\mathcal{Q}_e\qpe+\phisp\Qsp\qpa+\qpr\\
\vrcs\pderiv{\Ts}{t}&=\vKs\Le\pderiv[2]{\Ts}{x}+\phies\mathcal{Q}_e\qse\\
\vrcn\pderiv{\Tn}{t}&=\vKn \Le\pderiv[2]{\Tn}{x}+\phie\mathcal{Q}_e\qne+\phisn\Qsn\qna+\alphan\qnr
\end{align}
\end{subequations}
with additional non-dimensional numbers
\begin{align}
\begin{aligned}
\Le=\frac{\kp}{\Rcpp D_{e}^0},\quad \mathcal{Q}_e=\frac{i_0^2F}{\sigma_e^0 \mnn{a}{p}\mnn{g}{p}^0RT_a},&\quad \Qsi=\frac{i_0^2F}{\sigmasi^0\mnn{a}{p}\mnn{g}{p}^0RT_a},\\
\vrci=\frac{\Rcpi}{\Rcpp},\quad \vKi=\frac{\ki}{\kp},&\quad \alphan=\frac{\mnn{a}{n}\mnn{g}{n}^0}{\mnn{a}{p}\mnn{g}{p}^0} = \frac{\Gn}{\Gp},
\end{aligned}
\end{align}
where $\Le$ is the Lewis number which represents the ratio of thermal diffusion to mass diffusion and $\mathcal{Q}_i$ is the ratio of Ohmic heat generated to that from surface reactions.  Furthermore, $\vrci$, $\vKi$, and $\alphan$ are the ratios of electrode and ionic conductivities, phase-averaged volumetric heat capacities, phase-averaged thermal conductivities, and surface electrode currents respectively.  The non-dimensional heat sources $\mnn{q}{i}$ are given by
\begin{subequations}
\begin{align}
\qia &= \frac{\isi^2}{\mn{\sigma}{a}{i}(\Ti)}\\
\qie &= \frac{\iei^2}{\sigma_e(\Ti)} +\frac{\DL D_L(\Ti) - \Da D_A(\Ti)}{\sigma_e(\Ti)}\,\iei\,\pderiv{\cli}{x}\\
% \qir &= \gibar \big\{\etai + (1 + \gammat \Ti)\left[\log(\Ui \Vi) + \DelGi \right]\big\},
\qir &= \gibar \left(\Phisi - \Phiei + \log \Ui + \DelGi \right). \label{nondim:qir}
\end{align}
\end{subequations}
%where $\DelGi = \mn{\Gamma}{L}{i} - \mn{\Gamma}{a}{i} = (\mn{E}{L}{i} - \mn{E}{a}{i}) / (R T_a)$.
% The dimensionless electrochemical heat generation \eqref{nondim:qir} can be decomposed into reversible and irreversible parts as
% \subeq{
%   \begin{align}
%     \qir^\text{rev} &= (1 + \gamma_T T) \gibar \left[\log(\Ui \Vi) + \DelGi\right], \\
%     \qir^\text{irr} &= \gibar\left[\Phisi - \Phiei - \gamma_T T \left(\log \Ui + \DelGi\right) - (1 + \gamma_T T)\log\Vi\right].
%   \end{align}
% }

\subsubsection{Boundary and initial conditions}
The dimensionless boundary conditions at the positive electrode-collector interface are
\begin{subequations}
\label{sys:BC}
\begin{alignat}{2}
\iep&=0, \quad &&x = 0, \label{bc:nd_current_2} \\
\pderiv{\clp}{x} &= 0, \quad &&x = 0, \label{bc:no_flux_1}\\
\pderiv{\Tp}{x}&=\Bi\,\Tp, \quad &&x=0, \label{bc:newton_1}
\end{alignat}
where $\Bi=\hp L/\kp$ is the Biot number and represents the ratio of convective to conductive heat transfer at the boundary.
% Due to the non-dimensional variables for the separator accounting for the differences in the porosity between the cell components,
The continuity conditions for the liquid-phase lithium concentration at electrode-separator interfaces are given by
\begin{alignat}{2}
  \cli - \cls &= 0, \quad &&x = x_p,\,x_n; \label{bc:cl_cont1}\\
  \pderiv{\cli}{x}-\pderiv{\cls}{x}&=0,\quad &&x = x_p,\,x_n. \label{bc:cl_cont2}
\end{alignat}
Similarly, the electrical potentials and currents satisfy
\begin{alignat}{2}
% \pderiv{\csi}{x}&=0,\quad &&x = x_p,\,x_n;\label{bc:no_flux_2} \\
\pderiv{\Phiei}{x}-\pderiv{\Phies}{x}&=0,\quad &&x = x_p,\,x_n.\\
\isi&=0,\quad &&x = x_p,\,x_n; \label{bc:nd_current_3} \\
\Phiei-\Phies &= 0,\quad &&x = x_p,\,x_n. \label{bc:phi_e_cont}
\end{alignat}
The continuity conditions for the temperature and thermal flux are
\begin{alignat}{2}
\Ti-\Ts&= 0, \quad && x= x_p,\ x_n; \label{bc:nd_T_cont}\\
\pderiv{\Tp}{x}-\vKs\pderiv{\Ts}{x}&= 0, \quad && x=x_p; \label{bc:nd_Txp_cont}\\
\vKn\pderiv{\Tn}{x}-\vKs\pderiv{\Ts}{x}&= 0, \quad && x=x_n.\label{bc:nd_Txn_cont}
\end{alignat}
The conditions at the negative electrode-collector interface are
\begin{alignat}{2}
\Phien&=0, \quad &&x = 1\label{bc:phi_e_ground} \\ 
\ien&=0, \quad &&x = 1, \label{bc:nd_current_4} \\
\pderiv{\cln}{x} &= 0, \quad &&x = 1, \label{bc:no_flux_3}\\
\vKn\pderiv{\Tn}{x}&=-\vHn\Bi\,\Tn, \quad &&x =1, \label{bc_newton2}
\end{alignat}
\end{subequations}
where $\vHn=\hn/\hp$ is the ratio of negative and positive electrode heat transfer coefficients.
Under galvanostatic or galvanodynamic conditions, we impose
\begin{align}
\phisp\isp=-\mathcal{I}(t), \quad x = 0. \label{bc:nd_current_1}
\end{align}
Alternatively, the dimensionless cell potential $\Delta v = (F \Delta V)/(R T_a) - \log \Up + \log \Un$ can be prescribed and $\I(t)$ solved for. 

Finally, the non-dimensional initial conditions are given by $\csi(x,0) = \cli(x,0) = \Phisi(x,0) - \Phiei(x,0) = \Ti=0$.
%$\csi(x,0) = 0$, $\cli(x,0) = 0$, $\Phiei(x,0) = 0$, $\Phisi(x,0) = 0$, and $\Ti=0$.

\section{Reduction of the cell model}
\label{sec:reduction}

Using the parameters in Tables \ref{tab:params1}--\ref{tab:thermal} results in the following non-dimensional numbers associated with the electrochemical model:
\subeq{
  \label{eqn:nondim_num_ec}
\begin{align}
  \DL \simeq 0.79, \quad \Da \simeq 1.4, \quad \nusp \simeq 0.032, \quad \nusn \simeq 6.9\tten{-4}, 
  \quad \nue \simeq 0.027, \\
  \Gp \simeq 5.2, \quad \Gn \simeq 4.9, \quad \Cp \simeq 5.1\tten{-3}, \quad \Cn \simeq 7.4\tten{-5}, \quad 
  \deltap \simeq 2.4, \quad \deltan \simeq 0.045, \\
  \quad \xip \simeq 0.022, \quad \xin \simeq 0.86, \quad \gammac \simeq 0.058, \quad
  \log \Up \simeq 137.9, \quad \log \Un \simeq 2.7.
\end{align}
}
The corresponding non-dimensional parameters for the thermal model are
\subeq{
  \label{eqn:nondim_num_thermal}
\begin{align}
  \gammat \simeq 2.2\tten{-3}, \quad \Bi \simeq 9.0\tten{-4}, \quad \Le \simeq 2.8\tten{3}, \\ 
  \Qe \simeq 5.1\tten{-3}, \quad \Qsp \simeq 6.1\tten{-3}, \quad \Qsn \simeq 1.3\tten{-4}, \\
  \Gammaeffp \simeq 0.018, \quad \Gammaeffn \simeq 0.026, \quad \DelGp \simeq -134, \quad \DelGn \simeq -3.0, \\
  \vrcn \simeq 1.1, \quad \vrcs \simeq 1.2, \quad \vKn \simeq 0.71, \quad \vKs \simeq 0.35, \quad \vHn \simeq 1.0, \quad \alphan \simeq 0.93.
\end{align}
}
The cell-scale model will now be simplified by exploiting the smallness of these parameters.

%Importantly, the smallness of $\gamma_T$, which is a measure of the typical temperature increase $\Delta T$ to the ambient temperature $T_a$, implies that the temperature dependence of many material parameters can be neglected. 
%Discuss non-dim parameter values, show $\gamma_T \ll 1$ so very small temperature changes. Implies that temperature-dependent of material parameters can be neglected. Electrochemical model therefore decouples from the thermal model and can be reduced following the strategy outlined in the JPS paper...

\subsection{Simplification of the thermal problem}
\label{sec:prelim_thermal}
The heat equation \eqref{sys:heat} can be simplified by neglecting terms relating to Ohmic heating ($\Qsi \ll 1$ and $\Qe \ll 1$). The large Lewis number, $\Le$, implies that rapid thermal diffusion will lead to a temperature profile that is roughly uniform across the cell.  Thus, the temperatures are written as $\Ti(x,t) = T(t) + \Le^{-1}\tilde{\Ti}(x,t)$. Integrating the heat equations across their respective domains, adding them together, and using the boundary conditions results in a single heat equation for the dimensionless cell temperature $T$ given by
\begin{align}
  \cellav{\varrho} \deriv{T}{t} = -\Le\, \Bi (1 + \vHn) T + x_p \posav{\qpr} + (1 - x_n) \alphan \negav{\qnr} + O(\Le^{-1}),
  %\qpi = \gbari(\Phisi - \Phiei + \log \Ui + \DelGi)
  \label{eqn:cell_T}
\end{align}
where $\cellav{\varrho} = x_p + (x_n - x_p) \vrcs + (1 - x_n) \vrcn$ is the cell-averaged dimensionless heat capacity and
\begin{align}
  \posav{\qpr} = \frac{1}{x_p}\int_{0}^{x_p} \qpr\,\d x, \qquad
  \negav{\qnr} = \frac{1}{1-x_n}\int_{x_n}^{1} \qnr\,\d x,
  \label{eqn:av_qri}
\end{align}
are the component-averaged volumetric heating terms due to electrochemical reactions. In deriving \eqref{eqn:cell_T}, we have taken $\Le\,\Bi = O(1)$ in size, as justified by the parameter values in \eqref{eqn:nondim_num_thermal}, which is a distinguished limit of the model. For $\Le\,\Bi \gg 1$ there is rapid heat exchange with the environment, leading to small temperature changes of size $O(\Le^{-1}\Bi^{-1})$.  For $\Le\,\Bi \ll 1$, heat exchange occurs slowly and the temperature can undergo substantial variations. The small terms of size $O(\Le^{-1})$ in the heat equation \eqref{eqn:cell_T} will be neglected in the analysis below.

%\mh{From \eqref{eqn:cell_T} we see that the steady-state temperature will scale like $\Le^{-1}\Bi^{-1}$ and, as expected, is set by a balance between thermal diffusion and heat exchange with the environment.}

\subsection{Simplification of the electrochemical kinetics}
\label{sec:prelim_kinetics}

The dimensionless Bulter--Volmer kinetics \eqref{sys:nondimbutler} can be simplified using the fact that $\gammat \ll 1$.  
%For the moment we suppose that $\Gammaeffi$ is not small.  
We also set $\betai = 1/2$ in order to write asymptotic solutions in terms of hyperbolic trigonometric functions. This is equivalent to the common assumption of symmetric reactions at the electrode surface \cite{Li2014,Biesheuvel2010,Zhao2016,Smith2017}. Thus, the simplified electrochemical kinetics are given by
\begin{subequations}\label{sys:nondimbutler_simple}
\begin{align}
%   \gibar=\ji\left(\exp\left[(1-\betai)\etai\right]-\exp\left[-\betai\etai\right]\right),\\
%   \ji&=\Keffi(T)\left(1+\deltai\gammac\csi\right)^{\betai}\left(1-\deltai\xii(1-\xii)^{-1}\gammac\csi\right)^{1-\betai}, \label{eqn:nd_j0i_simple}\\
  \gibar&=2\ji\sinh(\etai/2), \\
  \ji&=\Keffi(T)\left(1+\deltai\gammac\csi\right)^{1/2}\left(1-\deltai\xii(1-\xii)^{-1}\gammac\csi\right)^{1/2}(1+\gammac\cli)^{1/2}, \\
  \Keffi &= \exp(\Gammaeffi T), \label{red:Keffi}
\end{align}
\end{subequations}
%\im{where terms $\gamma\csi$ are retained in \eqref{eqn:nd_j0i_simple} but $\gamma\cli$ are not. This is due to the expected relaxation of the electrolyte concentration to an $\Ord{1}$ steady state consistent with the isothermal problem derived by Moyles \etal \cite{Moyles2019}.}
To simplify the overpotential given by \eqref{eqn:butleretadef}, we first note that the combination $\log \Ui + \DelGi$ is $O(1)$ in magnitude despite the individual contributions being large for the positive electrode. The term $\gammat T (\log \Ui + \DelGi)$ will therefore be small and can be neglected, leading to
\begin{align}
\etai=\Phisi - \Phiei -\log \Vi, \label{eqn:butleretadef_simple}
\end{align}
with $\Vi$ as in \eqref{eqn:UVdef}.
% \begin{align}
%   \Vi=\frac{[1-\deltai\xii(1-\xii)^{-1}\gammac\csi]}{(1+\deltai\gammac\csi)}.
%   %\quad
%   %\Wi = \exp\left(\gamma_T \DelGi\, T\right).
%   %\Wi = \frac{\mn{K}{L}{i}(T)}{\mn{K}{a}{i}(T)}.
% \label{eqn:UVdef_simple}
% \end{align}
The simplified surface heating terms are given by
\begin{align}
  \qir = \gibar \left[\Phisi - \Phiei + \log \Ui + \DelGi\right],
  \label{red:qir}
\end{align}
which is exactly \eqref{nondim:qir}, but with $\gibar$ defined by \eqref{sys:nondimbutler_simple}.

Equation \eqref{red:Keffi} implies that Arrhenius effects will become relevant if the dimensionless temperature reaches values given by $T \sim \Gammaeffn^{-1} \sim 38$, which is equivalent to a dimensional temperature rise of $\Delta T \sim 25$~K. As shown in Sec.~\ref{sec:homogenisation},  a temperature rise of this magnitude is achievable by coupling several LIB cells together.
%Therefore, the temperature dependence of $\Keffi$ will be retained despite small values of $\Gammaeffi$.

\subsection{Simplification of the electrochemical model}

Due to the spatial uniformity of the temperature, the problem for the concentration of lithium ions in the electrolyte simplifies to
\subeq{
  \label{simp:cli}
\begin{align}
  \pderiv{\cli}{t} &= D_e(T)\pderiv[2]{\cli}{x} + [1 - \theta(T)]\pderiv{\iei}{x}, \\
  \pderiv{\cls}{t} &= D_e(T)\pderiv[2]{\cls}{x},
\end{align}
}
together with the boundary conditions \eqref{bc:no_flux_1}, \eqref{bc:cl_cont1}, \eqref{bc:cl_cont2}, and \eqref{bc:no_flux_3}.

Reductions also apply to the electrochemical problem for the solid phase \eqref{sys:electrode_solid} but these depend on the type of discharge condition (fixed current or fixed voltage) used and will be discussed appropriately.

\section{Analysis of the cell problem}
\label{sec:asymptotics}
The reduced model is now used to understand the thermal behaviour of an LIB cell under various operation modes. Matched asymptotic expansions are used to calculate approximate solutions in key time regimes.
The parameters $\nusi$ and $\gammac$ will be assumed to have the same order of magnitude as $\nue$. By doing so, a regular asymptotic expansion in powers $\nue$ can be used in each time regime to systematically capture the effects of gradients in the lithium concentration and electrical potentials. %The strategy is to first asymptotically solve the electrochemical problem and then use the solution to derive the simplified thermal problem. This is possible due to the relatively weak coupling between the thermal and electrochemical problems.

\subsection{Galvanostatic discharge}
\label{sec:galvanostatic}

We first consider the case where the discharge current $\mathcal{I}$ is constant in time. The isothermal version of this problem has been solved using asymptotic methods by Moyles \etal\cite{Moyles2019}, who found four distinct time regimes, the most relevant of which for this mode of battery operation occurs when $t = O(\gammac^{-1})$.  On this time scale, large changes in the concentration of intercalated lithium occur, whereas the liquid-phase lithium concentration remains $O(1)$ in size due to a balance between diffusion and electrochemical consumption and generation.  Following Ref.~\cite{Moyles2019}, we rescale the variables as $t = \gamma_c^{-1} \hat{t}$, $\csi = \gamma_c^{-1} \csihat$.  It is also convenient to introduce a rescaled temperature $\hat{u}$ defined by $T = \mathcal{T} \hat{u}$, where $\mathcal{T} = \Le^{-1} \Bi^{-1}(1 + \vHn)^{-1} \Gp^{-1}$. All of the variables on this time scale are denoted with a hat and written as asymptotic expansions of the form $\hat{f} = \hat{f}^{(0)} + \nue \hat{f}^{(1)} + O(\nue^2)$. 
%As previously mentioned, we assume that $\nusi/\nue = O(1)$ and $\gammac/ \nue = O(1)$. 
Capacitance effects are negligible on this time scale and do not need to be considered in \eqref{eqn:div_isi}.

\subsubsection{The leading-order problem}
\label{sec:galvanostatic_lo}

The leading-order problem is obtained by setting $\nue = 0$, $\nusi = 0$, and $\gammac = 0$ in the rescaled system.
%The analysis is similar to in Moyles \etal\cite{Moyles2019b} so we only summarise the details here.
Equations \eqref{eqn:ohma}, \eqref{eqn:ohme}, and \eqref{eqn:ohmes} imply that the electric potentials are uniform in space to leading order. The grounding condition on the electrolyte potential gives that $\Phieihat^{(0)}(x,t) = 0$. Since the concentration of intercalcated lithium is initially uniform in space, the spatially independent potentials further imply that the leading-order contribution $\csihat^{(0)}$ remains uniform for all time. Thus, integration of \eqref{eqn:nd_csi} over its respective domain shows that the solid-phase concentration changes linearly in time according to
\begin{align}
  \csnhat^{(0)}(\hat{t}) = -\frac{\mathcal{I}}{\phisn(1 - x_n)}\,\hat{t},  \qquad
  \csphat^{(0)}(\hat{t}) = \frac{\mathcal{I}}{\phisp x_p}\,\hat{t}.
  \label{const_I:csi_0}
\end{align}
Similarly, the leading-order parts of \eqref{eqn:div_isi} can be integrated in space to yield nonlinear equations for the solid-phase potentials,
\begin{align}
  \gpbar^{(0)}(\Phisphat^{(0)}, \csphat^{(0)}, \hat{u}^{(0)}) = -\frac{\mathcal{I}}{\Gp x_p}, \qquad
  \gnbar^{(0)}(\Phisnhat^{(0)}, \csnhat^{(0)}, \hat{u}^{(0)}) = \frac{\mathcal{I}}{\Gn(1 - x_n)},
  % 2 \jp^{(0)} \sinh\left(\etap^{(0)}/2\right) = -\frac{\mathcal{I}}{\Gp x_p}, \quad
  % 2 \jn^{(0)}\sinh\left(\etan^{(0)}/2\right) = \frac{\mathcal{I}}{\Gn(1 - x_n)},
  \label{const_I:Phi}
\end{align}
where the electrochemical kinetics are given by
\begin{subequations}
  \label{const_I:BV}
\begin{align}
  \gibar^{(0)}&=2\ji^{(0)}\sinh\left(\etaihat^{(0)}/2\right), \label{const_I:gi_0} \\
  \ji^{(0)}&=\Keffi^{(0)}\left(1+\deltai\csihat^{(0)}\right)^{1/2}\left(1-\deltai\xii(1-\xii)^{-1}\csihat^{(0)}\right)^{1/2}, \label{const_I:ji_0}\\
  \Keffi^{(0)} &= \exp\left(\Gammaeffi\mathcal{T}\hat{u}^{(0)}\right), \label{const_I:K_eff}
\end{align}
\end{subequations}
with $\etaihat^{(0)} =\Phisihat^{(0)}-\log\hat{\Vi}^{(0)}$ and
\begin{align}
  \hat{\Vi}^{(0)} &=\frac{1-\deltai\xii(1-\xii)^{-1}\csihat^{(0)}}{1+\deltai\csihat^{(0)}}.
\end{align}
The current densities in the solid-phase of the electrodes are given by
\begin{align}
  \isphat^{(0)} = -\frac{\mathcal{I}}{\phisp x_p}(x_p - x), \quad
  \isnhat^{(0)} = -\frac{\mathcal{I}}{\phisn(1-x_n)}(x - x_n),
  \label{red:isi}
\end{align}
and the current densities in the electrolyte are
\begin{align}
  \iephat^{(0)} = -\frac{\mathcal{I}}{\phie x_p}\,x,
  \quad
  \ieshat^{(0)} = -\frac{\mathcal{I}}{\phie},
  \quad
  \ienhat^{(0)} = -\frac{\mathcal{I}}{\phie(1-x_n)}(1 - x).
  \label{red:iei}
\end{align}
The leading-order liquid-phase lithium concentration is given by its quasi-steady profile, $\clihat^{(0)} = \mathcal{I}[1-\theta(\mathcal{T}\hat{u}^{(0)})] / [D_e(\mathcal{T}\hat{u}^{(0)})\phie] \cli^\infty$, where
\begin{align}
  \clp^\infty = \frac{1}{2}\left[\frac{x^2}{x_p} + x_p + \mathcal{B}\right],
  \quad
  \cls^{\infty} = x + \frac{\mathcal{B}}{2},
  \quad
  \cln^{\infty} = \frac{1}{2}\left[1+x_n + \mathcal{B} - \frac{(1-x)^2}{(1-x_n)}\right],
  \label{const_I:c_L_inf}
\end{align}                    
and $\mathcal{B} = (1/3)[(1-x_n)^2 - x_p^2] - 1$.

We are now in a position to solve for the leading-order contribution to the temperature. The electrochemical heating terms \eqref{red:qir} can be simplified using \eqref{const_I:Phi}. Upon taking the limit $\gamma_c \to 0$ in the rescaled heat equation, we find that the leading-order temperature evolves in a quasi-static manner so that
\begin{align}
  \begin{split}
  % \hat{T}(\hat{t}) = \Le^{-1} \Bi^{-1}(1 + \vHn)^{-1} \mathcal{I} \left[-\Gp^{-1}(\Phisphat(\hat{t})+ \DelGp) + \Gn^{-1}\alpha (\Phisnhat(\hat{t}) + \DelGn)\right].
    \hat{u}^{(0)}(\hat{t}) &= \mathcal{I}\left[\Phisnhat^{(0)} - \Phisphat^{(0)} + \log \Un - \log \Up + \DelGn
  - \DelGp\right].
%&+ \nue \mathcal{I}\left[\posav{\Phisp^{(1)} - \Phiep^{(1)}} - \negav{\Phisn^{(1)}-\Phien^{(1)}}\right] + O(\nue^2).
  \label{const_I:temp_0}
\end{split}
\end{align}
Due to the temperature dependence of the reaction constant in \eqref{const_I:K_eff}, the expression in \eqref{const_I:temp_0} must be regarded as a nonlinear equation for the temperature that must be solved at each point in time. If the temperature dependence is neglected, then the leading-order electrochemical problem completely decouples from the thermal problem. In this case, there is no positive feedback mechanism that could lead to thermal runaway and unsafe overheating of the LIB cell. Even if there was a strong dependence of $\Keffi$ on temperature, then from the electrochemical kinetics \eqref{const_I:BV} it can be seen that a large increase in $\Keffi$ only acts to push the electric potentials closer to the open-circuit potentials ($\Phisihat^{(0)} \to \log \Vi^{(0)}$). Consequently, there is no mechanism for auto-catalytic heat generation and thermal runaway.  However, it is important to point out that the terms associated with electrochemical heat generation become singular as the battery becomes fully charged or fully drained.  This singularity is due to an incompatibility with the constant-current boundary condition and results in the solid-phase potentials blowing up at the end of the (dis)charge process \cite{Moyles2019}.

% which has a minimal impact on the temperature as calculated from \eqref{const_I:temp_0}.

\subsubsection{The first-order problem}

The corresponding equations for the electrode solid phase are
\subeq{
\begin{align}
  \pderiv{\csihat^{(1)}}{\hat{t}} &= \pderiv{\isihat^{(1)}}{x}, \label{const_I:csi_1_de} \\
  (\nusi/\nue)\isihat^{(0)} &= -\sigmasi(\mathcal{T}\hat{u}^{(0)})\pderiv{\Phisihat^{(1)}}{x}, \\
  \phisi\pderiv{\isihat^{(1)}}{x} &= -\Gi\gibar^{(1)}, \label{const_I:div_I_1}
\end{align}
}
which are supplemented with the boundary and initial conditions $\isphat^{(1)}(0,\hat{t}) = \isphat^{(1)}(x_p,\hat{t}) = 0$,  $\isnhat^{(1)}(x_n,\hat{t}) = \isnhat^{(1)}(1,\hat{t}) = 0$, and $\csihat^{(1)}(x,0) = 0$. The electrolyte potential can be obtained from
\begin{align}
  \ieihat^{(0)} = -[\DL D_L(\mathcal{T}\hat{u}^{(0)}) - \Da D_A(\mathcal{T}\hat{u}^{(0)})]\pderiv{\clihat^{(0)}}{x} - \sigma_e(0,\mathcal{T}\hat{u}^{(0)})\pderiv{\Phieihat^{(1)}}{x},
\end{align}
with $\Phienhat^{(1)}(1,t) = 0$ and continuity at $x=x_p$ and $x=x_n$.
The correction to the Butler--Volmer reaction current is given by
\subeq{
  \label{eqn:BV_corr}
\begin{align}
  \gibar^{(1)} &= 2\ji^{(1)}\sinh\left(\etai^{(0)}/2\right) + \ji^{(0)} \etai^{(1)}\cosh\left(\etai^{(0)}/2\right), \\
  \ji^{(1)} &= (1/2)\ji^{(0)}\left\{(\gammac/\nue)\cli^{(0)} + \deltai\left[\frac{1}{1+\deltai\csihat^{(0)}} - \frac{\xii(1-\xii)^{-1}}{1-\deltai\xii(1-\xii)^{-1}\csihat^{(0)}}\right]\csihat^{(1)} + 2 \Gammaeffi \mathcal{T} \hat{u}^{(1)}\right\},
\end{align}
}
where $\etai^{(1)} = \Phisi^{(1)} - \Phiei^{(1)} - \Vi^{(1)}/\Vi^{(0)}$ and
\begin{align}
  \Vi^{(1)} = \Vi^{(0)}\left\{(\gammac/\nue)\cli^{(0)} - \deltai\left[\frac{1}{1+\deltai\csihat^{(0)}} + \frac{\xii(1-\xii)^{-1}}{1-\deltai\xii(1-\xii)^{-1}\csihat^{(0)}}\right]\csihat^{(1)}\right\}.
\end{align}
In order to solve this problem, we first note that integration of \eqref{const_I:csi_1_de} over the electrode domain and using the boundary and initial conditions reveals that the net corrections to the solid-phase concentrations are zero:
\begin{align}
  \int_{0}^{x_p}\csphat^{(1)}\,\d x = 0, \qquad \int_{x_n}^{1}\csnhat^{(1)}\,\d x = 0.
  \label{eqn:zero_net}
\end{align}
The correction to the steady-state electric potential of the electrolyte can be written as
\subeq{
  \label{const_I:Phi_e}
\begin{align}
  \Phieihat^{(1)} = \left\{\frac{\Da D_A(\mathcal{T}\hat{u}^{(0)})\mathcal{I}}{2D_e(\mathcal{T}\hat{u}^{(0)})\phie\sigma_e(0,\mathcal{T}\hat{u}^{(0)})}\right\}\Phiei^{\infty}
\end{align}
where
\begin{align}
  \Phiep^\infty = \frac{x^2 - (1 + x_n)x_p + x_p^2}{x_p},
  \quad
  \Phies^{\infty} = 2x - 1 - x_n,
  \quad
  \Phien^{\infty} = -\frac{(1-x)^2}{1-x_n}.
\end{align}
}
The corrections to the solid-phase potential are given by $\Phisphat^{(1)} = \Phisp^\infty + P_p$ and $\Phisnhat^{(1)} = \Phisn^{\infty} + P_n$, where
\begin{align}
  \Phisp^{\infty} = \frac{1}{2}\left(\frac{\nusp}{\nue}\right)\frac{\mathcal{I}(2 x_p - x)x}{\phisp\sigmasp(\mathcal{T}\hat{u}^{(0)})x_p},
  \quad
  \Phisn^{\infty} = -\frac{1}{2}\left(\frac{\nusn}{\nue}\right)\frac{\mathcal{I}(1-x)(1 + x - 2x_n)}{\phisn\sigmasn(\mathcal{T}\hat{u}^{(0)})(1-x_n)},
  \label{const_I:Phi_s_1}
\end{align}
and $P_i$ are the constants of integration.  The constants $P_i$ are determined by solvability conditions that are obtained through the integration of \eqref{const_I:div_I_1}, yielding
\begin{align}
  \int_{0}^{x_p} \gpbar^{(1)}\,\d x = 0, \qquad \int_{x_n}^{1}\gnbar^{(1)}\,\d x = 0.
  \label{eqn:solvability}
\end{align}
Due to the conditions \eqref{eqn:zero_net} and the linear dependence of $\csi^{(1)}$ in $\gibar^{(1)}$, the contributions from $\csi^{(1)}$ to the integrals in \eqref{eqn:solvability} become zero, implying that the correction to the concentration of intercalated lithium does not influence the correction to the electrical problem. Having determined the constants $P_i$, the correction to the (dimensionless) cell potential is simply given by $\Delta v^{(1)} = P_p - P_n$.

The correction to the temperature simplifies due to the contributions from $\gibar^{(1)}$ integrating to zero due to \eqref{eqn:solvability}. Thus, we find that
\begin{align}
  \hat{u}^{(1)} = \mathcal{I}\left[\negav{\Phisnhat^{(1)}-\Phienhat^{(1)}} - \posav{\Phisphat^{(1)} - \Phiephat^{(1)}}\right] - (\gammac/\nue) \Gp \mathcal{T}\cellav{\varrho}\deriv{\hat{u}^{(0)}}{\hat{t}}.
\end{align}

\subsubsection{Comparison with numerics}

\begin{figure}
  \centering
  \subfigure[Leading-order]{\includegraphics[width=0.48\textwidth]{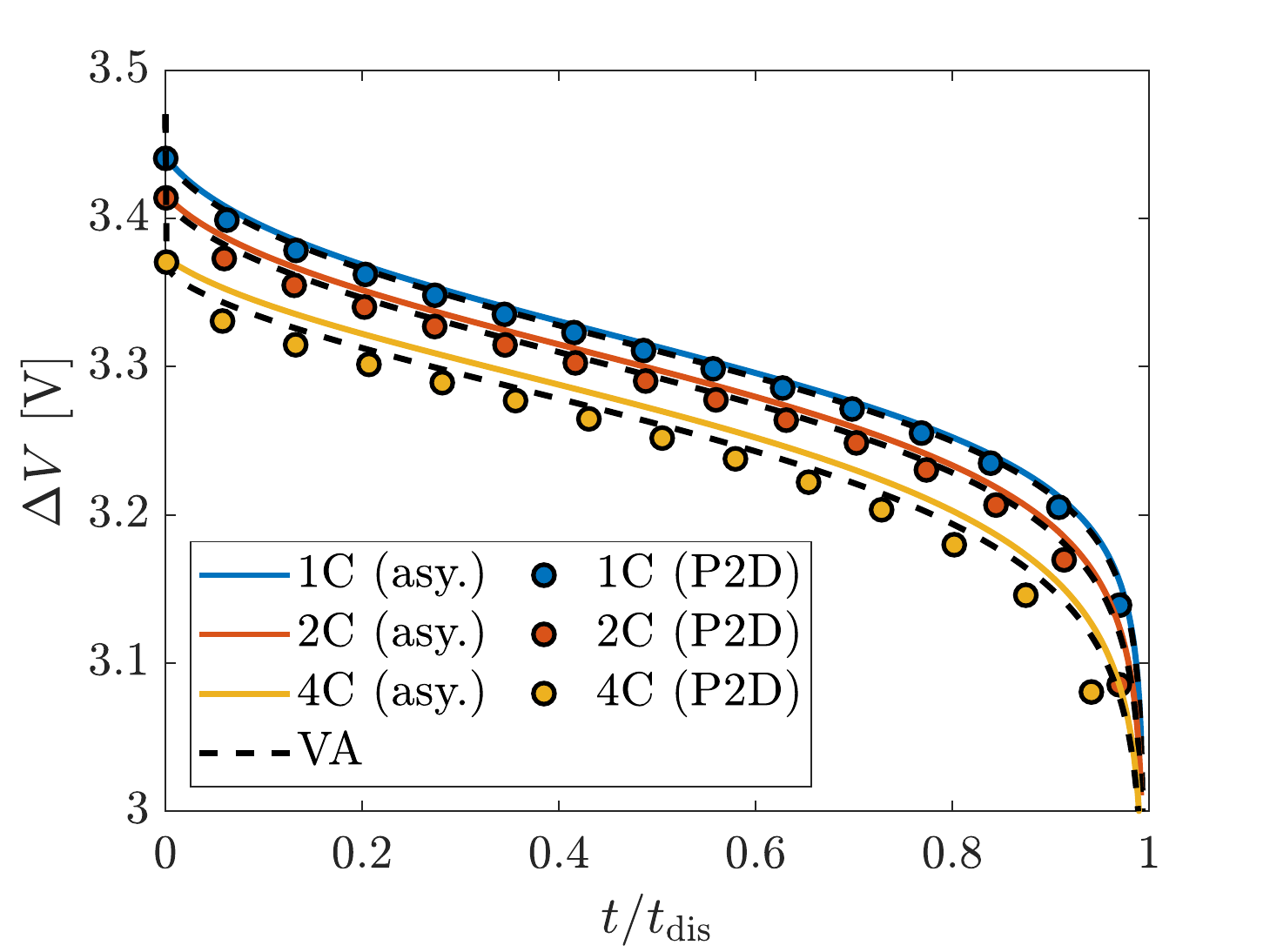}}
  \subfigure[Leading-order]{\includegraphics[width=0.48\textwidth]{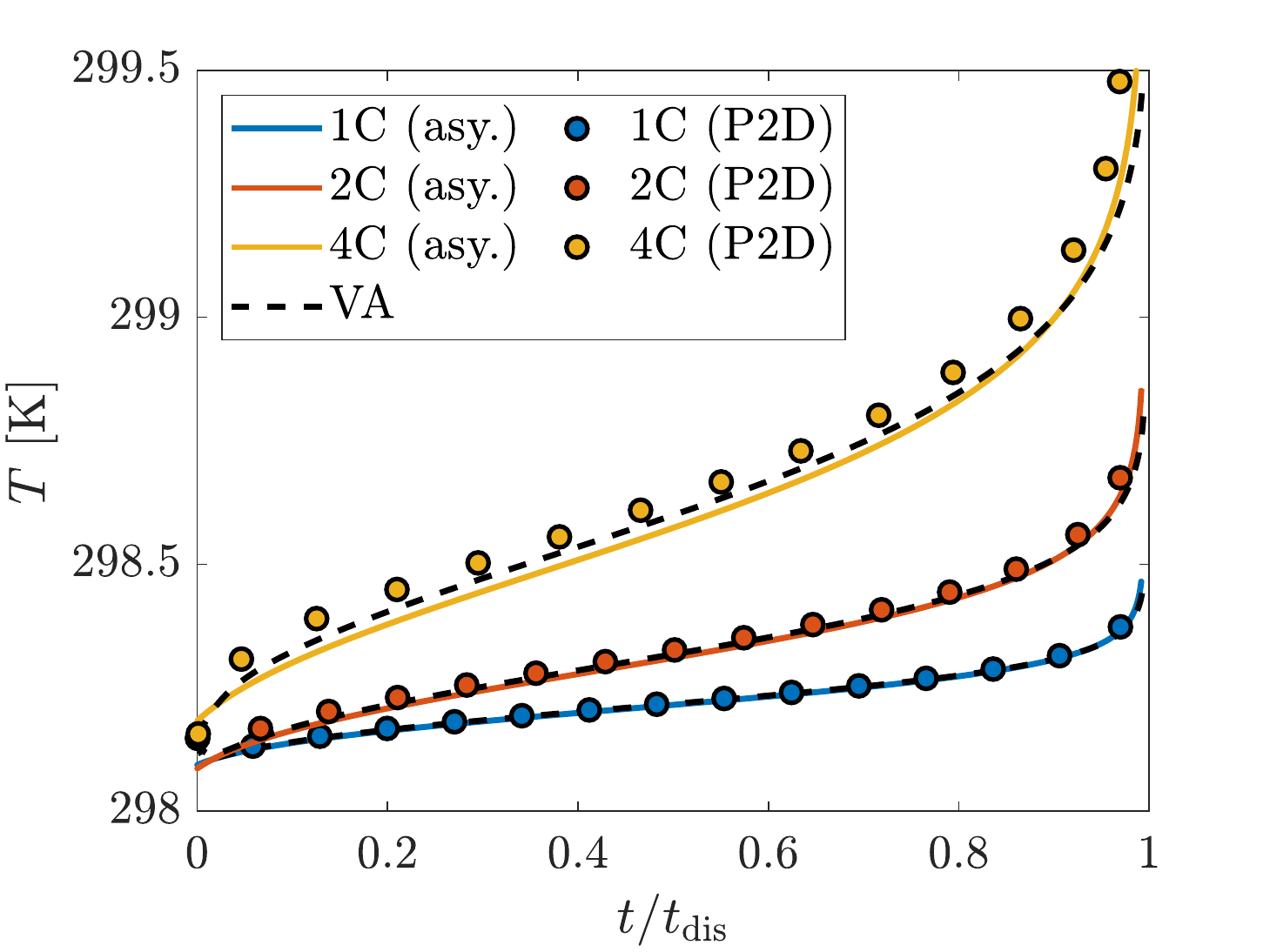}} \\
  \subfigure[Leading- and first-order]{\includegraphics[width=0.48\textwidth]{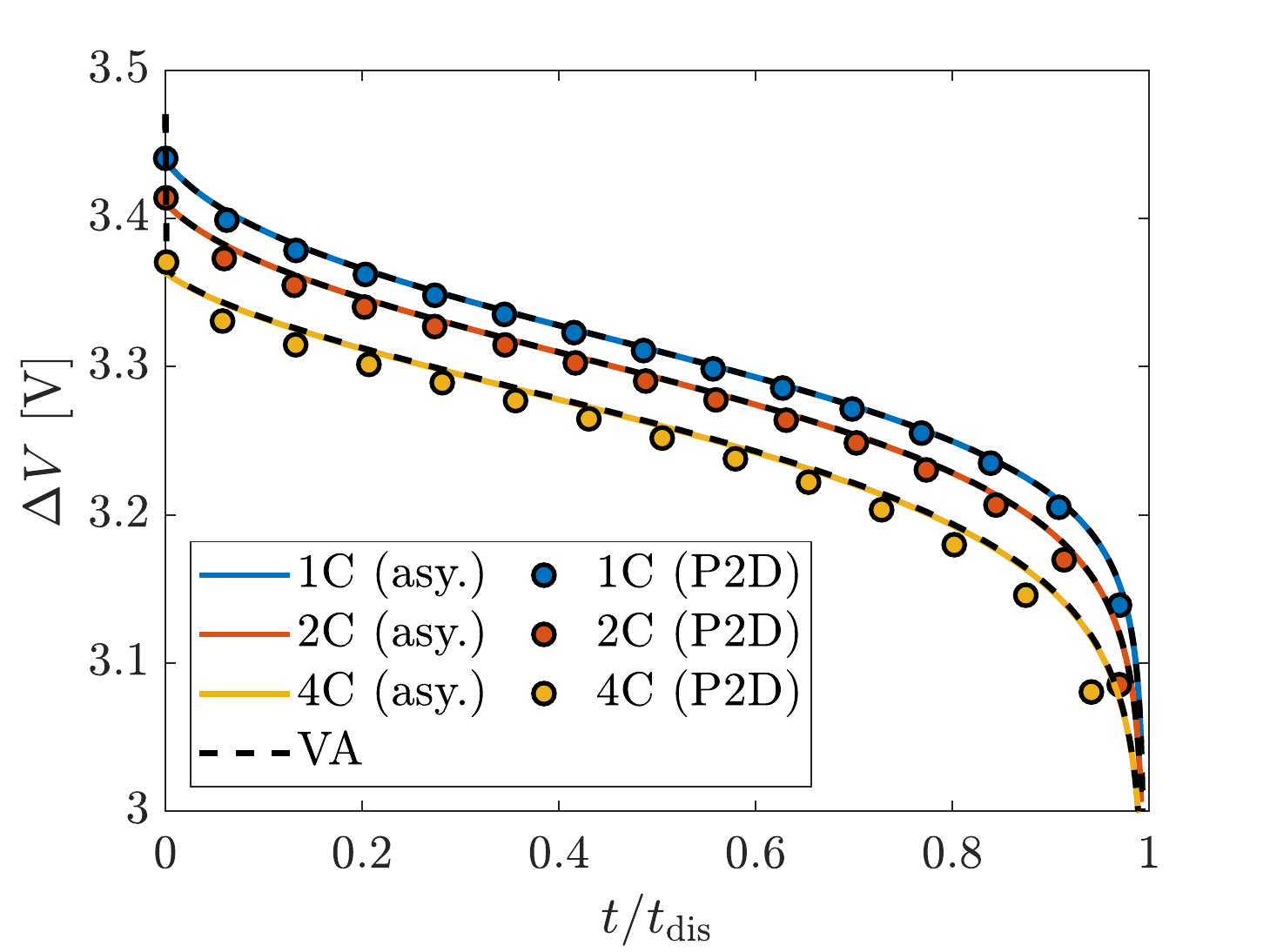}}
  \subfigure[Leading- and first-order]{\includegraphics[width=0.48\textwidth]{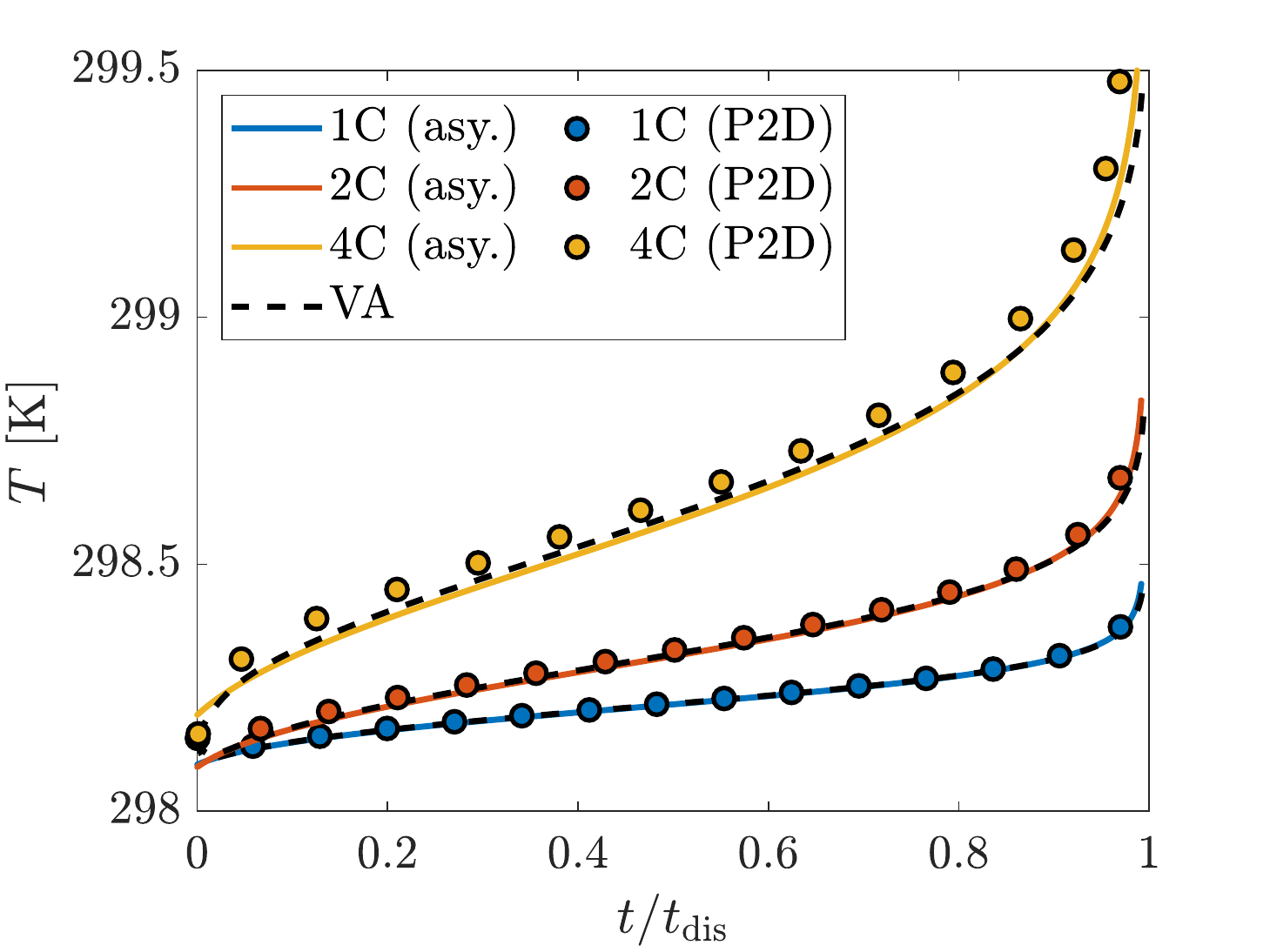}} \\
    \subfigure[Positive electrode]{\includegraphics[width=0.48\textwidth]{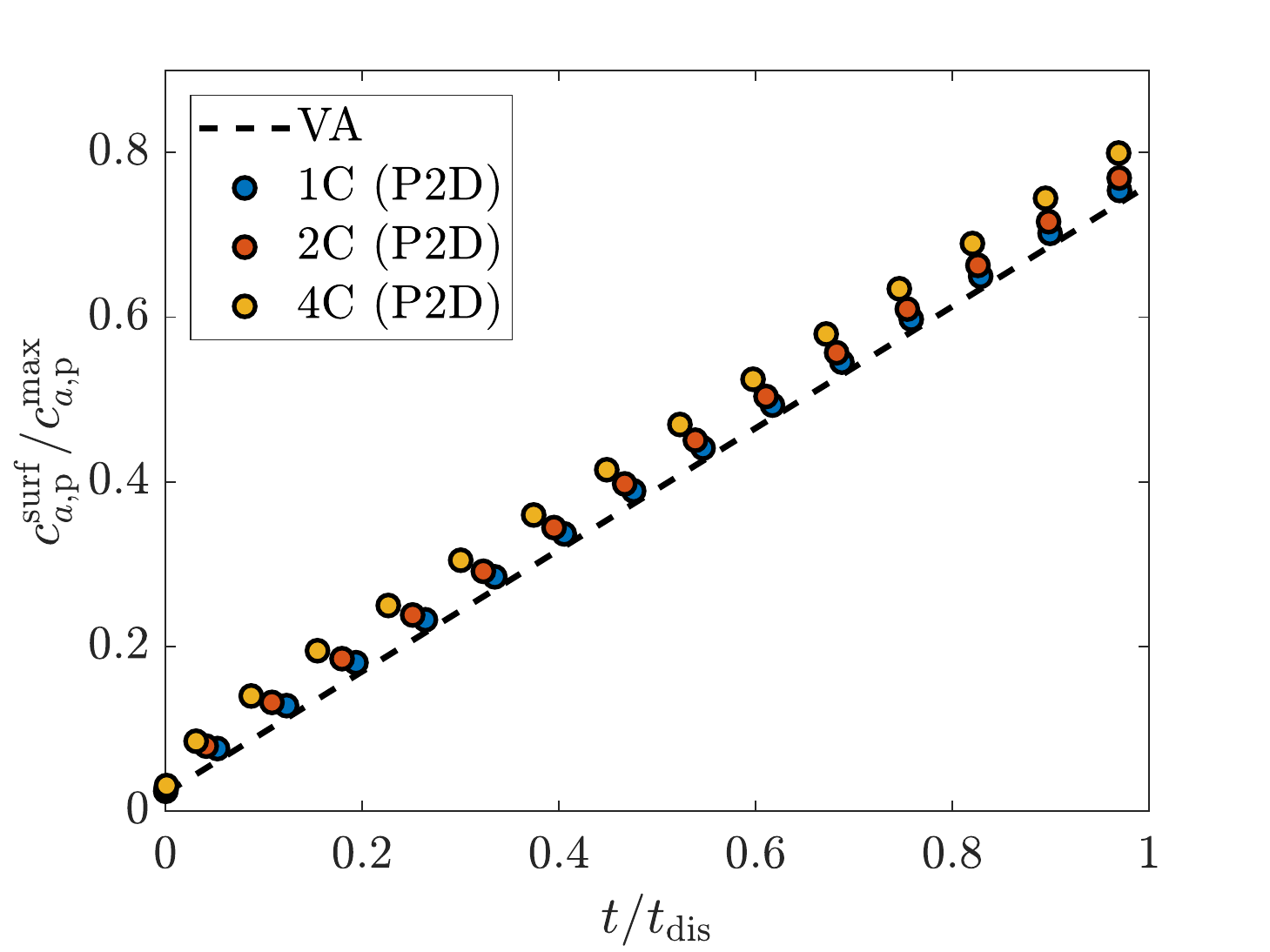}}
  \subfigure[Negative electrode]{\includegraphics[width=0.48\textwidth]{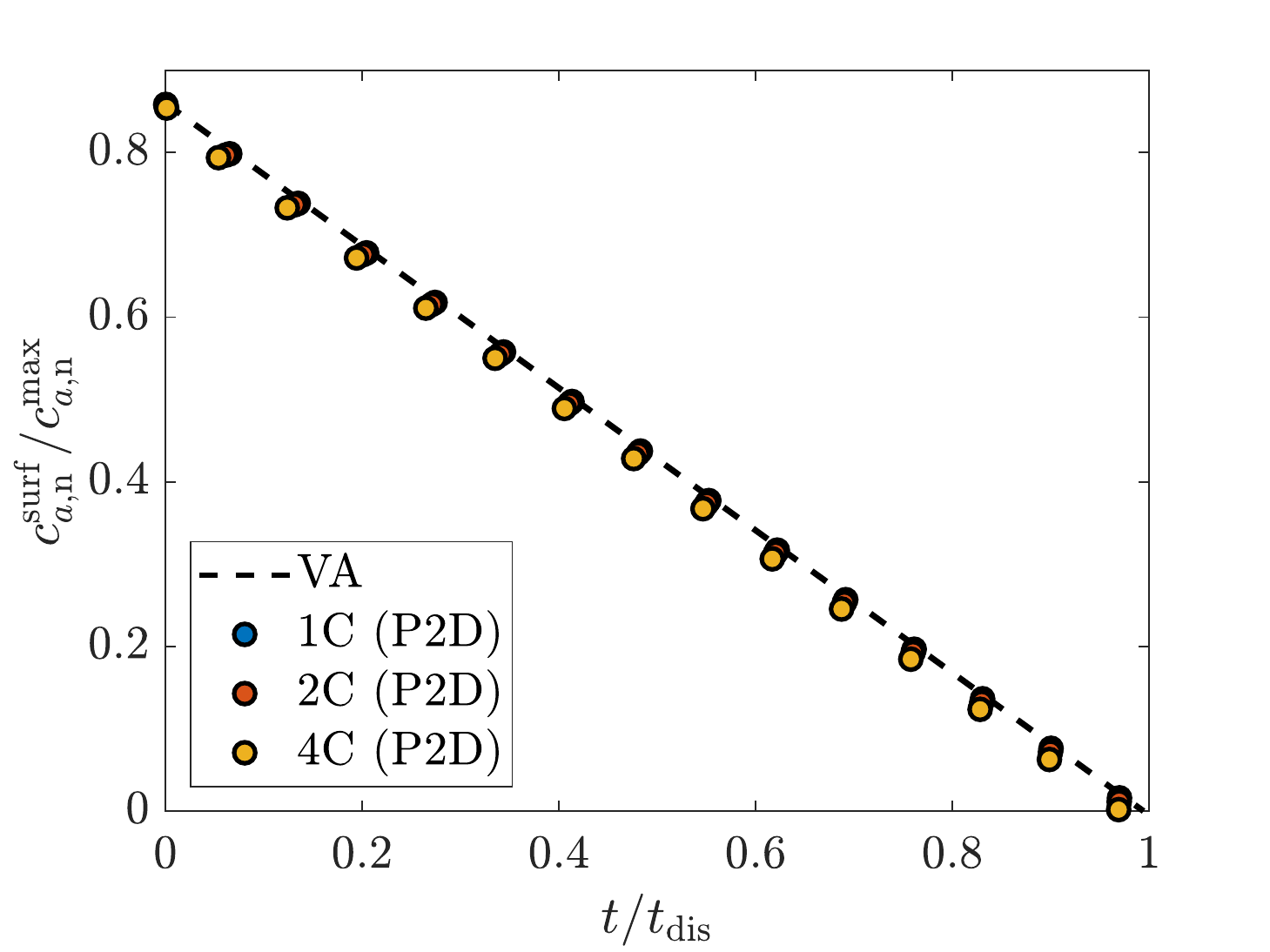}}

  \caption{Simulations of galvanostatic discharge processes at various C-rates $\mathcal{I} = 1$, $2$, $4$.  The asymptotic solutions in panels (a)--(b) are constructed using only the leading-order contributions; in panels (c)--(d) the leading- and first-order contributions are considered. The surface concentrations in panels (e)--(f) are averaged over each electrode.
    Dashed lines and circles denote numerical solutions of the VA and P2D model. Time is normalised using the discharge time $t_\text{dis} = (3600 / \mathcal{I})$~s. The temperature $T$ is averaged over a single cell. The surface concentrations $\csi^{\text{surf}}$ have been averaged over their respective electrode domains.}
  \label{fig:galvanostatic}
\end{figure}

The asymptotic solutions are compared with numerical simulations of the VA and P2D models by considering discharge processes of 1C, 2C, and 4C. These correspond to $\I = 1$, $2$, and $4$, respectively. The resulting discharge curves (cell potential as a function of time) and cell temperatures are shown in Fig.~\ref{fig:galvanostatic}. From the parameters in \eqref{eqn:nondim_num_thermal}, $\mathcal{T} \simeq 3.8\tten{-2}$ and therefore we neglect the temperature dependence of parameters other than the reaction rate. The asymptotic solutions that are constructed from only the leading-order contributions (see panels (a)--(b)) are in good agreement with the numerical solution of the VA model across all of the C-rates.  When the first-order contributions are included in the asymptotic solutions (see panels (c)--(d)), the agreement is nearly perfect. There is also good agreement between the cell potentials and temperatures computed from VA and P2D models, even at a high discharge rate of 4C.
The VA-P2D model divergence occurs near the end of the discharge process, which induces a singularity in the open-circuit potential. This divergence can be explained in terms of the different ways in which the models treat the solid concentrations. The VA model uses a bulk concentration which is equivalent to a uniform particle concentration. The P2D model however captures particle-scale diffusion which induces a concentration gradient between the surface and center of the particle. For slow diffusion in the particles, the surface concentration and hence the particle concentration gradients are amplified.  Near a singularity in the open-circuit potential, a change in the surface concentration due to slow diffusion can have a large impact on the electrochemical kinetics and drive the P2D model away from the VA model.
%Large concentration gradients increase the effect of concentration on the open-circuit potential near the singularities by virtue of equation \eqref{eqn:UVdef}, leading to a stronger deviation from the VA model at the singular tails of the open-circuit potential. 
A divergence between the models at the start of the discharge process, which is also close to singular point, is not observed in Fig.~\ref{fig:galvanostatic} because the time scale over which the concentration rapidly leaves the singular zone is fast.
%The differences are attributed to the finite rate of solid-phase lithium diffusion in the P2D model, which leads to concentration gradients near the particle surface. 
The electrode-averaged surface concentrations $\csi^{\text{surf}}$ are plotted as functions of time in Fig.~\ref{fig:galvanostatic} (e)--(f). In the VA model, the surface and bulk solid-phase concentrations are the same. Furthermore, when time is normalised by the discharge time $t_\text{dis} = (3600/\I)$~s, the results lie on a master curve (shown as the black dashed line) that is independent of the C-rate. The positive (negative) electrode (de)lithiates on discharge. The surface concentrations in the P2D model are therefore greater (smaller) compared to the bulk, which adds resistance and causes a greater voltage drop relative to the VA model. 

As predicted by the asymptotic analysis, the temperature change that occurs during discharge is small. This is due to the smallness of the Biot number, which implies there is excellent heat exchange with the environment. The strong agreement between asymptotic and numerical solutions also confirms the dominant mechanism of heat generation is due to electrochemical reactions rather than Ohmic heating.

% \begin{figure}
%   \centering
%   \subfigure[Positive electrode]{\includegraphics[width=0.48\textwidth]{figs/galvano_csurfp.pdf}}
%   \subfigure[Negative electrode]{\includegraphics[width=0.48\textwidth]{figs/galvano_csurfn.pdf}}
%   \caption{Simulations of galvanostatic discharge processes at various C-rates $\mathcal{I} = 1$, $2$, $4$.  The asymptotic solutions in panels (a)--(b) are constructed using only the leading-order contributions; in panels (c)--(d) the leading- and first-order contributions are considered.
%     Dashed lines and circles denote numerical solutions of the VA and P2D model. Time is normalised using the discharge time $t_\text{dis} = (3600 / \mathcal{I})$~s. The temperature $T$ is averaged over a single cell. The surface concentrations $\csi^{\text{surf}}$ have been averaged over their respective electrode domains.}
%   \label{fig:galvanostatic}
% \end{figure}

\subsection{Potentiostatic hold}

We now consider the behaviour of the battery during a potentiostatic hold, whereby the dimensionless cell potential is held at a constant value of $\Delta v$.  There are four time regimes that occur which are similar to case of the galvanostatic discharge.  The first two, defined by $t = O(\Cn \Gn \nusn / \phisn)$ and $t = O(\Cp)$, are extremely short and capture the formation of double layers and the onset of electrochemical reactions.  The third, defined by $t = O(1)$, involves the dynamics on the diffusive time scale and captures the transient behaviour of the lithium concentration and voltage profiles in the electrolyte.  The fourth regime occurs on the saturation time scale, $t = O(\gammac^{-1})$, and describes the approach to a quiescent state. 

The majority of heat generation occurs in the third and fourth time regimes, which will be the focus of our asymptotic analysis.  The governing equations in these regimes are solved using matched asymptotic expansions in time and regular expansions in powers of $\nue$.  The two capacitance regimes are briefly discussed and further details can be found in \ref{app:cap_regimes}.  The temperature dependence of the parameters can be ignored due to the small amount of heat that is generated over these short time regimes.

%Matched asymptotic expansions in time are used to solve the problem on the diffusive time scale, \emph{i.e.} when $t = O(1)$, and then on the saturation time scale, \emph{i.e.} when $t = O(\gammac^{-1})$. Since the cell potential is fixed, the C-rate is now time dependent, $\mathcal{I} = \mathcal{I}(t)$, and must be determined as part of the solution.

\subsubsection{Dynamics on the capacitance time scales}

The first capacitance time regime occurs when $t = O(\Cn \Gn \nusn / \phisn)$. 
%\im{We note this is different than the first short-time regime for the galvanostatic case with further explanation in \ref{app:cap_regimes}.}  
A small-time analysis in this regime reveals that instantaneously changing the cell potential results in a large initial C-rate given by 
\begin{align}
\I = -\cellav{\nu}^{-1}\,\Delta v,
\quad
\cellav{\nu} = x_p \mn{\nu}{\mathrm{eq}}{p} + (x_n - x_p)\mn{\nu}{\mathrm{eq}}{s} + (1-x_n)\mn{\nu}{\mathrm{eq}}{n},
\label{const_V:I0}
\end{align}
where $\cellav{\nu}$ is the cell-averaged resistivity, which is based on the equivalent resistivities of the cell components
\begin{align}
\mn{\nu}{\mathrm{eq}}{p} = (\phisp \nusp^{-1} + \phie \nue^{-1})^{-1}, \quad
\mn{\nu}{\mathrm{eq}}{s} = \nue/\phie, \quad
\mn{\nu}{\mathrm{eq}}{n} = (\phisn \nusn^{-1} + \phie \nue^{-1})^{-1}.
\label{const_V:nu_eq}
\end{align}
The battery resistivity, $\cellav{\nu}$, shows that the electrodes and separator behave as if connected in series, whereas the phase resistivities, $\mn{\nu}{\mathrm{eq}}{i}$, show that the porous phases behave as if connected in parallel.
%These resistivities show that the electrodes and separator behave as if connected in series, whereas the liquid- and solid-phases act as if connected in parallel.  
The initial current is thus set by the most resistive contribution to $\cellav{\nu}$, which in this case comes from the positive electrode.  For larger times, the electrical resistance in the negative electrode significantly increases due to the onset of electrochemical reactions.  Consequently, the magnitude of the C-rate undergoes a marked decrease and plateaus at an approximate value given by the solution of the nonlinear equation
\begin{align}
2\sinh^{-1}\left(\frac{\I}{2\Gn(1-x_n)}\right)
+ \left[x_p \mn{\nu}{\mathrm{eq}}{p} + (x_n-x_p)\mn{\nu}{\mathrm{eq}}{s} + (1-x_n)(\mn{\nu}{\mathrm{eq}}{s}/2)\right]\I
= -\Delta v.
\label{const_V:I1}
\end{align}
%Since the equivalent electrical resistances are small, the dominant contribution to the left-hand side of \eqref{const_V:I1} is from the first term, which greatly reduces the C-rate.

The second capacitance time regime is defined by $t=\Ord{\Cp}$.
%which is reminiscent of the second time regime for the galvanostatic case because there are no large currents to resolve.}
%$t = O(\Cp \Gp \nusp / \phisp)$, 
This regime marks the onset of electrochemical reactions in the positive electrode which leads to further increases in the resistance and decreases in the C-rate.  Eventually, the C-rate reaches another plateau and settles to a value given by the solution of
\begin{align}
2\sinh^{-1}\left(\frac{\I}{2\Gn(1-x_n)}\right)
+ 2\sinh^{-1}\left(\frac{\I}{2\Gp x_p}\right)
+ (\mn{\nu}{\mathrm{eq}}{s}/2)(1+x_n-x_p)\I
= -\Delta v.
\label{const_V:I2}
\end{align}

Despite the large currents that can arise in these two regimes, the change in lithium concentration and temperature remains negligible.  Therefore, we impose the original initial conditions in the third (diffusive) time regime. 

% As shown in Fig.~\ref{fig:cap_regimes_temp} of \ref{app:cap_regimes}, the change in temperature during the two capacitance time regimes is negligible.  The initial increase due to Ohmic heating, which is amplified by the large C-rates, can be estimated using scaling arguments and is found to be
% \begin{align}
% \Delta T \sim \left(\frac{\Ci\Gi\nusi}{\phisi}\right)\left(\frac{\Delta v}{\cellav{\nu}}\right)^2
% \end{align}

\subsubsection{Dynamics on the diffusive time scale}
\label{sec:potentio_diffusive}

The concentration-dependent contribution to the open-circuit potential and the exchange current density are expanded in powers of $\nue$ and the leading- and first-order terms are given by $\Vi^{(0)} = 1$, $\ji^{(0)} = 1$,
\subeq{
  \label{const_V:ec_01}
\begin{align}
  %\Vi^{(0)} &= 1, \\
  \Vi^{(1)} &= (\gammac/\nue)\left[\cli^{(0)} - \deltai \left(1 + \xii(1-\xii)^{-1}\right)\csi^{(0)}\right], \\
  %\ji^{(0)} &= 1, \\
  \ji^{(1)} &= (1/2)(\gammac/\nue)\left[\cli^{(0)} + \deltai \left(1 - \xii(1-\xii)^{-1}\right)\csi^{(0)}\right].
\end{align}
}
%\im{We note that these expansions differ from \eqref{eqn:BV_corr} because we need to include the $\gamma\ll1$ perturbation of the base states.} 
In the leading-order problem, the potentials and concentration of intercalated lithium are spatially uniform, implying that $\Phiei^{(0)} = 0$ and $\Phisp^{(0)} = \Phisn^{(0)} + \Delta v$. Following a similar strategy as in Sec.~\ref{sec:galvanostatic_lo}, we find that the leading-order electrical problem is given by
\subeq{
  \label{const_V:e_00}
\begin{align}
  \Gp x_p \sinh\left(\frac{\Phisn^{(0)} + \Delta v}{2}\right) + \Gn (1 - x_n)\sinh\left(\frac{\Phisn^{(0)}}{2}\right) &= 0, \label{const_V:phi_00}\\
  2\sinh\left(\frac{\Phisn^{(0)}}{2}\right) - \frac{\mathcal{I}^{(0)}}{\Gn(1 - x_n)} &= 0.
\end{align}
}
The solution for $\I^{(0)}$ matches with the leading-order solution (in terms of $\nue$) to \eqref{const_V:I2}.  The leading-order solid-phase potentials and C-rate are constant in time, implying that the solid-phase concentrations evolve linearly in time according to \eqref{const_I:csi_0}. Furthermore, the leading-order current densities are given by \eqref{red:isi}--\eqref{red:iei}. Having solved the electrical problem, the leading-order thermal problem for the scaled temperature $u^{(0)} = \mathcal{T}^{-1}T^{(0)}$ is given by
\begin{align}
  % \bar{\varrho} \deriv{T^{(0)}}{t} &= -\Le\, \Bi (1 + \vHn) T^{(0)} - \Gp^{-1} \mathcal{I}^{(0)}\left[\Delta v + \log \Up + \DelGp - \log \Un - \DelGn\right]
                                       \Gp \mathcal{T} \cellav{\varrho} \deriv{u^{(0)}}{t} &= -u^{(0)} - \mathcal{I}^{(0)}\left[\Delta v + \log \Up + \DelGp - \log \Un - \DelGn\right]
                                                                                             \label{const_V:u_10}
\end{align}                              
with $u^{(0)}(0) = 0$, which has a simple analytical solution that tends to a constant value. 

The first-order problem requires the leading-order solution for the liquid-phase lithium concentration, which satisfies \eqref{simp:cli}. Since the electrolyte current density $\iei^{(0)}$ is constant in time, the solution is straightforwardly obtained using separation of variables. By writing the concentration in terms of the deviation from the quasi-steady profile, $\cli^{(0)} = [\mathcal{I}^{(0)}(1-\theta^0)/\phie](\cli^\infty + C_L)$, it is possible to consider a single liquid-phase concentration $C_L$ for the entire cell, which is given by
\subeq{
  \label{const_V:trans_cli}
\begin{align}
  C_L &= -2 \sum_{n=1}^{\infty} b_n \e^{-n^2\pi^2t}\cos(n \pi x), \\
  b_n &= \int_{0}^{x_p}\clp^{\infty}\cos(n \pi x)\,\d x + \int_{x_p}^{x_n}\cls^\infty \cos(n \pi x)\,\d x + \int_{x_n}^{1}\cln^\infty \cos(n \pi x)\,\d x,
\end{align}
}
where we recall that we have ignored the temperature dependence of the parameters. The correction to the electrolyte potential satisfies \eqref{eqn:ohme} and \eqref{eqn:ohmes}, and can be found in a similar manner by writing $ \Phiei^{(1)} = [(\mathcal{I}^{(0)}\Da)/(2\phie)](\Phiei^{\infty} + \Psi_e)$, where $\Psi_e(x,t) = 2(1-\theta^0)(\DL/\Da - 1)[C_L(1,t)-C_L(x,t)]$. The solutions for the correction to the solid-phase potential is given by \eqref{const_I:Phi_s_1}; however, the constants of integration $P_p$ and $P_n$ that appear in these expressions must now be equal in order to satisfy the voltage condition $\Phisp^{(1)}(0,t) - \Phisn^{(1)}(1,t) = 0$. Thus, we let $P_p = P_n = P$. Solvability conditions that are analogous to \eqref{eqn:solvability} can be derived by integration of \eqref{eqn:div_isi}, and these provide two equations for the correction to the current $\mathcal{I}^{(1)}$ and the constant $P$ in the solid-phase potentials,
\begin{align}
  \Gp\int_{0}^{x_p} \gpbar^{(1)}\,\d x = -\mathcal{I}^{(1)}(t),
  \qquad
  \Gn\int_{x_n}^{1} \gnbar^{(1)}\,\d x = \mathcal{I}^{(1)}(t),
  \label{const_V:solvability}
\end{align}
where $\gibar^{(1)}$ is given by \eqref{eqn:BV_corr} with $\etai^{(0)} = \Phisi^{(0)}$, $\etai^{(1)} = \Phisi^{(1)} - \Phiei^{(1)} - \Vi^{(1)}$, and using the expressions in \eqref{const_V:ec_01}. Having, in principle, determined $\mathcal{I}^{(1)}$, it can be shown that the corrections to the (integrated) concentration of intercalated lithium satisfy
\begin{align}
  \deriv{}{t}\int_{0}^{x_p}\csp^{(1)}\,\d x = \frac{\mathcal{I}^{(1)}(t)}{\phisp},
  \qquad
  \deriv{}{t}\int_{x_n}^{1}\csn^{(1)}\,\d x = -\frac{\mathcal{I}^{(1)}(t)}{\phisn},
  \label{const_V:csi_1}
\end{align}
and thus do not vanish like in the galvanostatic case. The correction to the temperature satisfies
\begin{align}
  \begin{split}    
  \Gp \mathcal{T} \cellav{\varrho} \deriv{u^{(1)}}{t} = -u^{(1)} &- \mathcal{I}^{(1)}\left[\Delta v + \log \Up + \DelGp - \log \Un - \DelGn\right] \\
  &-\mathcal{I}^{(0)}\left[\posav{\Phisp^{(1)} - \Phiep^{(1)}}-\negav{\Phisn^{(1)}-\Phien^{(1)}} \right]
\end{split}
\end{align}                              
with $u^{(1)}(0) = 0$.

\subsubsection{Dynamics on the saturation time scale}

Following the analysis in Sec.~\ref{sec:galvanostatic}, on the saturation time scale we write $t = \gammac^{-1}\hat{t}$, $\csi = \gammac^{-1}\csihat$, and use hats on the all of the other variables. The analysis is very similar to that of the galvanostatic case; however, we must now account for the fact that the leading-order contribution to the current density, given by $\Ihat^{(0)}$, is now time dependent. As discussed in Sec.~\ref{sec:galvanodyn}, this only leads to minor modifications in the solutions. At leading-order, the solid-phase potentials and lithium concentrations are uniform. We thus have $\Phieihat^{(0)} = 0$, $\Phisphat^{(0)} = \Phisnhat^{(0)} + \Delta v$, and find that the leading-order electrical problem is
\subeq{
  \label{const_V:ec_20}
\begin{align}
  \Gp x_p\gnbar^{(0)}(\Phisnhat^{(0)} + \Delta v, \csnhat^{(0)}) + \Gn (1-x_n) \gnbar^{(0)}(\Phisnhat^{(0)}, \csnhat^{(0)}) &= 0, \\
  \gnbar^{(0)}(\Phisnhat^{(0)}, \csnhat^{(0)}) - \frac{\Ihat^{(0)}(\hat{t})}{\Gn(1-x_n)} &= 0,
\end{align}
}
where $\gibar^{(0)}$ is given by \eqref{const_I:BV} and the leading-order concentrations of intercalated lithium evolve according to
\subeq{
  \label{vary_I:csi}
  \begin{align}
    \deriv{\csphat^{(0)}}{\hat{t}} = \frac{\Ihat^{(0)}(\hat{t})}{\phisp x_p}, \qquad 
    \deriv{\csnhat^{(0)}}{\hat{t}} = -\frac{\Ihat^{(0)}(\hat{t})}{\phisn(1 - x_n)};\qquad \csihat^{(0)}(0)=0.
\end{align}
}
% \eqref{vary_I:csi} with $\mathcal{I}$ replaced by $\Ihat^{(0)}$.
The leading-order current densities are given by \eqref{red:isi} and \eqref{red:iei} and the corresponding solution to the thermal problem is
\begin{align}
  % \bar{\varrho} \deriv{T^{(0)}}{t} &= -\Le\, \Bi (1 + \vHn) T^{(0)} - \Gp^{-1} \mathcal{I}^{(0)}\left[\Delta v + \log \Up + \DelGp - \log \Un - \DelGn\right]
                                       \hat{u}^{(0)} = - \Ihat^{(0)}(\hat{t})\left[\Delta v + \log \Up + \DelGp - \log \Un - \DelGn\right].
                                       \label{const_V:u_20}
\end{align}

To derive the correction, we first note that in this time regime, both the liquid-phase concentration $\clihat^{(0)}$ and potential $\Phieihat^{(1)}$ are known and given by their quasi-steady profiles \eqref{const_I:c_L_inf} and \eqref{const_I:Phi_e}, respectively. The corrections to the solid-phase potential are again given by \eqref{const_I:Phi_s_1} with $P_p = P_n = P$ and the solvability conditions \eqref{const_V:solvability} still apply. However, in this case the electrochemistry is more tightly coupled and the solution to the electrical problem cannot be obtained without the correction to the solid-phase concentration. This is because the expressions for $\Vi^{(1)}$ and $\ji^{(1)}$ are given by \eqref{eqn:BV_corr} and the contributions to these from $\csihat^{(1)}$ do not integrate to zero unlike in the galvanostatic case analysed in Sec.~\ref{sec:galvanostatic}. Thus, obtaining the correction amounts to simultaneously solving the differential-algebraic system given by \eqref{const_V:solvability}--\eqref{const_V:csi_1}, which determines the constant $P$ in the solid-phase potentials, the correction to the current $\mathcal{I}^{(1)}$, and the correction to the net concentration of intercalated lithium. After the electrochemical problem has been solved, the correction to the temperature can be evaluated via the expression
\begin{align}
  \begin{split}    
\hat{u}^{(1)} = &- \Ihat^{(1)}(\hat{t})\left[\Delta v + \log \Up + \DelGp - \log \Un - \DelGn\right] \\
&\-\Ihat^{(0)}(\hat{t})\left[\posav{\Phisphat^{(1)} - \Phiephat^{(1)}} - \negav{\Phisnhat^{(1)}-\Phienhat^{(1)}}\right]
-(\gammac/\nue)\Gp \mathcal{T} \cellav{\varrho} \deriv{\hat{u}^{(0)}}{\hat{t}}.
\end{split}
\end{align}                              

\subsubsection{Construction of the composite solution}

The leading-order composite asymptotic solution to the electrochemical problem is straightforward to obtain because this is given by the leading-order solution on the saturation time scale, \emph{i.e.} the solution to \eqref{const_V:ec_20} along with \eqref{vary_I:csi}. The leading-order composite solution to the thermal problem is given by
\begin{align}
  u_\text{composite}^{(0)} = u^{(0)} + \hat{u}^{(0)} - u_\text{overlap}^{(0)},
\end{align}
where $u^{(0)}$ solves \eqref{const_V:u_10}, $\hat{u}^{(0)}$ is given by \eqref{const_V:u_20}, and $u_\text{overlap}^{(0)}$ corresponds to the temperature in the overlap region, which is given by the steady-state solution to \eqref{const_V:u_10},
\begin{align}
  u_\text{overlap}^{(0)} = -\mathcal{I}^{(0)}\left[\Delta v + \log \Up - \log \Un + \DelGp - \DelGn\right].
\end{align}
Constructing the composite solution for the leading- and first-order problems is more involved but can be done using the solutions in the overlap region given in \ref{app:matching}.

\subsection{Comparison with numerics}
\label{sec:potentio_num}

\begin{figure}
  \centering
  \subfigure[Leading-order]{\includegraphics[width=0.48\textwidth]{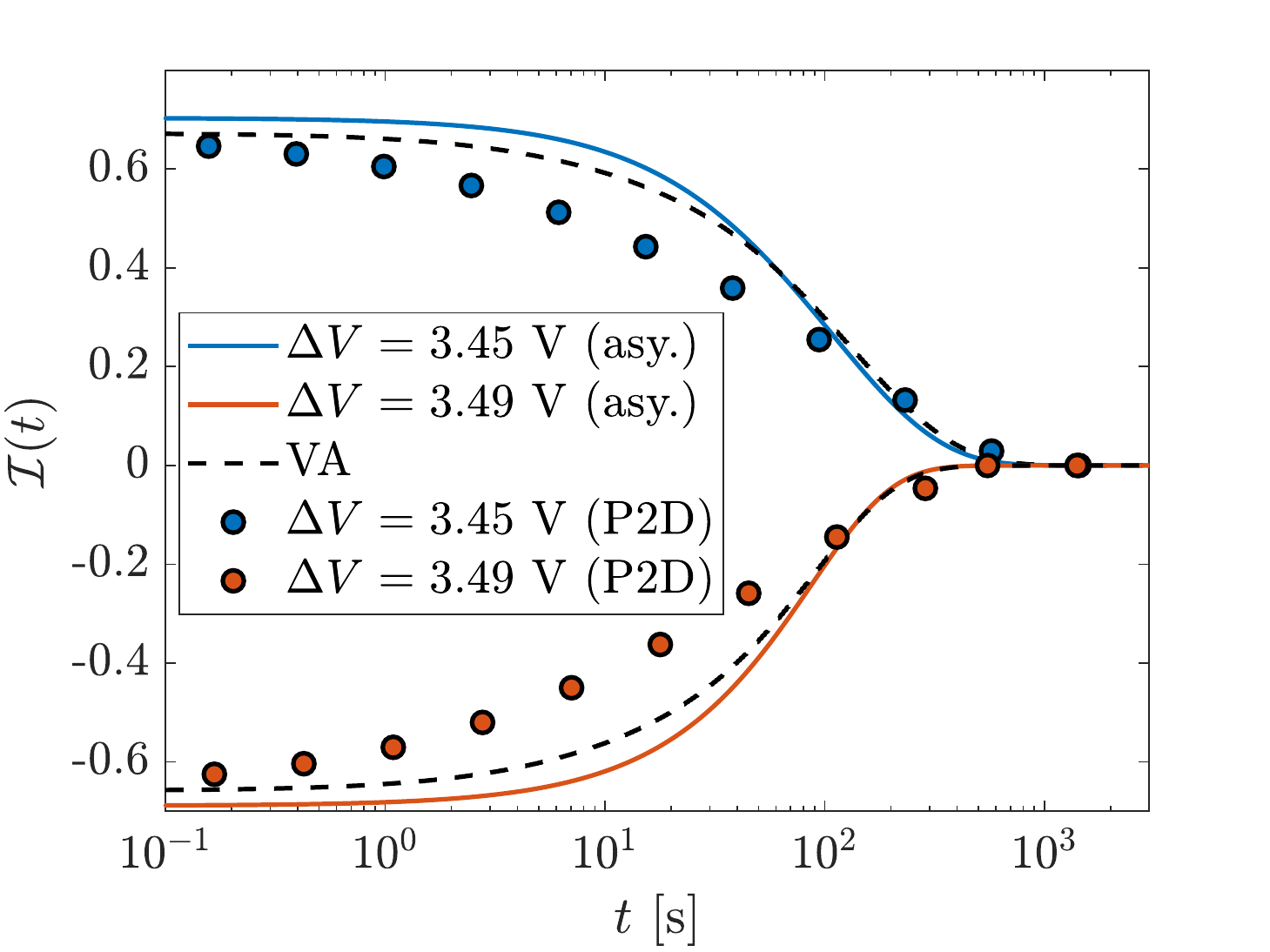}}
  \subfigure[Leading-order]{\includegraphics[width=0.48\textwidth]{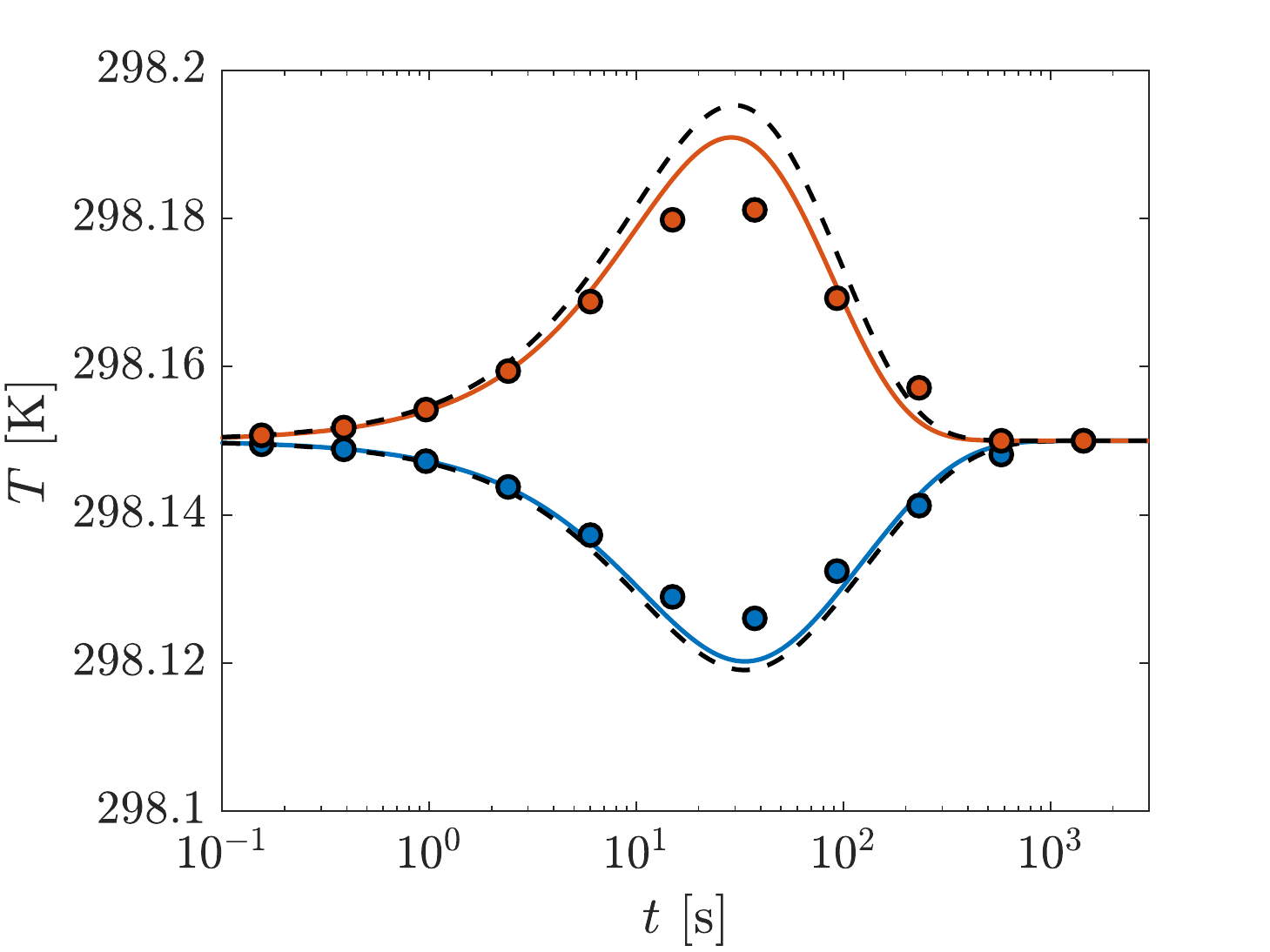}} \\
  \subfigure[Leading- and first-order]{\includegraphics[width=0.48\textwidth]{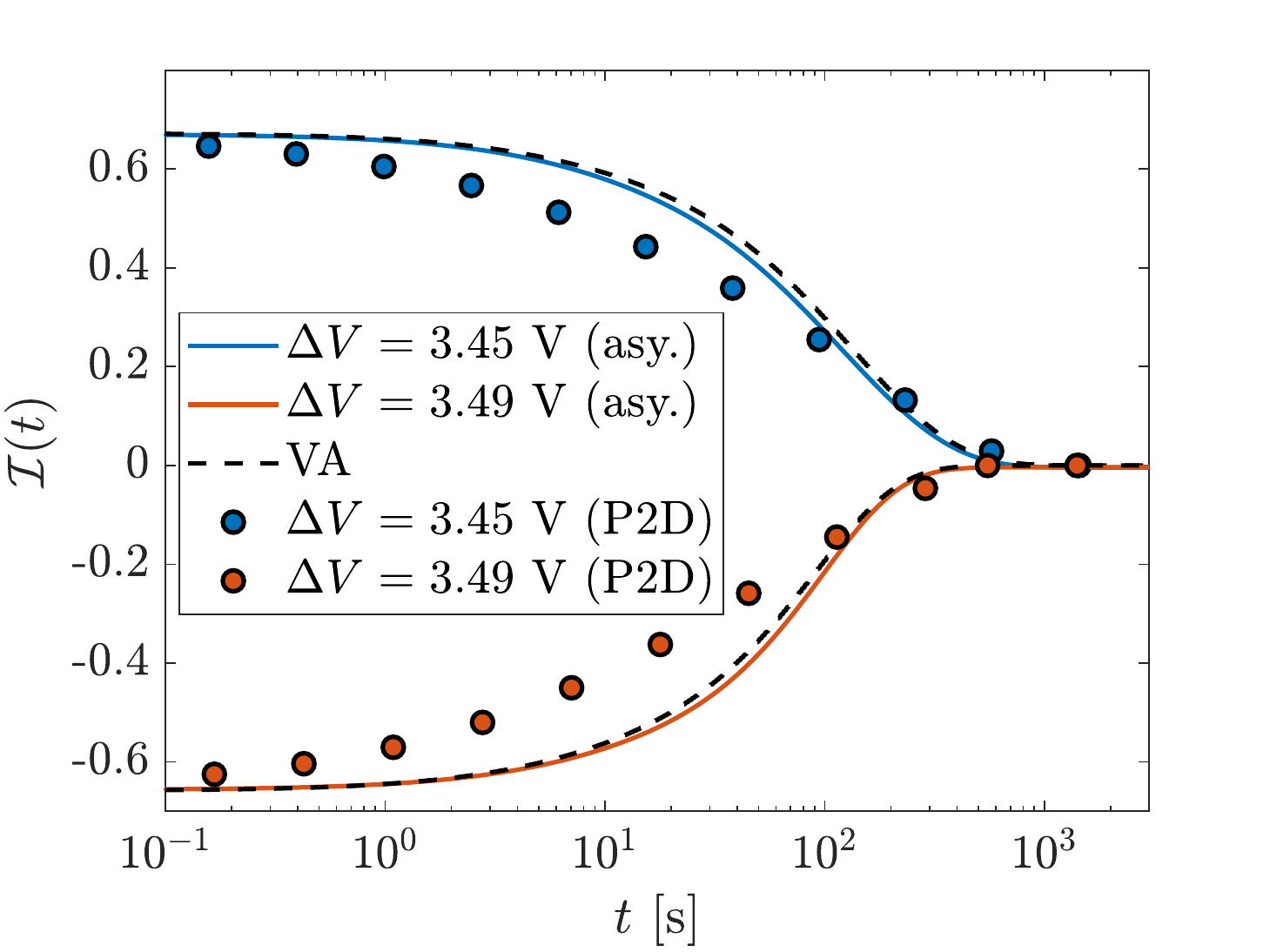}} 
  \subfigure[Leading- and first-order]{\includegraphics[width=0.48\textwidth]{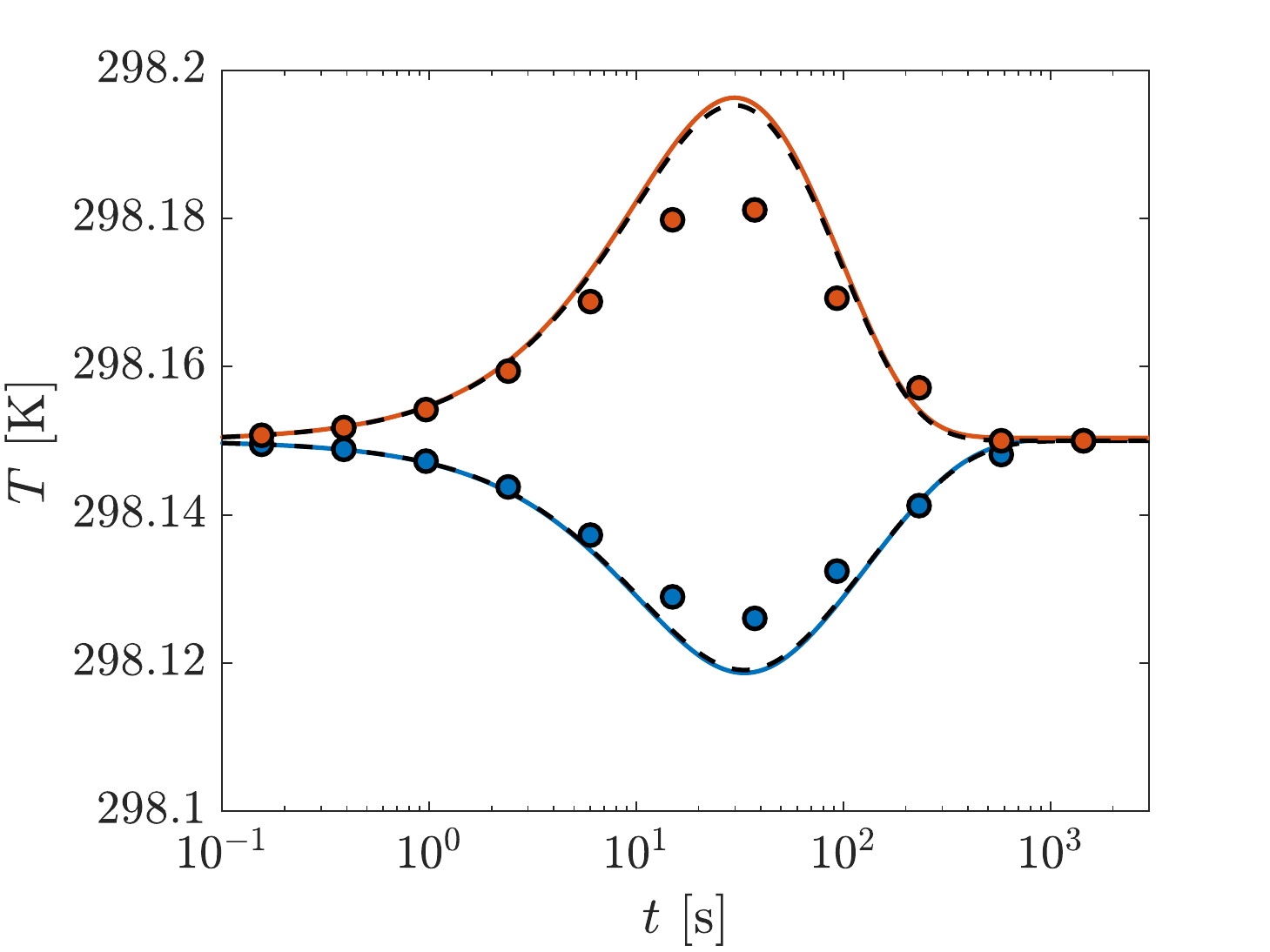}}
  \caption{Evolution of the C-rate $\mathcal{I}$ and the cell temperature $T$ during a potentiostatic discharge. The asymptotic solutions in panels (a)--(b) are constructed using only the leading-order contributions; in panels (c)--(d) the leading- and first-order contributions are considered. The resting potential of the cell is $\Delta V = 3.47$~V. The legends apply to all of the panels.}
  \label{fig:potentio2}
\end{figure}

We simulate the potentiostatic case using the parameters in \eqref{eqn:nondim_num_ec} and \eqref{eqn:nondim_num_thermal} which corresponds to a (dimensional) open-circuit potential of approximately $\Delta V=3.47$~V. Therefore we take $\Delta V=3.45$~V and $\Delta V=3.49$~V to discharge and charge the cell respectively with the results in Fig.~\ref{fig:potentio2} for the C-rate and temperature. The asymptotic and numerical solutions are compared to the simulations, with the focus being on the diffusive and saturation time regimes.  The dynamics in the capacitance time regimes are shown in Fig.~\ref{fig:cap_regimes} (a)--(b) of \ref{app:cap_regimes}.

Figure \ref{fig:potentio2} (a)--(b) shows the C-rate and cell temperature using the leading-order composite solutions and numerical simulations of the VA and P2D models. As with the galvanostatic case at these parameters, the leading-order correction is quite accurate with a near-perfect match when the first-order correction is considered. However, the agreement between the VA and P2D models has now diminished. As previously stated, the two models can diverge when the system is close to a singularity in the open-circuit potential due to the sensitive dependence on the surface concentration. For the simulations in Fig.~\ref{fig:potentio2}, the solid-phase concentrations do not vary significantly and thus the system remains close to the singular points of the open-circuit potential. However, agreement holds for $t\ll1$ because the solid concentration stays at the initial value while agreement for $t\gg1$ is due to the current decreasing to zero leading to small surface concentrations.

\begin{figure}
  \centering
  \subfigure[Leading-order]{\includegraphics[width=0.48\textwidth]{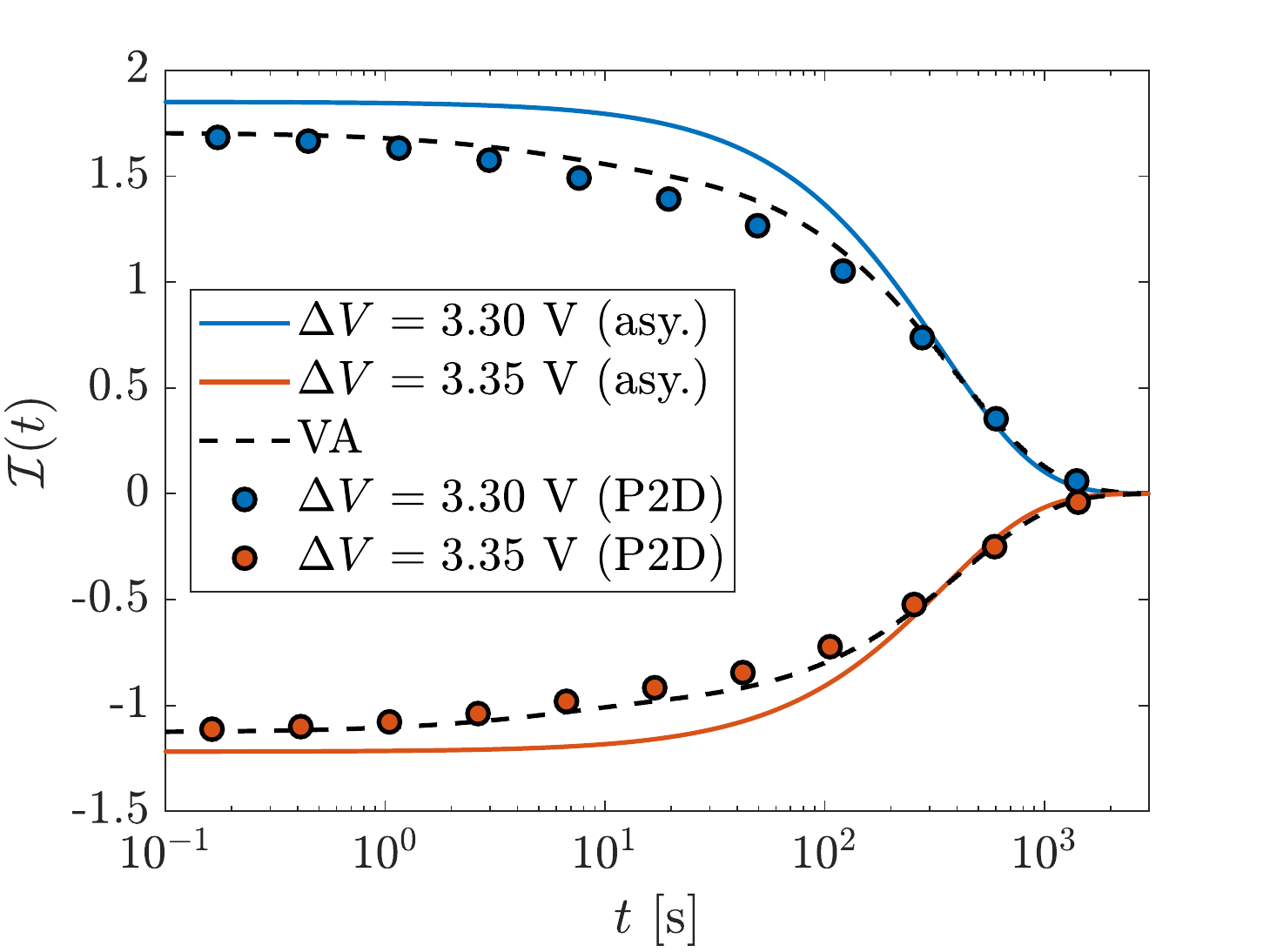}}
  \subfigure[Leading-order]{\includegraphics[width=0.48\textwidth]{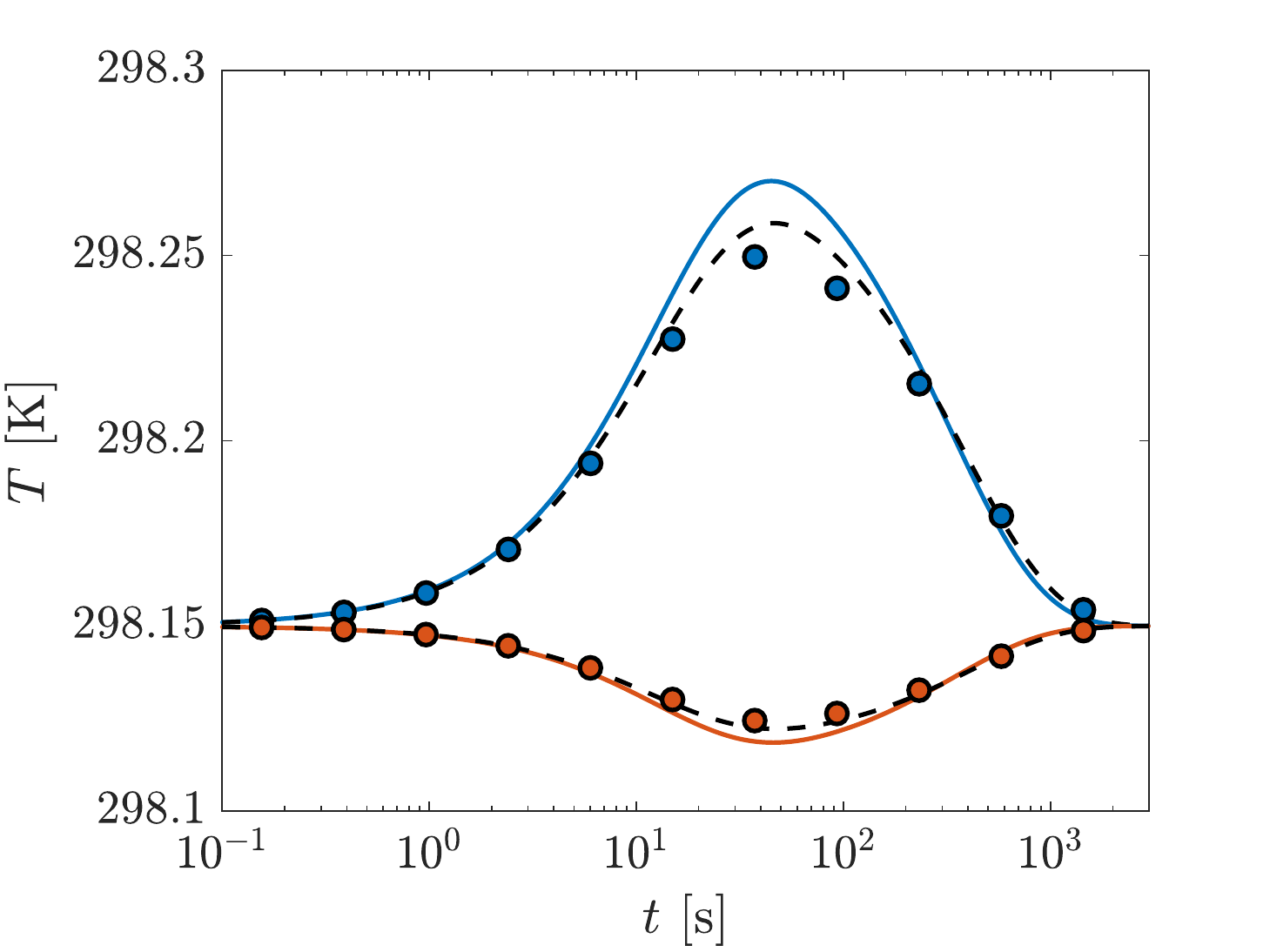}} \\
  \subfigure[Leading- and first-order]{\includegraphics[width=0.48\textwidth]{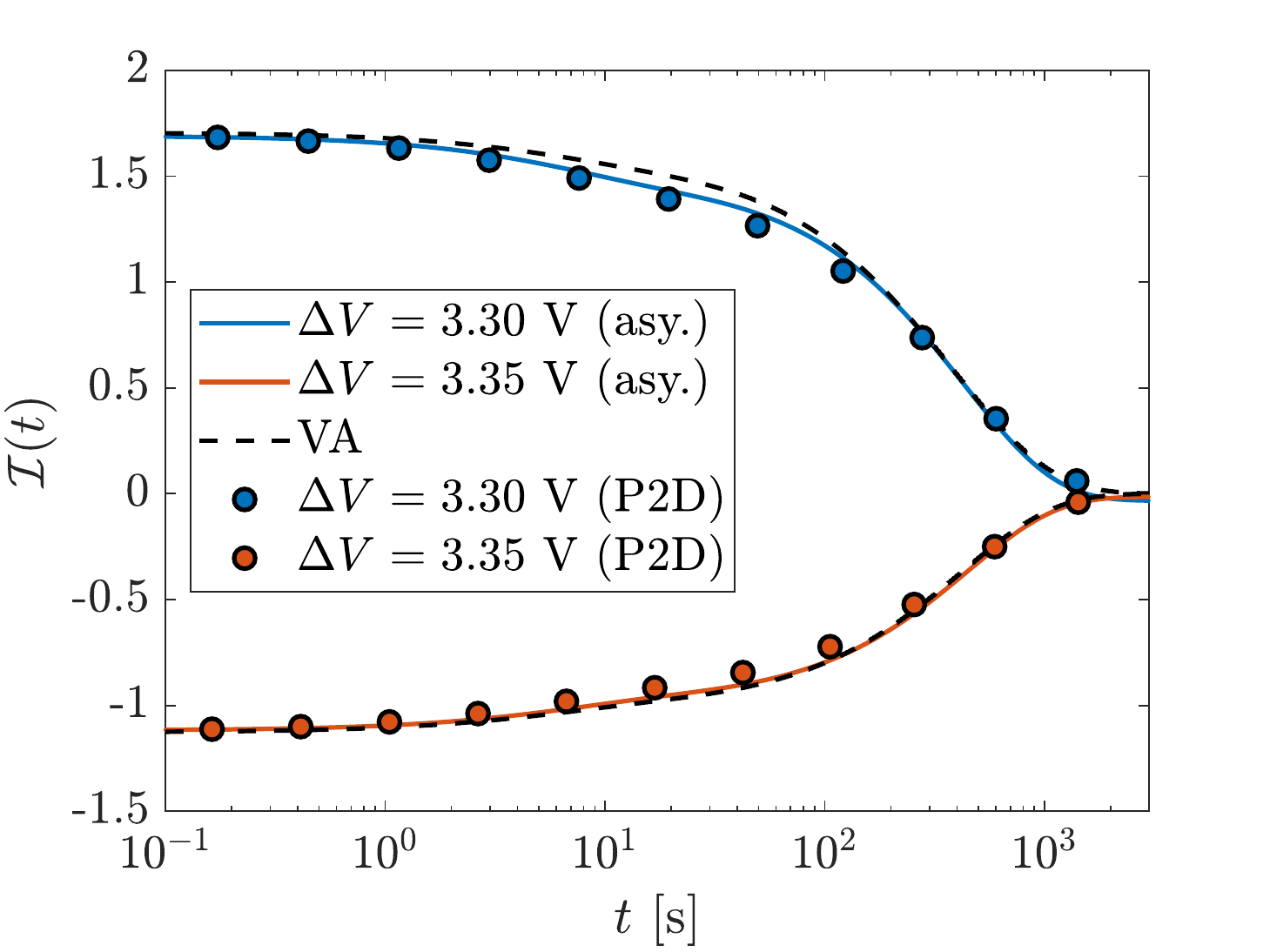}} 
  \subfigure[Leading- and first-order]{\includegraphics[width=0.48\textwidth]{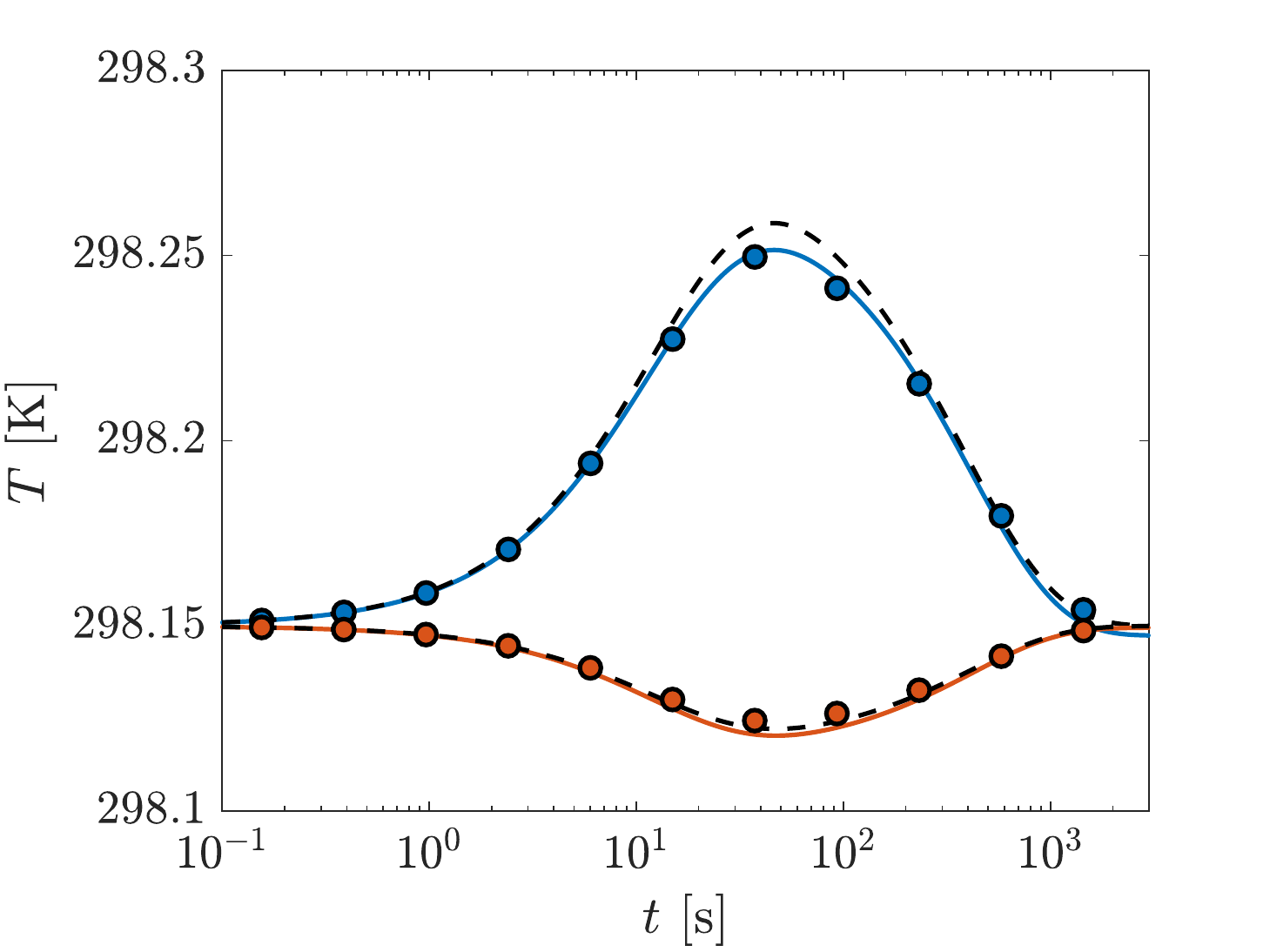}}
  \caption{Evolution of the C-rate $\mathcal{I}$ and the cell temperature $T$ during a potentiostatic discharge. The asymptotic solutions in panels (a)--(b) are constructed using only the leading-order contributions; in panels (c)--(d) the leading- and first-order contributions are considered. The resting potential of the cell is $\Delta V = 3.33$~V. The legends apply to all of the panels.}
  \label{fig:potentio}
\end{figure}

If we alter the initial conditions to be away from the singular points, the disagreement between the VA and P2D models should diminish. Therefore, we take the initial concentration of intercalated lithium to be $\csp^0 = 0.39\,\cspmax$ and $\csn^0 = 0.43\,\csnmax$. These values can be obtained by simulating a discharge at a low C-rate for half of the total discharge time. We note that by changing these initial concentrations, the parameters in \eqref{eqn:nondim_num_ec} and \eqref{eqn:nondim_num_thermal} associated with $\csi^0$ and $\gibar^0$ change as well. At these concentrations, the (dimensional) resting potential of the cell is about $\Delta V = 3.33$~V. We then prescribe cell potentials of $\Delta V = 3.30$~V and $3.35$~V.  We once again compare the asymptotic and numerical solutions for the C-rate and temperature, displaying the results in Fig.~\ref{fig:potentio} (a)--(b). The dynamics in the capacitance time regimes are shown in Fig.~\ref{fig:cap_regimes} (c)--(d). We see that the error between VA and P2D has indeed been reduced and excellent agreement is found.

Interestingly, the matching between the VA and P2D models has come at the expense of a disagreement in the leading-order asymptotics. This is due to the increased role of the electrolyte at small values of the open-circuit potential. If we instead compare with a composite solution including the leading- and first-order contributions (Fig.~\ref{fig:potentio} (c)--(d)) then we recover excellent agreement between the three cases.
%Some errors in the temperature remain however.  This is unsurprising given that the relative temperature change is about $10^{-4}$, which corresponds to a second-order correction in $\nue$.
%The composite asymptotic solutions in Fig.~\ref{fig:potentio} (c)--(d) have been constructed using the leading- and first-order contributions and are in closer agreement with the numerics compared to when only the leading-order terms are considered (as seen from Fig.~\ref{fig:potentio} (a)--(b)).

The necessary inclusion of first-order corrections in Fig.~\ref{fig:potentio2} for agreement between the asymptotic and numerical solutions can be rationalised by analysing \eqref{const_V:phi_00}, which can be interpreted as current balance involving two resistors connected in parallel, where the parameters $\Gi$ act as inverse resistances (i.e.~conductivities). Thus, \eqref{const_V:phi_00} describes how the electrical load is distributed across the two electrodes and determines where the drop in potential occurs. Recalculating the values of $\Gi$ using the new values for the initial solid-phase concentrations reveals that $\Gp x_p \simeq 9.5$ but $\Gn(1-x_n) \simeq 1.8$, in contrast to the previous values $\Gp x_p \simeq 2.8$ and $\Gn (1-x_n)=1.3$. %Therefore $\Gp x_p$ is no longer $O(1)$ making the leading order solution less reliable.
%While the asymptotic solution is able to accurately capture the cell temperature, there are substantial deviations in the C-rate. This is surprising given that these cell potentials do not induce large C-rates and our previous results showed that the leading-order asymptotic solutions were accurate up to 4C.
%The effect of a larger $\Gp$ on the results can be rationalised in terms of \eqref{const_V:phi_00}, which can be interpreted as current balance involving two resistors connected in parallel, where the parameters $\Gi$ act as inverse resistances (i.e.~conductivities). Thus, \eqref{const_V:phi_00} describes how the electrical load is distributed across the two electrodes and determines where the drop in potential occurs. 
For the new initial conditions, most of the potential drop occurs in the negative electrode and the electrical potential in the positive electrode remains close to zero. Consequently, the higher-order terms which have been neglected from the electrical problem in the positive electrode play much a stronger role. For the original parameters, the load is equally distributed across both electrodes.

\subsection{Galvanodynamic charging/discharging}
\label{sec:galvanodyn}

% The analysis can be extended to the case when the C-rate is a function of time. The only change that occurs in the problem structure is related to the leading-order contribution to the concentration of intercalacted lithium, $\csihat^{(0)}$. Rather than evolving linearly in time according to \eqref{const_I:csi_0}, these concentrations now satisfy ordinary differential equations given by

% The remainder of the leading- and first-order problems remain unchanged. 

The analysis in Sec.~\ref{sec:galvanostatic} for a galvanostatic discharge can be extended to the case when the C-rate is a function of time.  If the variations in current occur over the saturation time scale, then the asymptotic solutions presented in Sec.~\ref{sec:galvanostatic}
%for the galvanostatic case
can be used by replacing the expressions for the solid-phase concentrations \eqref{const_I:csi_0} with the differential equations \eqref{vary_I:csi}.  If the C-rate varies on the diffusive time scale, then the framework developed in Sec.~\ref{sec:potentio_diffusive} can be applied. One caveat, however, is that the transient solution for the liquid-phase concentration \eqref{const_V:trans_cli} needs to be adapted to account for a time-dependent current if the first-order correction is used.

To compare the asymptotic solutions with numerical simulations in the case of a time-dependent applied current, we consider an oscillating C-rate of the form $\I(t) = A \sin (2 \pi t / \tau)$, which is used in electrochemical impedance spectroscopy. We fix the amplitude at $A = 2$ so that the maximum (dis)charge rate is 2C. We consider three sets of initial conditions for the solid-phase concentrations, each of which corresponds to a different initial state of charge (SOC). We define the SOC to be the fraction of total capacity that is available to use. To generate new initial conditions, we solve the leading-order galvanostatic model with the initial conditions in Table \ref{tab:params2} to full discharge, which define the 100\% and 0\% SOC points, respectively.  We then determine the solid concentrations in each electrode at a given SOC along the voltage curve and use these values to re-initialise the model. Since the solid concentrations at a given SOC are dependent on the initial conditions used in solving the galvanostatic model, there is not a one-to-one correspondance between a given SOC and the new initial solid concentrations. The new initial conditions we take are:
$\csp^0 = 0.21\,\cspmax$, $\csn^0 = 0.64\,\csnmax$ (75\% SOC); $\csp^0 = 0.39\,\cspmax$, $\csn^0 = 0.43\,\csnmax$ (50\% SOC); and $\csp^0 = 0.58\,\cspmax$, $\csn^0 = 0.21\,\csnmax$ (25\% SOC). We recall that each time the initial conditions are changed, the parameters associated with $\csi^0$ and $\gibar^0$ in \eqref{eqn:nondim_num_ec} and \eqref{eqn:nondim_num_thermal} are changed as well. The period of oscillation is chosen to be either $\tau = 1200$~s or $\tau = 60$~s. In the former, the current varies on the saturation time scale, in the latter, it varies on the diffusive time scale.

\begin{figure}
  \centering
  \subfigure[]{\includegraphics[width=0.48\textwidth]{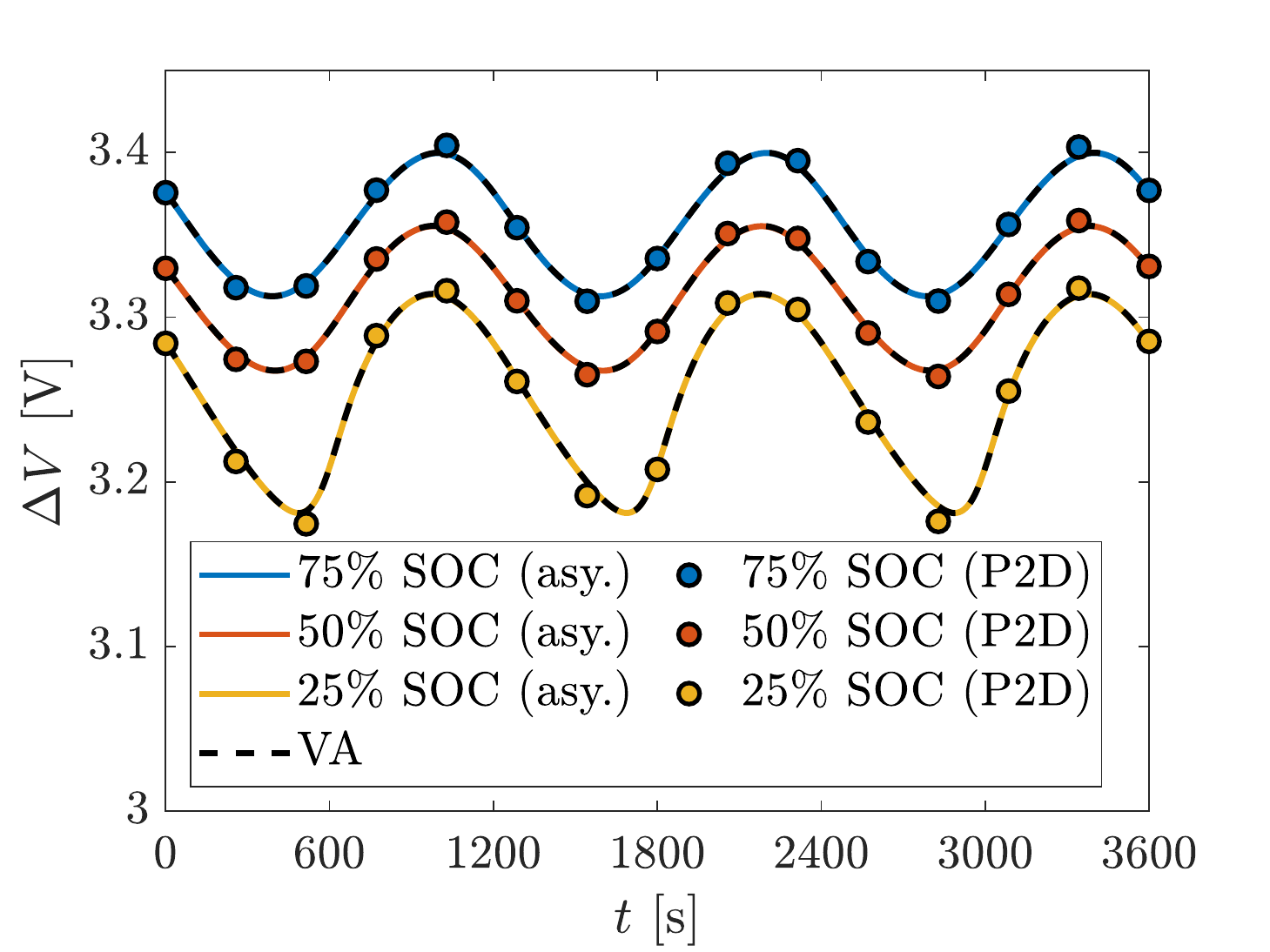}}
  \subfigure[]{\includegraphics[width=0.48\textwidth]{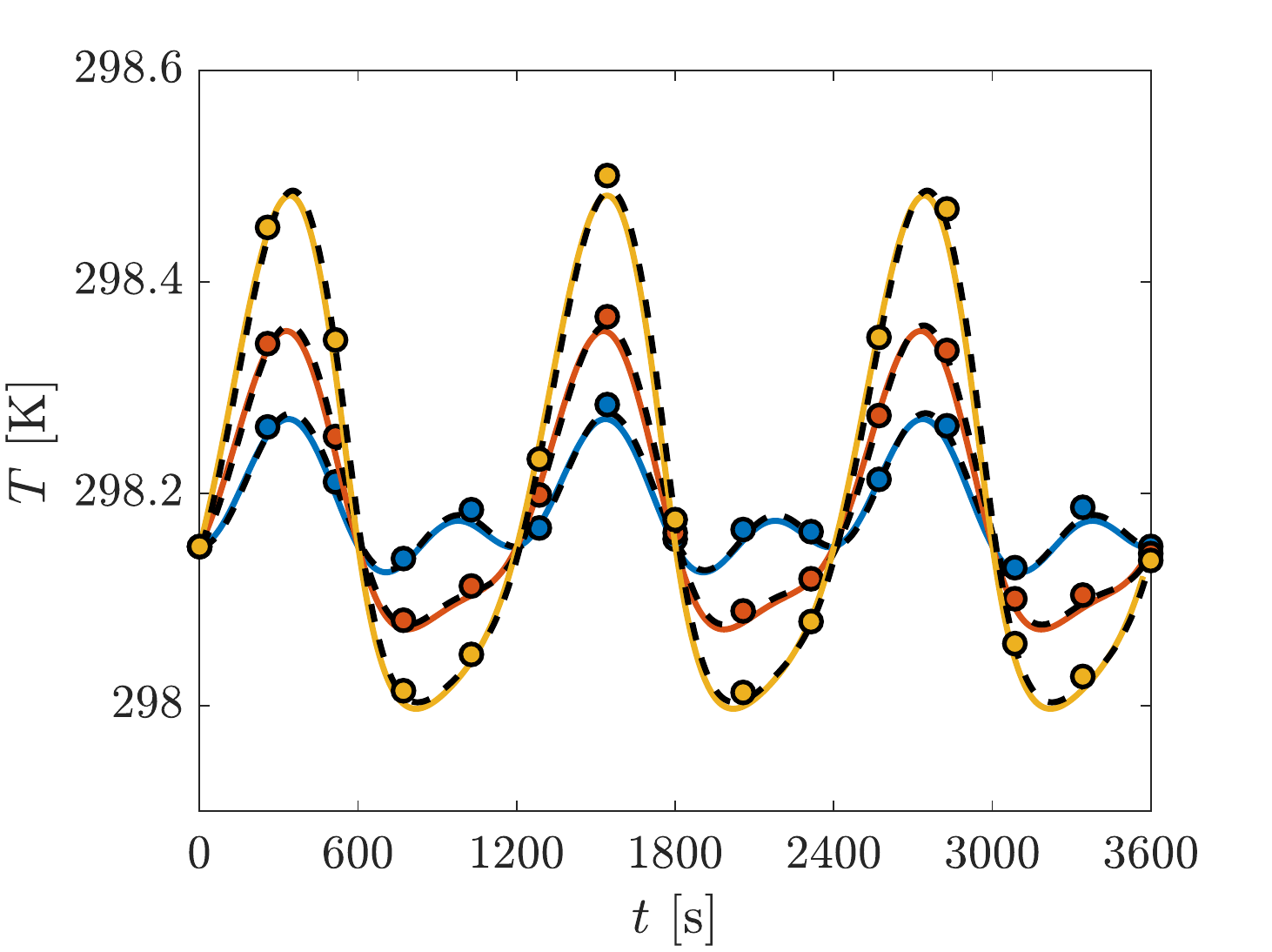}} \\
  \subfigure[]{\includegraphics[width=0.48\textwidth]{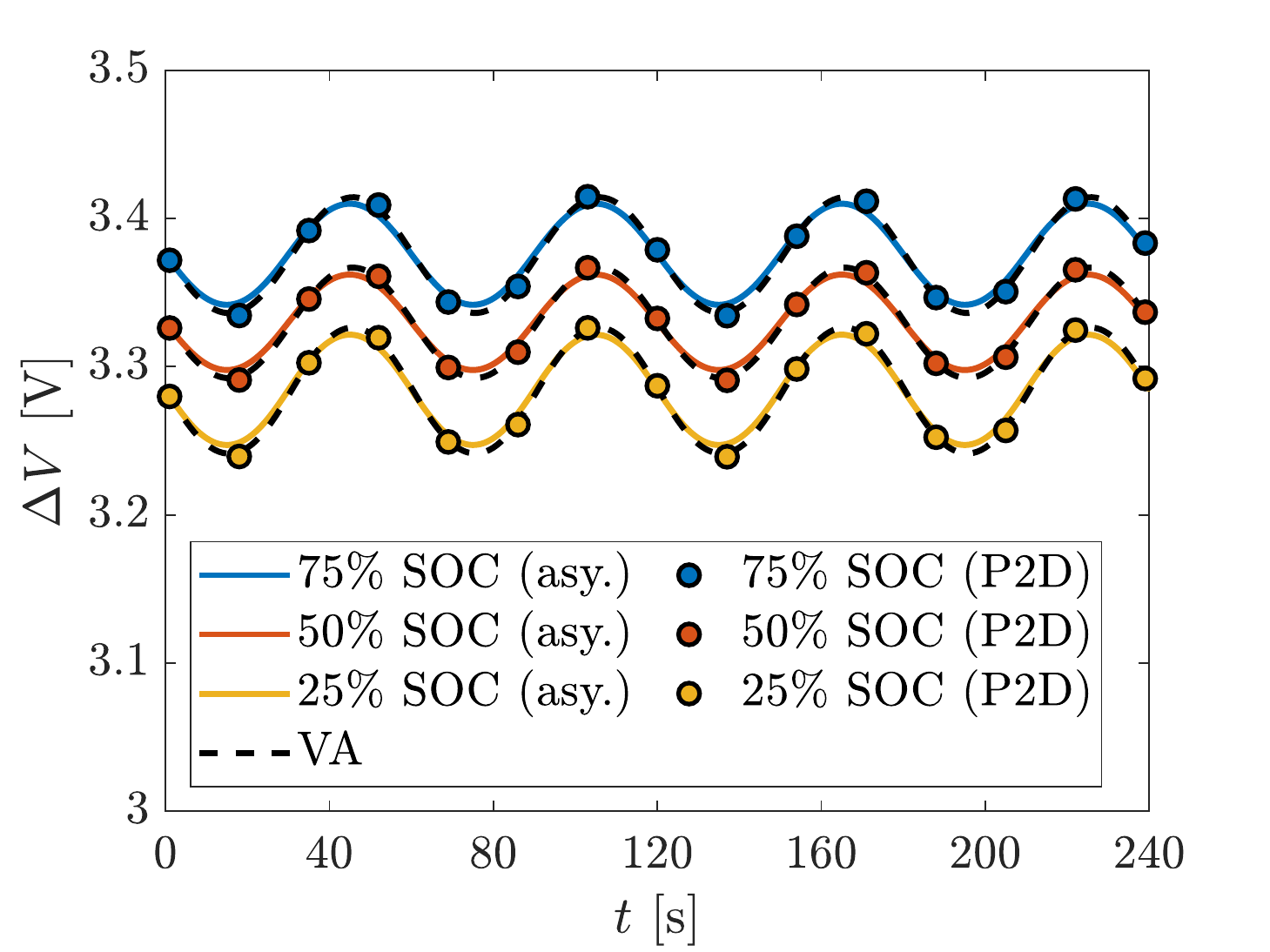}}
  \subfigure[]{\includegraphics[width=0.48\textwidth]{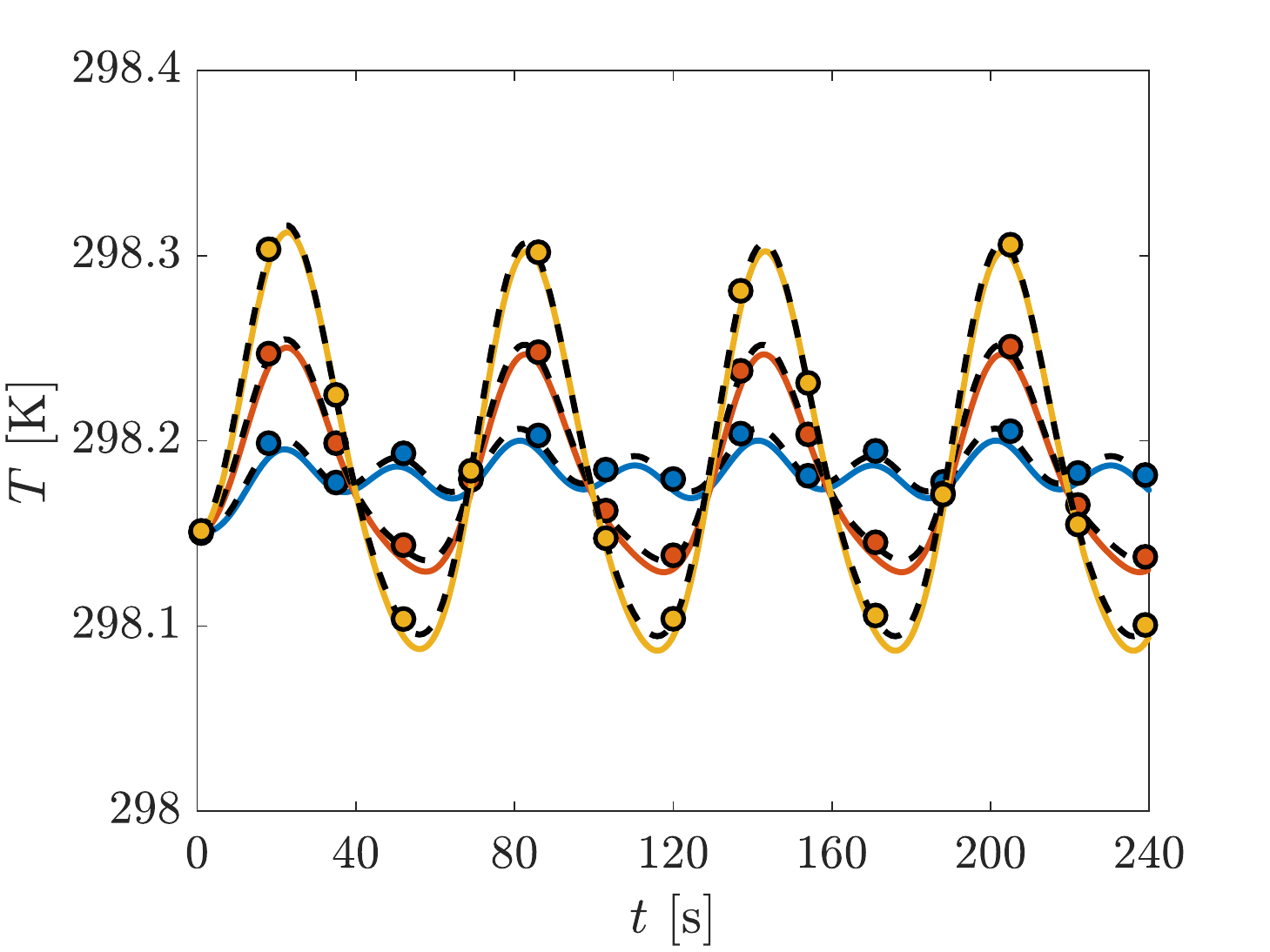}}
  \caption{Evolution of the cell potential and temperature for an oscillating C-rate
    of the form $\I = A \sin(2 \pi t / \tau)$ with $A = 2$. The state of charge (SOC)
    is defined in the main text.  Panels (a)--(b) correspond
    to $\tau = 1200$~s and the asymptotic solution consists of leading- and first-order contributions. Panels (c)--(d) correspond to $\tau = 60$~s and the asymptotic solution consists of only leading-order contributions.}
  \label{fig:galvanodyn}
\end{figure}

The results are shown in Fig.~\ref{fig:galvanodyn}, which plots the cell potential and temperature for the different SOC values using the two different frequencies. In the low-frequency case ($\tau = 1200$~s; panels (a)--(b)), the asymptotic solution is constructed using the leading- and first-order contributions. The agreement between all three models is excellent. The temperature exhibits an oscillatory behaviour as well as a result of the electrochemical heat generation being dominated by the reversible contribution. For the high-frequency case ($\tau = 60$~s; panels (c)--(d)), the asymptotic solution is only based on the leading-order expressions for simplicity but it still leads to an accurate approximation of the numerical solution.

\section{Upscaling to a multi-cell model}
\label{sec:homogenisation}
\begin{figure}
  \centering
  \includegraphics[width=\textwidth]{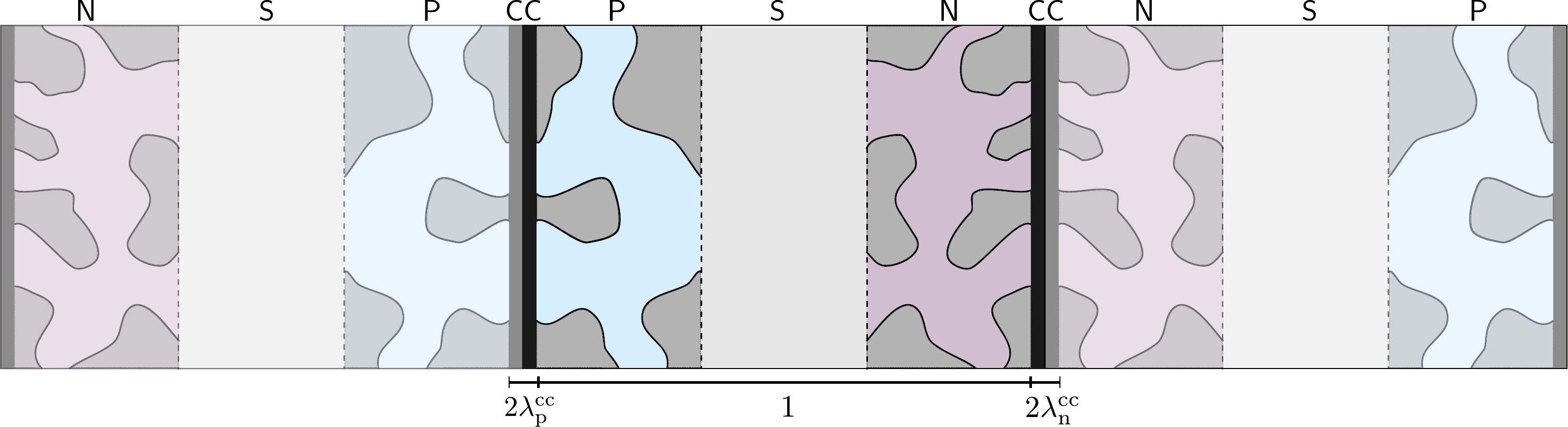}
  \caption{Schematic diagram of an LIB consisting of several repeating identical cells connected by metal current collectors (CC). The non-dimensional thicknesses of the cells and current collectors are illustrated. The unit cell is shown is opaque and its periodic extensions are slightly transparent.}
  \label{fig:coupled_model}
\end{figure}

While a single-cell model provides important insights into the coupling that occurs between the electrochamical and thermal processes, a real battery consists of many cells that are coupled together. The heat that is generated within a cell can accumulate due to the cooled exterior of the battery being further away, which can cause a greater increase in temperature than predicted by the single-cell model. To capture this behaviour, we now consider a multi-cell model consisting of repeating LIB cells that are separated by metal current collectors (Fig.~\ref{fig:coupled_model}). Current collectors are required for a multi-cell model to prevent short-circuiting between adjacent electrodes. Using homogenisation theory, we can derive a single heat equation that holds across the array of cells, where the volumetric heating term is calculated using the cell-scale asymptotic solutions to the electrochemical problem derived in the previous section.

Homogenisation of the heat equation requires a consideration of the current collectors, which were neglected in the single-cell model. The non-dimensional half-thickness of the current collectors adjoining the positive and negative electrodes are denoted by $\lambdap$ and $\lambdan$, respectively. Each current collector is adjacent to the same type of electrode, as shown in Fig.~\ref{fig:coupled_model}. Thus, for periodicity on the cell scale, we take the cell problem to involve two half current collectors for the positive electrode, two positive and negative electrodes, two separators, and one current collector for the negative electrode. The total length of this geometry is denoted by $2 L_c$ where $L_c = 1 + \lambdap + \lambdan$. The centerlines of the positive-electrode current collectors are taken to be at $x = -\lambdap$ and $x = 2(1+\lambdan)+\lambdap$. At the edge of the periodic array, each end is connected to a positive half current collector. The dimensionless volumetric heat capacity and thermal conductivities of the current collectors are denoted by $\rhocci$ and $\Kcci$. Although the current collectors can generate heat through Ohmic heating, their large electrical conductivities makes this negligible compared to the electrochemical reaction heat \cite{Li2014}.

The LIB is assumed to consist of $N \gg 1$ (physical) cells, where $N$ is even so that there are $N/2$ unit cells. To carry out the multiple-scales expansion, we introduce the slow variable $X = \epsilon (x + 2\lambdap)$, where $\epsilon = (N L_c + 2\lambdap)^{-1} \ll 1$, which describes spatial variations across the battery scale. This definition of $X$ and $\epsilon$ is chosen so that $0 \leq X \leq 1$. We are interested in the thermal diffusion over the length of the array and therefore the analysis is performed in the distinguished limit where $\Le = \epsilon^{-2} \oLe$ with $\oLe = O(1)$. However, the resulting model can also be used to study the subcases $\oLe \ll 1$ and $\oLe \gg 1$. The details of the homogenisation procedure are given in Ref.~\cite{Chapman2015} so we will only briefly discuss them here. It is convenient to group the five heat capacities and thermal conductivies into piecewise-constant functions, $\varrho(x)$ and $\mathcal{K}(x)$, that vary on the cell scale. After neglecting sub-dominant Ohmic heat terms, the non-dimensional heat equation describing the temperature in the ``unit cell'' is given by
\begin{align}
  \epsilon^2 \varrho \pderiv{T}{t} = \oLe \left(\pderiv{}{x} + \epsilon \pderiv{}{X}\right)\left[\mathcal{K}\left(\pderiv{T}{x} + \epsilon \pderiv{T}{X}\right)\right] + \epsilon^2 q_r,
\end{align}
with $q_r = q_r(x,X,t)$ denoting a piecewise function describing the electrochemical heat generation in each electrode. The Newton cooling conditions \eqref{bc:newton_1} and \eqref{bc_newton2} are now applied at $X=0$ and $X=1$ while periodic boundary conditions are imposed at the edges of the unit cell.
% \begin{align}
% T(-\lambdap,X,T)-T(2(1+\lambdan)+\lambdap,X,t)=\left.\pderiv{T}{x}\right|_{x=-\lambdap}-\left.\pderiv{T}{x}\right|_{x=2(1+\lambdan)+\lambdap}=0.
% \end{align}
Continuity of temperature and flux is maintained at the electrode-collector boundaries. The temperature is expanded as $T = T^{(0)} + \epsilon T^{(1)} + \epsilon^2 T^{(2)} + O(\epsilon^3)$. At leading-order we find that $T^{(0)} = T^{(0)}(X,t)$ and thus the temperature is constant across individual cells, in agreement with the analysis in Sec.~\ref{sec:prelim_thermal}. From the $O(\epsilon)$ problem we can deduce that
\begin{align}
  \mathcal{K}\left(\pderiv{T^{(1)}}{x} + \pderiv{T^{(0)}}{X}\right) = -\oQ, \quad
  \oQ(X,t) = -\oK\, \pderiv{T^{(0)}}{X}, \quad
  \oK = L_c \left[\int_{-\lambdap}^{1+\lambdan} \frac{\d x}{\mathcal{K}(x)}\right]^{-1},
\end{align}
which corresponds to an effective form of Fourier's law based on the harmonic mean of the thermal conductivity. The $O(\epsilon^2)$ problem leads to the battery-scale model for the temperature $T^{(0)}$, which is given by
\begin{align}
  \ovrho \pderiv{T^{(0)}}{t} = \oLe\, \oK\,\pderiv[2]{T^{(0)}}{X}
  + (x_p/L_c)\posav{\qpr} + [(1-x_n)/L_c]\alphan\negav{\qnr},
  \label{hom:final_heat}
\end{align}
where $\ovrho = L_c^{-1}\int_{-\lambdap}^{1+\lambdan} \varrho(x)\,\d x$.
The averages of the electrochemical heating terms, $\posav{\qpr}$ and $\negav{\qnr}$, are defined in \eqref{eqn:av_qri}. However, these heating terms are now functions of the battery-scale variable $X$ through their dependence on the temperature $T^{(0)}$, which leads to all of the cell-scale electrochemical problems being coupled together.

As all of the cell components are effectively lumped together on the battery scale, and only the positive-electrode current collectors are in contact with the environment, the correct form of the Newton cooling conditions are
\subeq{
\begin{alignat}{2}
  \oK\,\pderiv{T^{(0)}}{X} &= \oBi\, T^{(0)}, &\quad X &= 0, \\
  \oK\,\pderiv{T^{(0)}}{X} &= -\oBi\, T^{(0)}, &\quad X &= 1,
\end{alignat}
}
where $\oBi = \epsilon^{-1} \Bi$. 

\subsection{Heat generation during a galvanostatic discharge}

To illustrate the thermal response of several connected LIB cells, we consider a situation where a constant current $\mathcal{I}_\text{tot}$ is drawn from the battery. The cells are assumed to be identical and connected in parallel. A similar configuration has been used by An \etal\cite{An2018} to model heat generation in a commercial LIB. The potential difference $\Delta v(t)$ across all of the cells will be the same but the local applied current density (i.e. the local C-rate) $\mathcal{I}(X,t)$ can vary between cells. Thus, currents must satisfy
\begin{align}
  \mathcal{I}_\text{tot} = \int_{0}^{1} \mathcal{I}(X,t)\,\d X.
\end{align}
Since the electrical potentials are approximately uniform on the cell scale, we also have that $\Phisp(X,t) - \Phisn(X,t) = \Delta v(t)$.

The leading-order analysis in Sec.~\ref{sec:galvanostatic} is used to construct and simplify the thermal-electrochemical problem. By writing $t = \gammac^{-1} \hat{t}$, $\csi = \gammac^{-1}\csihat$, and taking the limit as $\Ci \to 0$ and $\gammac \to 0$, we find that the corresponding problem is given by the following system of partial-differential-integral equations
\subeq{
  \label{hom:galvano}
\begin{align}
  &\gpbar^{(0)}(\Phisnhat^{(0)}+\Delta v, \csphat^{(0)}, T^{(0)}) = -\frac{\mathcal{I}(X,t)}{\Gp x_p}, \label{hom:local_ec_1} \\
  &\gnbar^{(0)}(\Phisnhat^{(0)}, \csnhat^{(0)}, T^{(0)}) = \frac{\mathcal{I}(X,t)}{\Gn(1 - x_n)}, \label{hom:local_ec_2} \\
  &\int_{0}^{1} \gpbar^{(0)}(\Phisnhat^{(0)}+\Delta v, \csphat^{(0)}, T^{(0)})\, \d X = -\frac{\mathcal{I}_\text{tot}}{\Gp x_p}, \label{hom:glocal_ec}\\
  &\oLe\, \oK\, \pderiv[2]{T^{(0)}}{X} - (L_c \Gp)^{-1} \mathcal{I}(X,t) \left[\Delta v(t) + \log \Up - \log \Un + \DelGp - \DelGn\right] = 0, \label{hom:heat_reduced}
\end{align}
}
where the solid-phase concentrations are given by \eqref{vary_I:csi} and electrochemical kinetics are given by \eqref{const_I:gi_0}--\eqref{const_I:ji_0} with  $\Keffi^{(0)} = \exp\left(\Gammaeffi T^{(0)} \right)$. Equations \eqref{hom:local_ec_1} and \eqref{hom:local_ec_2} determine the local potential in the negative electrode and the local current density.  Equation \eqref{hom:glocal_ec} is a global constraint that determines the potential drop across the battery. Finally, Eqn~\eqref{hom:heat_reduced} describes the evolution of the local temperature.

The spatial variation that occurs in the electrical problem across the battery scale is due to gradients in the temperature. However, the effective Biot number $\oBi$ will generally be small and thus temperature gradients will be weak. In this case, it is possible to carry out an asymptotic expansion in terms of $\oBi \ll 1$ to find that the leading-order contributions are independent of $X$ and thus only functions of time. The leading-order battery temperature is given by
\begin{align}
  T^{(0)}(t) = \,(2\,\oBi\,\oLe\,L_c \Gp)^{-1} \mathcal{I}_\text{tot}\left[\Delta v(t) + \log \Up - \log \Un + \DelGp - \DelGn\right].
  \label{eqn:T_hom}
\end{align}
Equation \eqref{eqn:T_hom} is the battery-scale analogue of \eqref{const_I:temp_0} and it shows that the temperature rise scales roughly with the number of cells $N$. Using $\mathcal{I} = \mathcal{I}_\text{tot}$, the battery potential $\Delta v$ can be obtained be solving \eqref{hom:local_ec_1}--\eqref{hom:local_ec_2} and the solid-phase lithium concentrations are given by \eqref{const_I:csi_0}.

\begin{figure}
  \centering
  \subfigure[]{\includegraphics[width=0.48\textwidth]{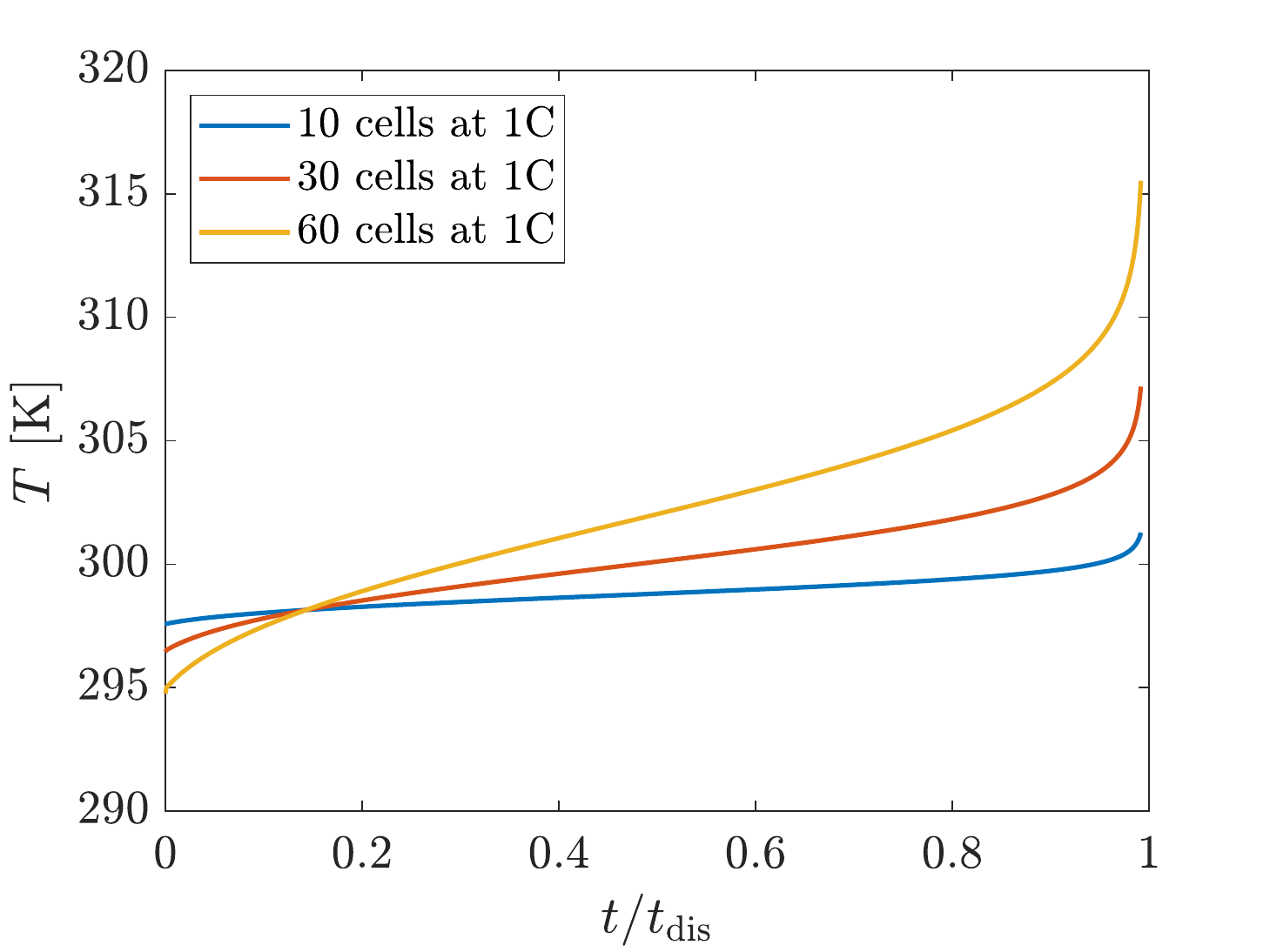}}
  \caption{Evolution of the maximum temperature in a battery pack at a discharge rate of 1C.
    Time as been normalised by the discharge time of one hour.}
  \label{fig:hom_T_N}
\end{figure}

We numerically solve the battery problem \eqref{hom:galvano} for a 1C discharge and explore how the number of cells $N$ affects the evolution of the temperature. The results are shown in Fig.~\ref{fig:hom_T_N}. In all the cases considered, the effective Biot number $\oBi$ is small and so the temperature is approximately uniform across the battery pack. As expected, a battery that consists of more cells becomes hotter during discharge. For a 60-cell battery, the temperature rise is roughly 20~K, which is close to the threshold at which Arrhenius reaction kinetics begin to play a role.

\begin{figure}
  \centering
  \subfigure[]{\includegraphics[width=0.48\textwidth]{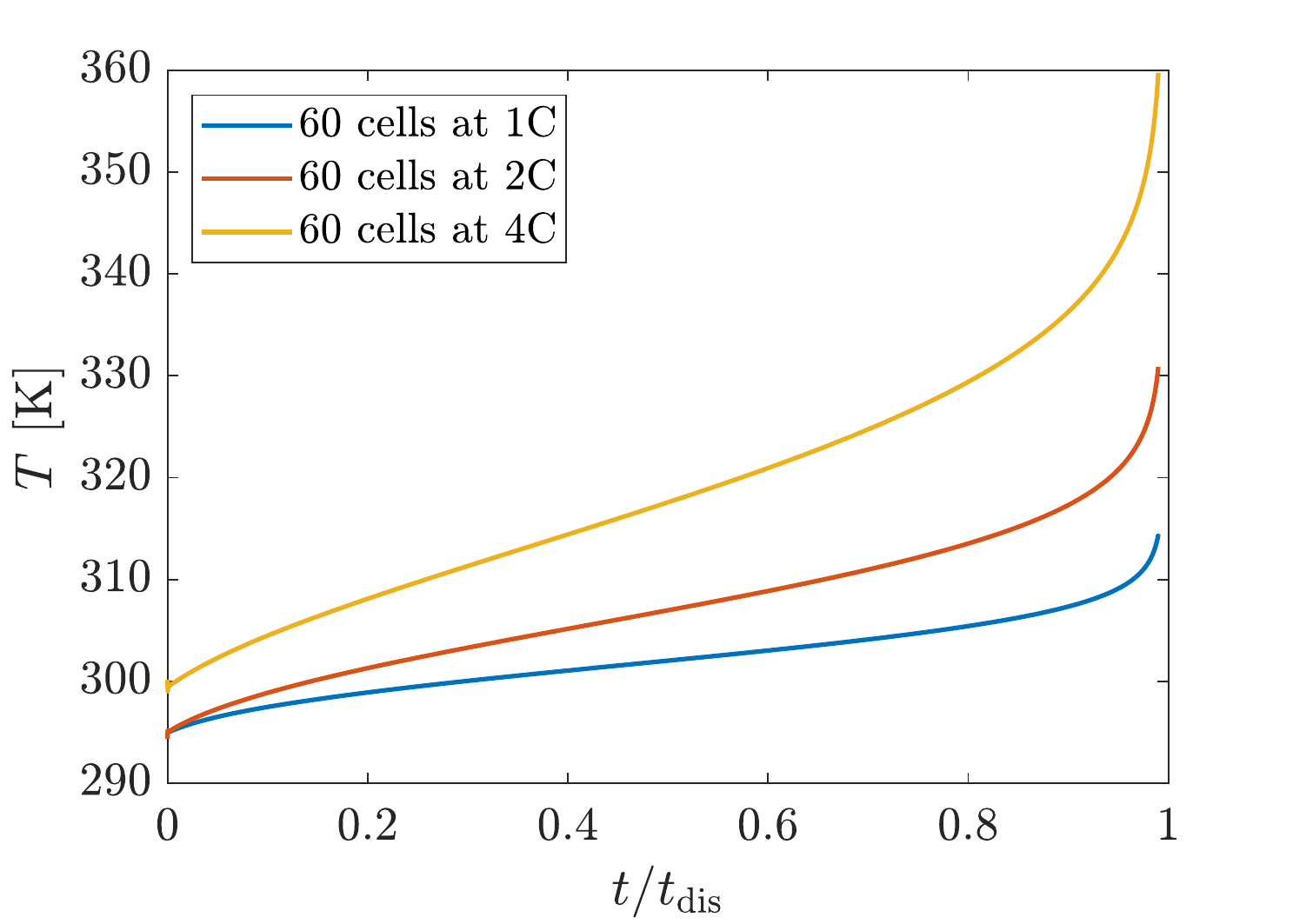}}
  \subfigure[]{\includegraphics[width=0.48\textwidth]{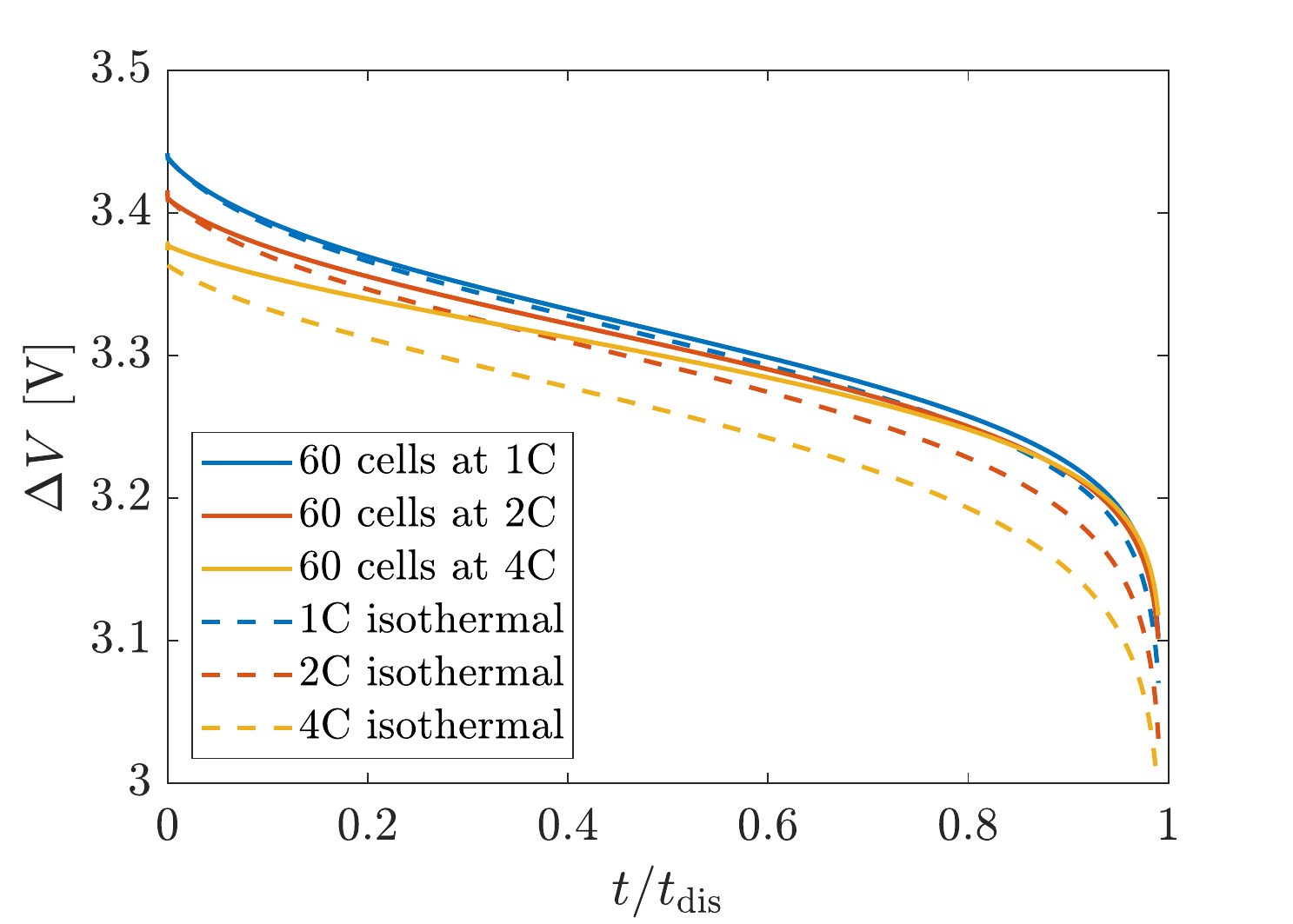}}
  \caption{Temperature increase and cell potential in a battery pack consisting of 60
    cells at various discharge rates. The isothermal voltage curves are shown as dashed
    lines in panel (b). Time is normalised by the battery discharge time $t_\text{dis} = (3600/\I_\text{tot})$~s.}
  \label{fig:hom_C}
\end{figure}

In Fig.~\ref{fig:hom_C}, we focus our attention on the heat generation in a 60-cell battery at higher discharge rates extending up to 4C. The temperature increase in these cases, shown in Fig.~\ref{fig:hom_C} (a), can be substantial and exceed 50~K. The corresponding voltage curves are plotted as solid lines in Fig.~\ref{fig:hom_C} (b), with the isothermal curves shown as dashed lines for reference. The increase in battery temperature leads to a greater cell potential relative to the isothermal case. This is due to the increase in the effective reaction rate. There is less electrical resistance caused by the finite rate of electrochemical reactions driving (de)lithiation and thus the cell potentials are pushed closer to the open-circuit potentials. Such behaviour was predicted by analysing the thermal-electrochemical problem at the cell level in Sec.~\ref{sec:galvanostatic_lo}. Importantly, we see that even with a temperature rise of roughly 60~K and the onset of Arrhenius-dominated reaction kinetics, the model does not predict the occurrence of thermal runaway. We emphasise that the rapid temperature rise near the end of the discharge process is associated with the sharp change in potential caused by the electrodes becoming fully saturated or fully depleted of lithium. 

\section{Discussion and conclusion}
\label{sec:conclusions}
We present thermo-electrochemical models for an LIB which treat the electrode microstructure in two different ways. In one case, volume averaging is used to homogenise the microstructure. In the second, the electrode is treated as a collection of spherical particles following the P2D framework. The two approaches become equivalent in the limit of fast diffusion in the solid particles. The VA model is mathematically simpler and, consequently, asymptotic solutions can be obtained from it for common modes of battery operation. The VA model for a single LIB cell is then homogenised to obtain a model for a battery consisting of $N \gg 1$ cells. 

In the case of a single LIB cell, numerical simulations show strong agreement between the VA and P2D models for the moderate discharge rates considered here. The asymptotic solutions give excellent approximations to the VA model and thus to the P2D model as well. The P2D framework has become a standard modelling approach in the battery community and the results presented here show that the simpler and computationally advantageous VA model can yield the same predictions of macroscopic quantities such as cell temperature and potential. However, the loss of information about the microstructure can prevent the VA model from being used in applications involving phase-separating electrodes, degradation, and active material utilisation. Furthermore, the VA model is unable to capture the effective loss of capacity that occurs in some cells at high C-rates \cite{Srinivasan2004}, which is attributed to non-uniform lithium distribution in the electrode particles \cite{Moyles2019}. Thus, the best model to use is situation dependent. Establishing the conditions under which the P2D model can be asymptotically reduced to the VA model can serve as a useful guide for model selection and is a key area of future research. 

%. For instance, many electrode materials such as LiFePO$_4$ undergo complex phase transformations during (dis)charging \cite{bazantpaper}, resulting in (de)lithiation occuring in isolated groups of particles rather than uniformly across the cell as our model predicts. The small fraction of actively (de)intercalating particles leads to high current densities which can accelerate particle degradation. 

An important area of battery modelling concerns model validation. Typically, models are compared with experimental voltage curves that are obtained by subjecting a battery (or a cell) to a given current. The good agreement between the leading-order asymptotic solutions and the numerical simulations of the full models in the case of fixed or alternating currents at moderate C-rates indicates that spatial effects and the electrolyte play almost no role in determining the cell potential. Thus, these aspects of the model may be difficult to experimentally validate using galvanostatic or galvanodynamic modes of operation. However, these models are ideal for validating the thermal aspects of the model because a sustained current held at a moderate C-rate is required to raise the temperature of the battery to a detectable level. In the case of a potentiostatic hold, whereby the cell potential is fixed, obtaining good agreement between the asymptotic and numerical solutions for the current required the first-order contributions to be included, which captures effects relating the spatial gradients and the electrolyte. The current obtained from potentiostatic experiments therefore encodes useful information about the dynamics of the electrolyte which can assist with model validation. 

The scaling analysis revealed that the electrochemical reactions are the dominant source of heat generation. Given the Arrhenius dependence of the reaction coefficients and the large temperature rises that can occur when many LIB cells are coupled together, it is surprising to find that thermal runaway is not possible in our model. On one hand, this is reassuring, as it shows that LIBs are generally safe and thermal runaway must be due to mechanical abuse or poor design. On the other hand, our model has neglected the fact that above a critical temperature, the electrolyte, active materials, and the solid electrolyte interphase will decompose. These reactions are exothermic and generate flammable gases, which can lead to catastrophic battery failure.  In other LIB designs, Ohmic heating in the electrolyte can be a strong contributor to the total heat generation \cite{Li2014}.  The framework developed here can easily accommodate this case: the asymptotic solutions for the liquid-phase concentrations and current densities can be used to calculate the component-averaged Ohmic heating terms in reduced heat equations such as \eqref{hom:final_heat}.  Our reduction framework can be extended even further to account for empirical expressions for the open-circuit voltage and its derivative with respect to temperature.  
Asymptotically reduced and homogenised models such as those derived here are not only adaptable, but computationally inexpensive, making them ideal for use in on-board thermal management systems, thus paving the way towards safer battery operation.

\subsection*{Acknowledgements}
IRM gratefully acknowledges funding from the Charlemont scholar program of the Royal Irish Academy as well as an NSERC Discovery Grant (2019-06337). MGH has received funding from the European Union's Horizon 2020 research and innovation programme under the Marie Sk{\l}odowska-Curie grant agreement No.~707658. The authors thank Brian Wetton, Tim Myers, Jon Chapman, Colin Please, and Mohit Dalwadi for insightful discussions relating to this work.

\begin{appendix}

  \section{Parameter determination}
  \label{app:parameters}

  The parameters values are adapted from Refs.~\cite{Amiribavandpour2015, Li2014}, in which thermal P2D models are developed for ANR26650m1-a lithium-ion cells \cite{nanobattery2}. The negative and positive electrodes of these cells are constructed from graphite and LiFePO$_4$ (LFP), respectively.   The physical dimensions of the cell, along with its capacity and common parameters, are given in Table \ref{tab:params1}. Values for the electrochemical and thermal parameters are listed in Tables \ref{tab:params2} and \ref{tab:thermal}, respectively. The parameters $I_{\rm app}$ and $A_{\rm cell}$ define the current density $i_0=I_{\rm app}/A_{\rm cell}$.

  The reaction constants at ambient temperature, $\mn{K}{a}{i}^0$ and $\mn{K}{L}{i}^0$, are determined by fitting the analytical expression for the open-circuit potential \eqref{eqn:Ui} to the empirical expressions for graphite and LFP given by
\begin{subequations}
	\begin{align}
	\mnn{U}{n} &= 0.6379+0.5416\exp(-305.5309\yn)+0.044\tanh\left(-\frac{\yn-0.1958}{0.1088}\right)\nonumber\\
	&-0.1978\tanh\left(\frac{\yn-1.0571}{0.0854}\right)-0.6875\tanh\left(\frac{\yn+0.0117}{0.0529}\right)-0.0175\tanh\left(\frac{\yn-0.5692}{0.0875}\right),\label{eqn:Undef}\\
	\mnn{U}{p} &= 3.4323-0.8428\exp(-80.2493(1-\yp)^{1.3198})-3.2474\tten{-6}\exp(20.2645(1-\yp)^{3.8003})\nonumber\\
	&+3.2482\tten{-6}\exp(20.2646(1-\yp)^{3.7995}),\label{eqn:Updef}
	\end{align}
\end{subequations}
where $\yi$ is the state of charge of the battery defined as $ \yi= \csi/\csimax$. Having determined $\mn{K}{a}{i}^0$ and $\mn{K}{L}{i}^0$, the difference in activation energies $\mn{E}{L}{i} - \mn{E}{a}{i}$ is then obtained by fitting the analytical expression for $\partial \mnn{U}{i}/\partial T$ obtained from \eqref{eqn:Ui} to empirical expressions given by
\subeq{
  \begin{align}
    \pderiv{\mnn{U}{n}}{T} &= 10^{-3} \cdot \bigg(\frac{344.1347148\exp(-32.9633287\yn + 8.316711484)}{1 + 749.0756003\exp(-34.79099646\yn + 8.887143624)} \nonumber
    \\ &\qquad- 0.8520278805\yn + 0.362299229\yn^2 + 0.2698001697\bigg), \\
    \pderiv{\mnn{U}{p}}{T} &= -0.35376\yp^8 + 1.3902\yp^7 - 2.2585\yp^6 + 1.9635\yp^5 - 0.98716\yp^4 \nonumber \\
    & \qquad + 0.28857\yp^3 - 0.046272\yp^2 + 0.0032158\yp
    - 1.9186\cdot10^{-5}.
  \end{align}  
}
The diffusivitity of lithium ions and anion in the electrolyte, $D_L^0$ and $D_A^0$, are obtained by simultaneously solving \eqref{eqn:De} and \eqref{eqn:theta} given known values of the effective diffusivity $D_e^0$ and the transference number $\theta^0$ at ambient temperature.

\begin{table}
  \centering
  \footnotesize
  \caption{Parameters applied to the battery}
    \begin{tabular}{cc}
    \hline
Parameter (Units) & Value (Reference) \\
\hline
$H$ (m) & 65$\tten{-3}$ \cite{Li2014} \\
%\hline
$x_p$ (m) & 70$\tten{-6}$ \cite{Li2014} \\
%\hline
$x_n$ (m) & 95$\tten{-6}$ \cite{Li2014} \\
%\hline
$L$ (m) & 129$\tten{-6}$ \cite{Li2014} \\
% \hline
      $\lambdap$ (m) & $1.0\tten{-6}$ \cite{Li2014} \\
      $\lambdan$ (m) & $6.2\tten{-6}$ \cite{Li2014} \\
$A_{\rm cell}$ (m\unit{2}) & 16.94$\tten{-2}$ \cite{Li2014} \\
%\hline
$F$ (C mol\unit{-1}) & 96487 \cite{Newman2004} \\
%\hline
$R$ (J mol\unit{-1} K\unit{-1}) & 8.314  \cite{Newman2004} \\
%\hline
$T_a$ (K) & 298.15 \cite{Li2014} \\
%\hline
$I_{\rm app}$ (A) & 2.3 \cite{nanobattery2} \\
%\hline
$i_0$ (A m\unit{-2}) & 13.6 \cite{Moyles2019} \\
\hline
    \end{tabular}%
  \label{tab:params1}%
  \vspace{1em}
% \end{table}%
% \begin{table}
%   \centering
%   \footnotesize
  \caption{Electrochemical parameters associated with the electrodes, separator, and electrolyte.}
    \begin{tabular}{ccccc}
    \hline
      %& \multicolumn{4}{c|}{Value (Reference)} \\
    %\hline
    Parameter (Units)       & Positive Electrode & Negative Electrode & Electrolyte & Separator \\
    \hline
    $\phiei$ & 0.33 \cite{Li2014} & 0.33 \cite{Li2014} &       & 0.33\\
      %\hline
    $\phisi$ & 0.43 \cite{Li2014} & 0.55 \cite{Li2014} &       &  \\
      %\hline
      $R_p$ (m) & $3.65\tten{-8}$ \cite{Li2014} & $3.5\tten{-6}$ \cite{Li2014} & & \\
      %\hline
      $\mnn{a}{i}$ (m\unit{-1}) & 3.53 $\tten{7}$ \cite{Li2014} & 4.71 $\tten{5}$ \cite{Li2014} &       &  \\
      %\hline
      $D_e^0$ (m\unit{2} s\unit{-1}) &       &       & $2.6\tten{-10}$  \cite{Amiribavandpour2015} &  \\
      %\hline
      $D_L^0$ (m\unit{2} s\unit{-1}) &       &       & $2.0\tten{-10}$  &  \\
      %\hline
    $D_A^0$ (m\unit{2} s\unit{-1}) &       &       & $3.6\tten{-10}$  &  \\
      %\hline
    $\mn{D}{a}{i}^0$ (m\unit{2} s\unit{-1}) & $1.18\tten{-18}$ \cite{Li2014} & 3.9 $\tten{-14}$ \cite{Li2014} &       &  \\
      %\hline
      $\theta^0$ &       &       & 0.363 \cite{Li2014,Amiribavandpour2015}  &  \\
      %\hline
    $\mu_L^0$ (m\unit{2} mol J\unit{-1} s\unit{-1}) &       &       & $8.23\tten{-14}$ [Eqn~\eqref{eqn:NE}] &  \\
    %\hline
    $\mu_A^0$ (m\unit{2} mol J\unit{-1} s\unit{-1}) &       &       & $1.44\tten{-13}$ [Eqn~\eqref{eqn:NE}] &  \\
    %\hline
    $\sigmasi^0$ (S m\unit{-1}) & 2.15 \cite{Moyles2019} & 100 \cite{Li2014} &       &  \\
    %\hline
    $\sigma_e^0$ (S m\unit{-1}) &       &       & 2.53 [Eqn~\eqref{eqn:sigma_e}] &  \\
    %\hline
    $\csimax$ (mol m\unit{-3}) & 22806 \cite{Li2014} & 31370 \cite{Li2014}  &       &  \\
    %\hline
    $\csi^0$ (mol m\unit{-3}) & 0.022 $\cspmax$ \cite{Li2014} & 0.86 $\csnmax$ \cite{Li2014} &       &  \\
    %\hline
    $c_{L}^0$ (mol m\unit{-3}) &       &       & 1200 \cite{Li2014} &  \\
    % $\Uiref$ (V) & 3.43 \tnote{c} & 0.116  \tnote{c} &       &  \\
    %\hline
    $\mnn{\beta}{i}$ & 0.5   & 0.5   &       &  \\
    %\hline
    % $\mnn{\hat{C}}{i}$ (m\unit{2.5} mol\unit{-0.5} s\unit{-1}) & 1.4$\tten{-12}$ \tnote{d} & 3$\tten{-11}$ \tnote{d} &       &  \\
      $\Keffi^0$ (m s\unit{-1}) & $2.11\tten{-10}$ \cite{An2018} & $5.13\tten{-9}$ \cite{Li2014} & & \\  %\hline
      $\mn{K}{a}{i}^0$ (m s\unit{-1}) & $3.71\tten{-40}$ & $1.05\tten{-10}$ & & \\  %\hline
      $\mn{K}{L}{i}^0$ (m s\unit{-1}) & $1.19\tten{20}$ & $2.69\tten{-7}$ & & \\  %\hline
    $\mn{g}{0}{i}$ (A m\unit{-2}) & 1.57$\tten{-2}$ [Eqn~\eqref{eqn:g0def}] & 1.09 [Eqn~\eqref{eqn:g0def}] &       &  \\
    %\hline
    $\mn{C}{\Gamma}{i}$ (F m\unit{-2}) & 0.2 \cite{Li2014} & 0.2 \cite{Li2014} &       &  \\
    \hline
    \end{tabular}%
  \label{tab:params2}%
% \end{table}%
% \begin{table}
%   \centering
%   \footnotesize
\vspace{1em}
  \caption{Thermal properties of the lithium-ion cell. All parameters values except those for the heat transfer coefficients $\hi$ are from Li \etal\cite{Li2014}. We abbreviate ``current collector'' by CC.}
  \label{tab:thermal}
  \begin{tabular}{ccccccc}
    \hline
    Parameter (Units)    & P.~Electrode & N.~Electrode & Separator & CC (P) & CC (N) \\
    \hline
    $\rho_{\text{solid,i}}$ (kg m\unit{-3}) & 1500 & 2223 & 900 & 2700 & 8700  \\
    $\rho_{\text{e},i}$ (kg m\unit{-3}) & 1210 & 1210 & 1210 & &\\
    $\mathsf{c}_{\text{solid,i}}$ (J kg\unit{-1} K\unit{-1}) & 800 & 641 & 1883 &897 & 396 \\
    $\mathsf{c}_{e,\text{i}}$ (J kg\unit{-1} K\unit{-1}) & 1518 & 1518 & 1518 & &\\
    $k_{\text{solid,i}}$ (W m\unit{-1} K\unit{-1}) & 1.48 & 1.04 & 0.5 & 237 & 398\\
    $k_{e,\text{i}}$ (W m\unit{-1} K\unit{-1}) & 0.099 & 0.099 & 0.099 & & \\
    $\hi$ (W m\unit{-2} K\unit{-1}) & 7.17 & 7.17 & & & \\
    $\Eeffi$ (kJ mol\unit{-1}) & 20 & 30 & & & \\
    $\mn{E}{L}{i} - \mn{E}{a}{i}$ (kJ mol\unit{-1}) & $-330$ & $-7.5$ & & &\\
    %$\mn{E}{L}{i}$ (kJ mol\unit{-1}) & \mh{????} & \mh{???} & \\
    \hline
  \end{tabular}
\end{table}

\section{Analysis of capacitance time regimes for the potentiostatic case}
\label{app:cap_regimes}

The first capacitance regime for the case of a galvanostatic discharge is defined by $t=\Cn\tilde{t}$ for C-rates that are $O(1)$ in size due to the singular nature of the time derivative in \eqref{sys:electrode_solid}.  For the potentiostatic problem, 
%Using this scaling for the potentiostatic case yields $\Phiei=\deriv{\Phisi}{t}=0$ which nullifies the singular contribution. 
a very small-time analysis demonstrates that $\Phisi(x)=\Phiei(x)$, corresponding to zero initial overpotential. If the C-rate is again $O(1)$ then these
potentials are spatially independent.  The grounding condition implies that $\Phisi = \Phiei = 0$, which prevents the condition on the cell potential, $\Phisp - \Phisn = \Delta v$, from being satisfied. The potentials at very small times must therefore be spatially inhomogeneous and this suggests that there is a distinguished small-time limit with large current densities. This large current is due to the lack of charge-transfer resistance at extremely small times. Therefore, to capture the dynamics in the first capacitance regime, we write $t = \tau_n \tilde{t}$ where $\tau_n = \Cn \Gn \nusn / \phisn$.  The C-rate, and thus the current densities, scale with $\cellav{\nu}^{-1}\Delta v$, where $\cellav{\nu}\ll1$ is defined by \eqref{const_V:I0}.  On this time scale, the change in solid- and liquid-phase lithium concentration scales like $\Delta c \sim \tau_n \cellav{\nu}^{-1}\Delta v$ and will be small provided that $\Delta v = O(1)$.  The large current amplifies Ohmic heating, which leads to temperature changes that scale like $\Delta T \sim \tau_n \Qsi (\cellav{\nu}^{-1}\Delta v)^2$ for the electrode contributions and $\Delta T \sim \tau_n \Qe (\cellav{\nu}^{-1}\Delta v)^2$ for the electrolyte.  Electrochemical reactions lead to changes in temperature that scale like $\Delta T \sim \tau_n$.  Based on these scaling estimates, we can assume that neither the concentration nor the temperature changes in this time regime, and thus we only need to consider the electrical problem.  By eliminating the current, the electrical problem takes the form
\begin{align}
\ell^2\pderiv[2]{\Phisp}{x} = \varepsilon_n \gpbar + \mathcal{C}\pderiv{}{\tilde{t}}\left(\Phisp - \Phiep\right), \label{eqn:taunPhip}\\
\pderiv[2]{\Phisn}{x} = \varepsilon_n \gnbar + \pderiv{}{\tilde{t}}\left(\Phisn - \Phien\right), \label{eqn:taunPhin}\\
\phisi\nusi^{-1}\pderiv{\Phisi}{x} + \phie \nue^{-1}\pderiv{\Phiei}{x} = \I(\tilde{t}).
\label{eqn:glob_pot}
\end{align}
where $\ell = [(\phisp \nusn \Gn) / (\phisn \nusp \Gp)]^{1/2} \ll 1$, $\mathcal{C} = \Cp / \Cn \gg 1$, and $\varepsilon_n = \Gn \nusn / \phisn \ll 1$.  Boundary conditions are given by
\begin{alignat}{2}
\pderiv{\Phisp}{x} &= \frac{\nusp \I(\tilde{t})}{\phisp}, &\quad x &= 0, \\
\pderiv{\Phisp}{x} &= 0, &\quad x &= x_p, \\
\pderiv{\Phisn}{x} &= 0, &\quad x &= x_n, \\
\pderiv{\Phisn}{x} &= \frac{\nusn \I(\tilde{t})}{\phisn}, &\quad x &= 1.
\end{alignat}
In the limit $\tilde{t} \to 0$, we find that 
%\im{From \eqref{eqn:taunPhip}, $\Phisp(x)=\Phiep(x)$ which is also true for the negative electrode in} the limit $\tilde{t} \to 0$ \im{satisfying the zero overpotential initial condition. These conditions can be used in \eqref{eqn:glob_pot} to show that for small time}
the solid- and liquid-phase potentials are equal and independent of time,
$\Phisi(x) = \Phiei(x)$, satisfying the zero initial overpotential condition.  The equality of solid- and liquid-phase potentials can be used in \eqref{eqn:glob_pot} to show that
\begin{align}
\pderiv{\Phiei}{x} = \mn{\nu}{\mathrm{eq}}{i}\,\I,
\label{eqn:parallel}
\end{align}
where the equivalent resistances are defined in \eqref{const_V:nu_eq}. Note that \eqref{eqn:parallel} also applies to the separator.  Integration of \eqref{eqn:parallel} across each of the respective domains, adding the results, and using continuity of the potential along with $\Phiep(0) - \Phien(1)=\Delta v$, leads to the expression for the initial C-rate given by \eqref{const_V:I0}.  As time increases (\emph{i.e.} for $\tilde{t} = O(1)$), the gradient in the solid-phase current in the negative electrode (the left-hand side of \eqref{eqn:taunPhin}) drives an overpotential (the second term on the right-hand side of \eqref{eqn:taunPhin}), resulting in the onset of electrochemical reactions.  The increase in electrical resistance due to charge transfer across the electrode-electrolyte interface leads to a marked decrease in the current.  For even longer times (\emph{i.e.} for $1 \ll t \ll \varepsilon_n^{-1}\mathcal{C}$), the current will relax an $O(1)$ value which reaches a plateau when the time derivative in \eqref{eqn:taunPhin} vanishes. When this happens, the current in the negative electrode is once again obtained from the electrochemistry through the equation $\gnbar=\mathcal{I}/[\Gn(1-x_n)]$. We can isolate the cell potential from this expression:
\begin{align}
-\Delta v = 2\sinh^{-1}\left(\frac{\mathcal{I}}{2\Gn(1-x_n)}\right)-\Phisp(0,t).
\end{align}
To leading order in $\nusp$ and $\nue$, we find $\Phiep = \Phisp=0$. However, we can easily obtain the corrections by noting that \eqref{eqn:parallel} still holds for the positive electrode and separator which can once again be integrated across each domain. We couple this with $\partial \Phien/\partial x = -\nue \ien=\nue\mathcal{I}[1-(x-x_n)/(1-x_n)]$ to produce $\Phisp(0,t)=-\mathcal{I}[\mn{\nu}{\mathrm{eq}}{p}x_p+\mn{\nu}{\mathrm{eq}}{s}(x_n-x_p)+\mn{\nu}{\mathrm{eq}}{s}/2(1-x_n)]$ yielding the plateau value \eqref{const_V:I1}.

In the second capacitance regime, we write $t = \tau_p \check{t}$, where $\tau_p=\Cp$ %$\tau_p = \Cp \Gp \nusp / \phisp$, 
and take $\mathcal{C} \to \infty$ to obtain
\begin{align}
%\pderiv[2]{\Phisp}{x} = \varepsilon_p \gpbar + \pderiv{}{\check{t}}\left(\Phisp - \Phiep\right), \\
\frac{\phisp}{\nusp}\pderiv[2]{\Phi}{x}=\Gp\left[\gpbar+\pderiv{}{\check{t}}\left(\Phisp-\Phiep\right)\right],\label{eqn:taupPhi}\\
 \frac{\phisn}{\nusn}\pderiv[2]{\Phisn}{x} = \Gn \gnbar, \\
\phisi\nusi^{-1}\pderiv{\Phisi}{x} + \phie \nue^{-1}\pderiv{\Phiei}{x} = \I(\check{t}),
\end{align}
%where $\varepsilon_p = \Gp \nusp / \phisp \ll 1$.  Taking $\check{t} \to 0$ shows that $\Phisp = \Phiep$.  
Due to the smallness of $\nusi$ and $\nue$, the potentials in both electrodes are approximately uniform in space.  Thus, the solid- and liquid-phase electrical potentials in the %negative
electrodes are related through the equations $\gnbar = \I/[\Gn(1-x_n)]$ and $\gpbar=\I/(\Gp x_p)$ while the currents are given by the expressions in \eqref{red:isi}--\eqref{red:iei}. Taking the limit $\check{t}\gg1$ yields
\begin{align}
-\Delta v = 2\sinh^{-1}\left(\frac{\I}{2\Gn(1-x_n)}\right)+2\sinh^{-1}\left(\frac{\I}{2\Gp x_p}\right)-\Phiep(0,t).
\end{align}
The positive electrolyte potential satisfies $\Phiep(0,t)=0$ to leading order, but the correction can be computed similarly to $\Phisp(0,t)$ in the first capacitive regime producing
\begin{align}
\Phiep(0,t)=-\frac{\mn{\nu}{\mathrm{eq}}{s}}{2}(1-x_p+x_n)\I,
\end{align}
resulting in the plateau value given by \eqref{const_V:I2}.
%By integrating \eqref{eqn:parallel} across the separator and positive electrode, $\partial \Phien/\partial x = -\nue \ien$ across the negative electrode, and using the boundary conditions results in the equation for the C-rate given by \eqref{const_V:I1}.  Equation \eqref{const_V:I2} can be obtained by considering the problem on the diffusive time scale, taking the limit as $t \to 0$, and following the same strategy. 

\begin{figure}
\centering
\subfigure[$\Delta V = 3.49~V$, $\csp^0 = 0.022 \cspmax$, $\csn^0 = 0.86 \csnmax$]{\includegraphics[width=0.48\textwidth]{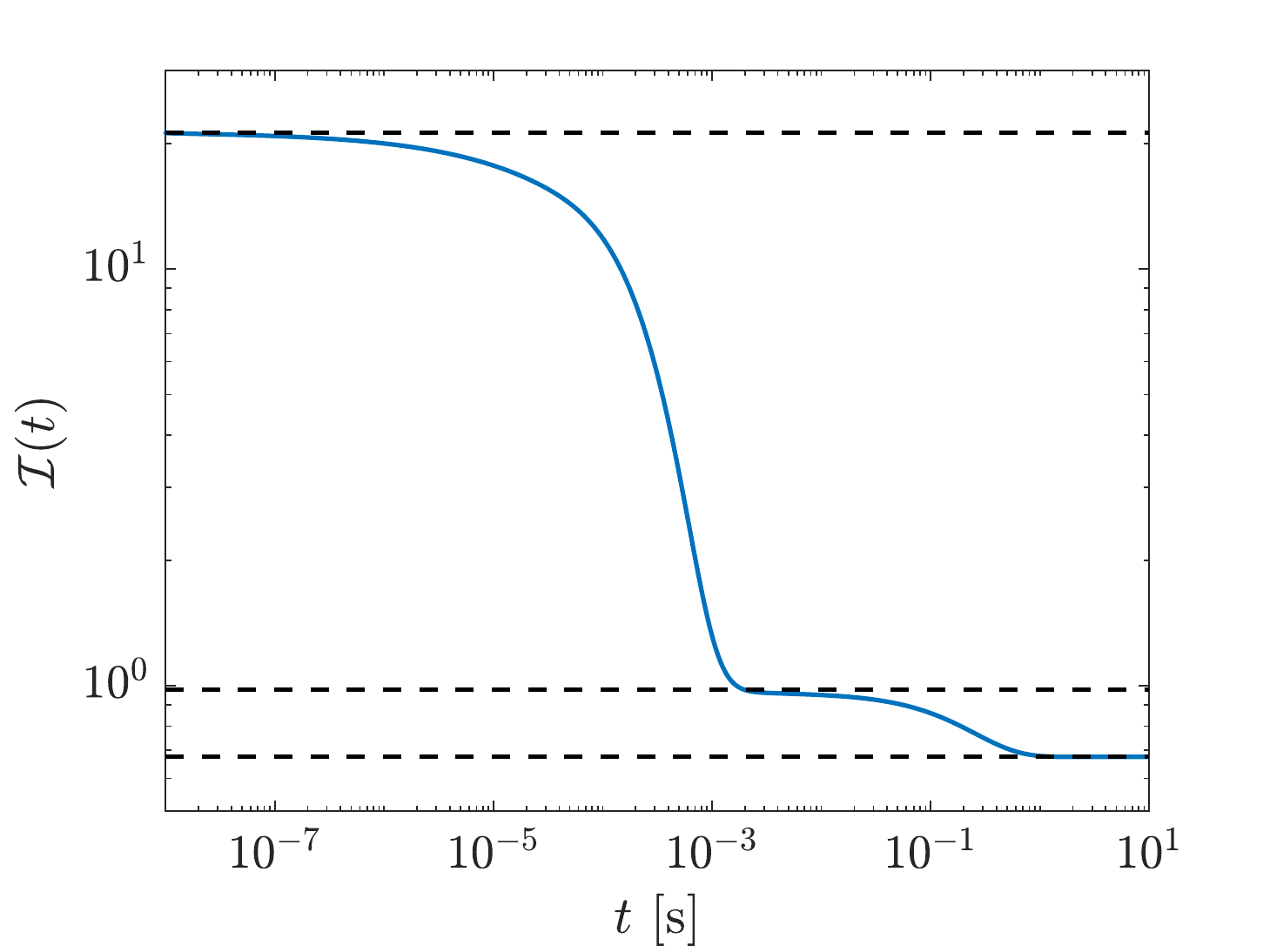}}
\subfigure[$\Delta V = 3.45~V$, $\csp^0 = 0.022 \cspmax$, $\csn^0 = 0.86 \csnmax$]{\includegraphics[width=0.48\textwidth]{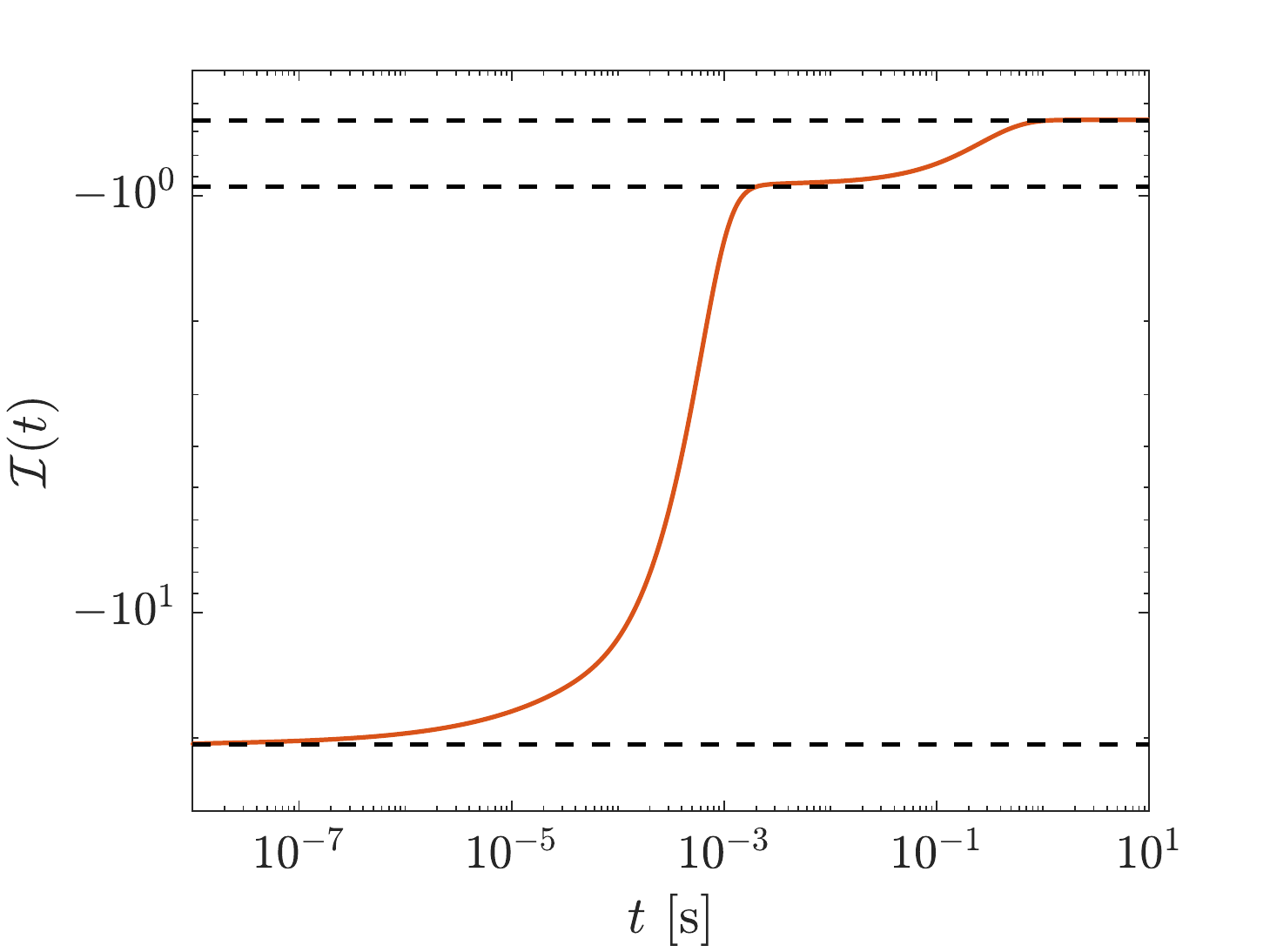}}
\subfigure[$\Delta V = 3.30~V$, $\csp^0 = 0.39 \cspmax$, $\csn^0 = 0.43 \csnmax$]{\includegraphics[width=0.48\textwidth]{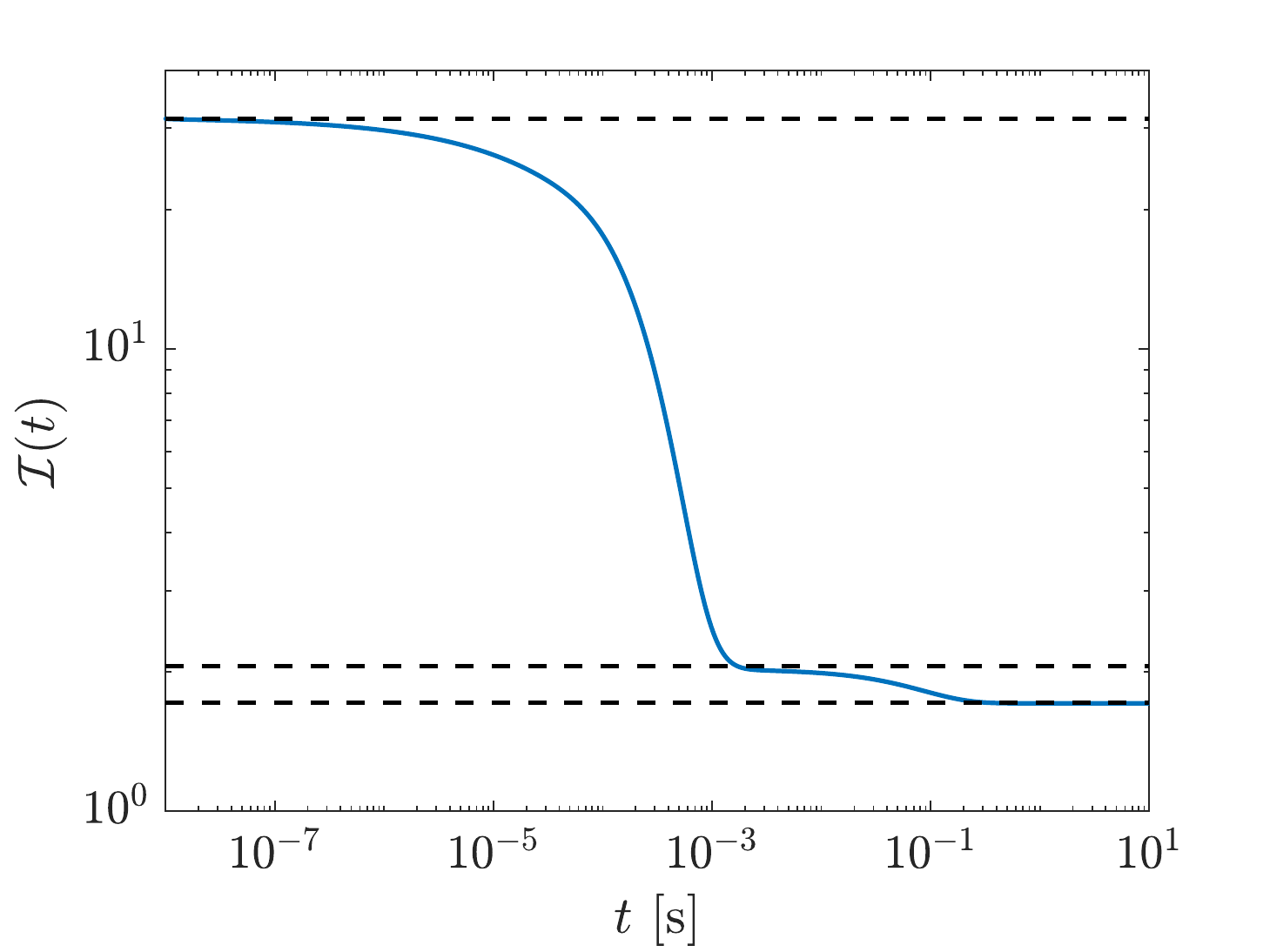}}
\subfigure[$\Delta V = 3.35~V$, $\csp^0 = 0.39 \cspmax$, $\csn^0 = 0.43 \csnmax$]{\includegraphics[width=0.48\textwidth]{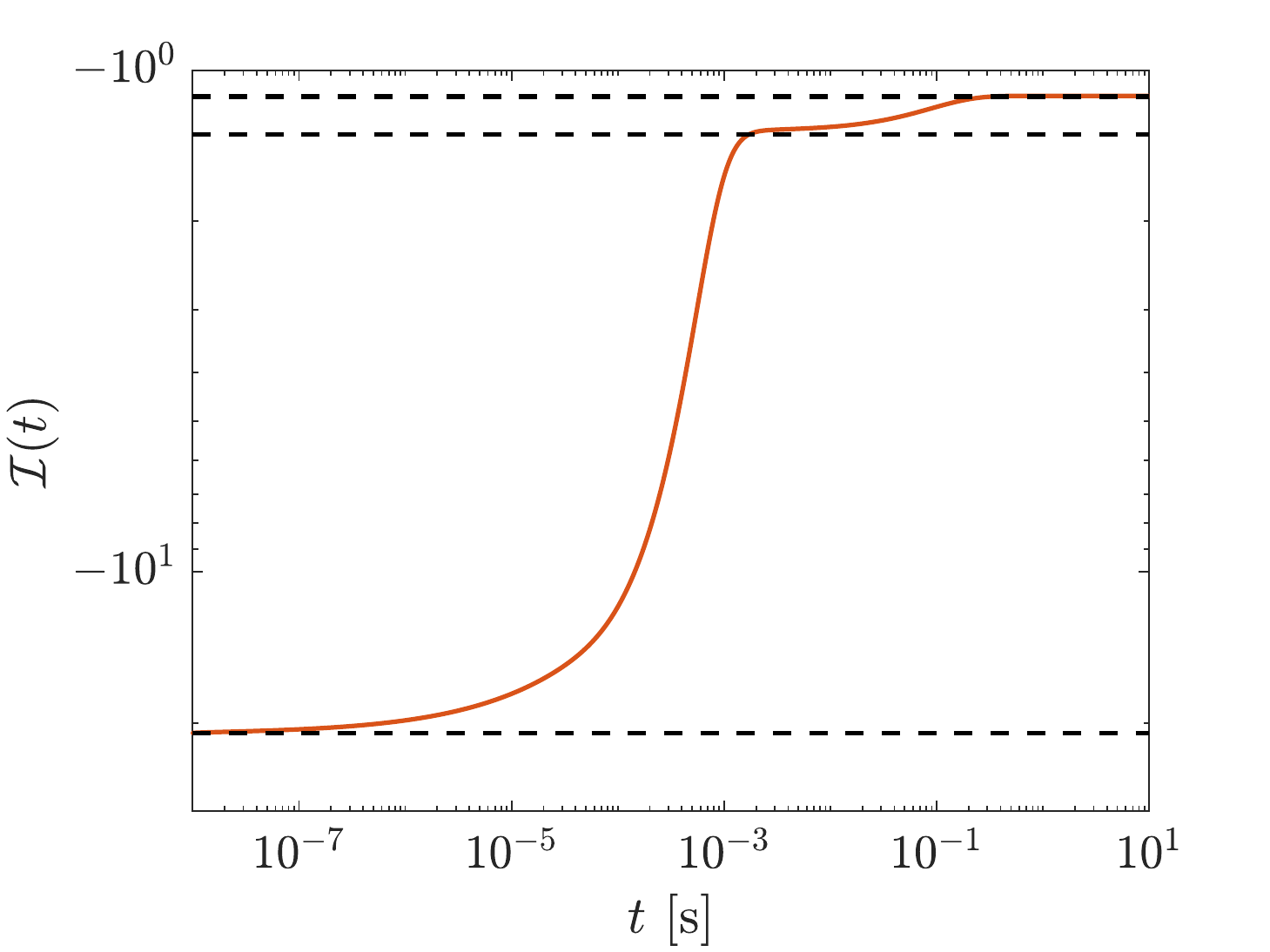}}
\caption{Numerical simulations of the C-rate in the capacitance time regimes in the case of a constant cell potential. The dashed lines correspond to the current plateaus given by \eqref{const_V:I0}, \eqref{const_V:I1}, and \eqref{const_V:I2}.}
\label{fig:cap_regimes}
\end{figure}

The transient evolution of the C-rate in the two capacitance time regimes is shown in Fig.~\ref{fig:cap_regimes}.  The curves are obtained by solving \eqref{eqn:glob_pot} using a fully implicit method based on a spectral discretisation in space and Newton iterations at each time step.  The initial conditions are the same as those discussed in Sec.~\ref{sec:potentio_num}.  Therefore, the curves in Fig.~\ref{fig:cap_regimes} capture the early dynamics that are not shown in Fig.~\ref{fig:potentio}.  The capacitance time scales are $\tau_n \simeq 3\tten{-5}$~s and $\tau_p \simeq 0.3$~s for panels (a)--(b) and $\tau_p = 0.1$ for panels (c)--(d).
%0.1$~s.
The three plateaus described by \eqref{const_V:I0}, \eqref{const_V:I1}, and \eqref{const_V:I2} as indicated by dashed lines.

\section{The composite solution in the case of a potentiostatic hold}
\label{app:matching}
We first construct the composite solution to the leading- and first-order electrical problem and then consider the solution to the thermal problem. To simplify the notation, we write the steady-state liquid-phase concentration and potential as
\subeq{
  \begin{align}
    \cli^{(0)}(x,t\to\infty) &= \clibar^{\infty} = [\I(1-\theta^0)/\phie]\cli^{\infty}, \\
    \Phiei^{(1)}(x,t\to\infty) &= \Phieibar^{\infty} = [(\Da \I)/(2\phie)]\Phiei^{\infty}.
  \end{align}
}
The large-time expansions of the corrections to the open-circuit potential $\Vi^{(1)}$ and exchange current density $\ji^{(1)}$ on the diffusive time scale are given by $\Vi^{(1)} \sim \mn{M}{V}{i} t + \mn{N}{V}{i}(x)$, $\ji^{(1)} \sim \mn{M}{j}{i} t + \mn{N}{j}{i}(x)$ as $t \to \infty$, where
\begin{align}
  \mn{M}{V}{i} &= -(\gammac / \nue) \deltai(1 + \xii(1-\xii)^{-1})\deriv{\csi^{(0)}}{t}, \\
  \mn{N}{V}{i} &= (\gammac / \nue) \clibar^{\infty}, \\
  \mn{M}{j}{i} &= (1/2)(\gammac/\nue)\deltai(1-\xii(1-\xii)^{-1})\deriv{\csi^{(0)}}{t}, \\
  \mn{N}{j}{i} &= (1/2)(\gammac / \nue) \clibar^{\infty},
\end{align}
where we have used the fact that the leading-order solid-phase lithium concentrations $\csi^{(0)}$ are linear functions of time. Consequently, the constant $P$ that appears in the correction to the solid-phase potentials $\Phisi^{(1)} \sim \Phisi^{\infty} + P$ can be written as $P \sim M_P t + N_P$ as $t \to \infty$, where
\begin{align}
  \mathcal{S} &= \Gp x_p \cosh(\Phisp^{(0)}/2) + \Gn(1-x_n)\cosh(\Phisn^{(0)}/2), \\
  \mathcal{S}M_P &= -\Gp x_p \left[2 \sinh(\Phisp^{(0)}/2) \mn{M}{j}{p} - \cosh(\Phisp^{(0)}/2)
                             \mn{M}{V}{p}\right] \nonumber \\ &\quad- \Gn(1-x_n)\left[2 \sinh(\Phisn^{(0)}/2) \mn{M}{j}{n} - \cosh(\Phisn^{(0)}/2)\mn{M}{V}{n}\right], \\
  \mathcal{S}N_P &= -\Gp x_p \left[2 \sinh(\Phisp^{(0)}/2) \posav{\mn{N}{j}{p}} + \cosh(\Phisp^{(0)}/2)
                   \posav{\Phisp^{\infty} - \Phiepbar^{\infty}-\mn{N}{V}{p}}\right] \nonumber \\ &\quad- \Gn(1-x_n)\left[2 \sinh(\Phisn^{(0)}/2) \negav{\mn{N}{j}{n}} + \cosh(\Phisn^{(0)}/2)\negav{\Phisn^{\infty} - \Phienbar^{\infty} - \mn{N}{V}{n}}\right].
\end{align}
The correction to the C-rate can also be expanded as $\I^{(1)} \sim M_I t + N_I$ as $t \to \infty$, where
\begin{align}
  M_I &= \Gn(1-x_n)\left[2 \sinh(\Phisn^{(0)}/2) \mn{M}{j}{n} + \cosh(\Phisn^{(0)}/2)(M_P - \mn{M}{V}{n})\right], \\
  N_I &= \Gn(1-x_n)\left[2 \sinh(\Phisn^{(0)}/2) \negav{\mn{N}{j}{n}} + \cosh(\Phisn^{(0)}/2)\negav{N_P + \Phisn^{\infty} - \Phienbar^{\infty} - \mn{N}{V}{n}}\right].
\end{align}
The solution to the C-rate in the overlap region is found by matching to the solutions on the saturation time scale in the limit as $\hat{t} \to 0$. We find that the leading-order contribution is given by $\I^{(0)}_\text{overlap} = \I^{(0)}$, where $\I^{(0)}$ is determined from \eqref{const_V:e_00}. The first-order contribution is given by $\I^{(1)}_\text{overlap} = M_I t + N_I$.

To construct the composite solution for the temperature, we use the fact that $u^{(1)} \sim M_u t + N_u$ as $t \to \infty$, where
\begin{align}
  M_u &= -M_{I}[\Delta v + \log \Up - \log \Un + \DelGp - \DelGn], \\
  N_u &= -N_{I}[\Delta v + \log \Up - \log \Un + \DelGp - \DelGn] \nonumber \\
  &\quad - \I^{(0)}[\posav{\Phisp^{\infty}-\Phiepbar^{\infty}} - \negav{\Phisn^{\infty} - \Phienbar^{\infty}}]
        - \Gp \mathcal{T}\cellav{\varrho} M_u
\end{align}
By matching to the asymptotic solution for the tempetaure on the saturation time scale in the limit as $\hat{t} \to 0$, we find that the first-order contribution in the overlap region is given by $u^{(1)}_\text{overlap} = M_u t + N_u$.

\end{appendix}

\section*{References}

\bibliographystyle{elsarticle-num}
\bibliography{bib}

\end{document}